\documentclass[conference=ieee,final,fontsize=10,savespace=false,savespacebib=true]{confpaper}

\usepackage{wasysym}
\usepackage{dblfloatfix}
\usepackage[bottom]{footmisc}
\usepackage{longtable}
\usepackage{supertabular,booktabs}
\usepackage{lscape}
\usepackage{rotating}
\usepackage{calc}
\usepackage{balance}
\usepackage{wrapfig}
\usepackage{afterpage}

\usepackage{threeparttable}
\newcolumntype{R}[1]{>{\raggedright\arraybackslash}p{#1}}
\newcolumntype{P}[1]{>{\raggedleft\arraybackslash}p{#1}}
\newcolumntype{C}[1]{>{\centering\arraybackslash}p{#1}}
\definecolor{Gray}{gray}{0.85}
\newcolumntype{g}{>{\columncolor{Gray}\raggedleft\let\newline\\\arraybackslash\hspace{0pt}}p{1.5em}}
\newcolumntype{w}{>{\raggedleft\let\newline\\\arraybackslash\hspace{0pt}}p{1.5em}}

\newcommand\UsrPopulationpoplocUnarkedTotal{eleven\xspace}

\newcommand\UsrPopulationextvalMenLimitTotal{30\xspace}

\newcommand\UsrPopulationextvalConsideredAddrTotal{seven\xspace}

\newcommand\UsrPopulationextvalNotDiscussedTotal{eleven\xspace}

\newcommand\UsrResearchfocusTestahypothesisMarkedTotal{seven\xspace}

\newcommand\UsrTheoryandFrameworkGTMarkedTotal{seven\xspace}
\newcommand\UsrTheoryandFrameworkGTMarkedTotalNumber{7\xspace}

\newcommand\UsrTheoryandFrameworkGTAnalyticalTotal{five\xspace}

\newcommand\UsrTheoryandFrameworkGTMiddlegroundTotal{one\xspace}

\newcommand\UsrTheoryandFrameworkGTFullTotal{one\xspace}

\newcommand\UsrTheoryandFrameworkDSBeforeAfterTotal{two\xspace}

\newcommand\UsrTheoryandFrameworkDSBeforeTotal{one\xspace}

\newcommand\UsrTheoryandFrameworkHFUnderstandingTotal{26\xspace}

\newcommand\UsrTheoryandFrameworkHFEliminatoryTotal{one\xspace}

\newcommand\ExpPopulationpoplocItemsTotal{48\xspace}
\newcommand\ExpPopulationpoplocUnarkedTotal{24\xspace}

\newcommand\ExpPopulationextvalMenLimitTotal{24\xspace}

\newcommand\ExpPopulationextvalConsideredAddrTotal{seven\xspace}

\newcommand\ExpResearchfocusTestahypothesisMarkedTotal{eight\xspace}

\newcommand\ExpResearchfocusExploreMarkedTotal{12\xspace}

\newcommand\ExpTheoryandFrameworkGTMarkedTotal{15\xspace}

\newcommand\ExpTheoryandFrameworkGTAnalyticalTotalNumber{10\xspace}

\newcommand\ExpTheoryandFrameworkGTMiddlegroundTotal{three\xspace}

\newcommand\ExpTheoryandFrameworkGTFullTotal{two\xspace}

\newcommand\ExpTheoryandFrameworkDSBeforeAfterTotal{three\xspace}

\newcommand\ExpTheoryandFrameworkDSBeforeTotal{four\xspace}

\newcommand\ExpTheoryandFrameworkHFUnderstandingTotal{20\xspace}

\newcommand\ExpTheoryandFrameworkHFEliminatoryTotal{eleven\xspace}

\makeatletter
\def\cdfootnote{\xdef\@thefnmark{$*$}\@footnotetext}
\def\kkfootnote{\xdef\@thefnmark{${\textdagger}$}\@footnotetext}
\def\kbfootnote{\xdef\@thefnmark{${\ddagger}$}\@footnotetext}
\def\tffootnote{\xdef\@thefnmark{$\S$}\@footnotetext}
\makeatother

\newcommand{\notext}[1]{}

\title{Human Factors in Security Research: \\ Lessons Learned from 2008-2018}
\author{Mannat Kaur, Michel van Eeten, Marijn Janssen, Kevin Borgolte, and Tobias Fiebig\\\{M.Kaur, M.J.G.vanEeten, M.F.W.H.A.Janssen, K.Borgolte, T.Fiebig\}@tudelft.nl \\ TU Delft}

\RequirePackage{listings} 
\RequirePackage{color} 

\begin{document}

\abstract{
Instead of only considering technology, computer security research now strives to also take into account the \emph{human factor} by studying regular users and, to a lesser extent, experts like operators and developers of systems.
We focus our analysis on the research on the crucial population of experts, whose human errors can impact many systems at once, and compare it to research on regular users.
To understand how far we advanced in the area of human factors, how the field can further mature, and to provide a point of reference for researchers new to this field, we analyzed the past decade of human factors research in security and privacy, identifying 557 relevant publications.
Of these, we found 48 publications focused on expert users and analyzed all in depth.
For additional insights, we compare them to a stratified sample of 48 end-user studies.

In this paper we investigate:~
\begin{enumerate*}[label=\emph{(\roman*)}, itemjoin={{; }}, itemjoin*={{; and }}, after={{. }}]
	\item The perspective on human factors, and how we can learn from safety science
	\item How and who are the participants recruited, and how this---as we find---creates a western-centric perspective
	\item Research objectives, and how to align these with the chosen research methods
	\item How theories can be used to increase rigor in the communities scientific work, including limitations to the use of Grounded Theory, which is often incompletely applied
	\item How researchers handle ethical implications, and what we can do to account for them more consistently
\end{enumerate*}
Although our literature review has limitations, new insights were revealed and avenues for further research identified.
}

\maketitle

\section{Introduction}

Traditionally, computer security concerned itself with understanding the technical properties of systems and networks in order to guarantee confidentiality, integrity, and availability.
With the rapid societal adoption of computer systems over the past decades, researchers identified new security and privacy issues, stemming from the interaction between users and systems.
This gave rise to the study of human factors.
In this context, most research emerged as part of either
\begin{enumerate*}[label=\emph{(\roman*)}, itemjoin={{, }}, itemjoin*={{, or }}, after={{. }}]
  \item designing secure and usable systems
  \item empirical studies of problems around how users interact with systems and services
\end{enumerate*}

The first approach is design-oriented.
Data on users is collected as part of a design process or an evaluation of an existing system.
Think of eliciting user requirements or validating the performance of the designed system with actual users.
Organized around the concepts of usability and human-computer-interaction, researchers have worked on creating secure \emph{and} usable systems.
A classic example is Whitten and Tygar’s usability analysis of GPG's user interface~\cite{whitten1999johnny}.

The second approach to human factor research is descriptive, i.e., focuses on soliciting empirical data on users' \emph{behavior}.
Users are studied in various security-relevant contexts, to learn more about their behavior in general, not directly tied to the design process of a specific system or service.
This work typically relies on experiments, surveys, and observational data.
For example, Krombholz et al. conducted experiments to see if system operators are able to properly deploy HTTPS~\cite{Krombholz2017}
Dietrich et al. surveyed system operators' perspectives on security misconfigurations~\cite{Dietrich2018},
and Golla et al.\@ collected behavioral data around password reuse notifications from a production system~\cite{golla2018site}.

Large-scale security incidents are often traced back to human error, like mistakes or forgetfulness~\cite{Au2019, Vizard2019, Hiskey2019, Rashid2019, Braue2019}.
The status quo approach to managing the human factors in cybersecurity says that humans are the weakest-link in security.
Numerous efforts are made to eliminate, control or train the human factor in order to improve security~\cite{zimmermann2019}.
Such human-factors studies have been a steady presence in the main security and privacy venues in the recent past.
In fact, the portion of published work that includes user research has more than doubled over the past years.
But what all constitutes human factor research? How has the field evolved in the recent past and what are the research gaps that still exist? 

We focus our analysis of the state of the art on one critical population in human factors research: experts. By experts we mean the people who develop, build and run systems (a more precise taxonomy is developed in \Fref{sec:method}. 
Their errors can be highly consequential, as they can impact many systems at once or impact critical systems, on which many users and organizations rely. 
To better locate the studies on experts in the overall field of human-factors research, we also analyze a sample of end-user studies and compare both types of research throughout our paper.

Investigating human factors is not one of our community's traditional areas of expertise and other disciplines have been studying human factors since much longer.
Research in these domains have shown that the "weakest-link" approach is not the only way to manage the human factor~\cite{dekker2010system}. 
This provides an opportunity for our community to learn from more mature areas which have investigated human behavior for many decades.
Valuable lessons can be gained from safety science, which is an engineering-dominated discipline that aims at preventing adverse outcomes, similar to security, but with a substantially longer track record of incorporating human factors (discussed further in Section \ref{sec:perspective} and Figure \ref{fig:hf-stages}).

Research that crosses over from computer security to these other fields is still rare. 
Examples include
Egelman and Peer, who developed a Security Behaviour Intentions Scale (SeBIS) that measures users' attitudes towards various computer security tasks \cite{Egelman2015},
and, H\'{a}mornik and Krasznay, who developed a research framework linking computer-supported collaborative-work (CSCW) and team cognition in high risk situations to better understand teamwork in security operation centers (SOCs)~\cite{Harmonik2018}.
However, our community can build more systematically upon the work in social and safety sciences to leverage their theories and methods and to increase scientific rigor and generalizability, which are issues plaguing ``security as a science'' as pointed out by Herley et al.~\cite{herley2017sok}.

Our research question for this paper is: \textit{What is the state of the art of human factors research on experts in the computer security domain and what lessons can we learn for future research?}
We analyze the current state of human factor research in computer security to serve as a point of reference for new and established researchers alike.
We review the literature on six aspects and answer the following sub questions:
\begin{enumerate}[nolistsep]
    \item What insights from the safety science domain can be applied to the computer security domain?
    \item What sample populations are being investigated and in what ways are they recruited?
    \item What is the objective of the research?
    \item What are the research methods that the researchers are using to study users?
    \item What kind of theories, if any, are the researchers using and how?
    \item How did the researchers evaluate the ethics of their work?
\end{enumerate}
Whether it be design-oriented or descriptive work, we first want to account for all the human factors research and create an overview of the state of the art.
We scope our work by identifying papers that directly involve people in the main computer security venues from the past ten years.
We end up with 557 publications in total.
Then we group these papers based on the population they investigate, i.e., whether the paper deals with end users or expert users.
End users are the focus of 91.4\% of the papers, while expert-user studies make up a mere 8.6\% of the publications. 
We systematize the state of the art for the expert user group and analyze all of the 48 papers in depth.
For comparison, we also review a sample of end user papers.
Since we cannot analyze all the 509 papers in depth, we have chosen a stratified random sample of 48 end-user publications.
Subsequently, for each category, we provide recommendations on how the field can further mature.
Our key contributions are:
\begin{enumerate}
	\item We find that expert users, different from end users, are an understudied population in terms of human factors in computer security, even though their behaviors and mistakes have higher stakes and more severe consequences.
	\item We also find that papers on expert users commonly treat human error as a root cause to be removed from the system. This is an opportunity to learn from safety science research where the focus is to better understand the human factor and in turn build resilient systems that produce the desired outcome despite human error.
	\item Similar to other fields, we find that the recruitment of study participants is dominated by convenience sampling and has a geographical bias towards the US and Europe, which threatens international generalizability.
	\item Most human factors research (78.12\%) lacks theory to inform research design and causal reasoning. Even research that utilizes Grounded Theory regularly stops before the step of building a theory from the empirical findings. The absence of theory limits the generalizability of the findings beyond the context of the study itself.
\end{enumerate}
In addition, we release the annotated version of our list with 557 human-factors studies in security and privacy as open data along with this paper.

\begin{figure*}[tb!]
\centering
\includegraphics[width=1.6\columnwidth,trim={0cm 0 0cm 0},clip]{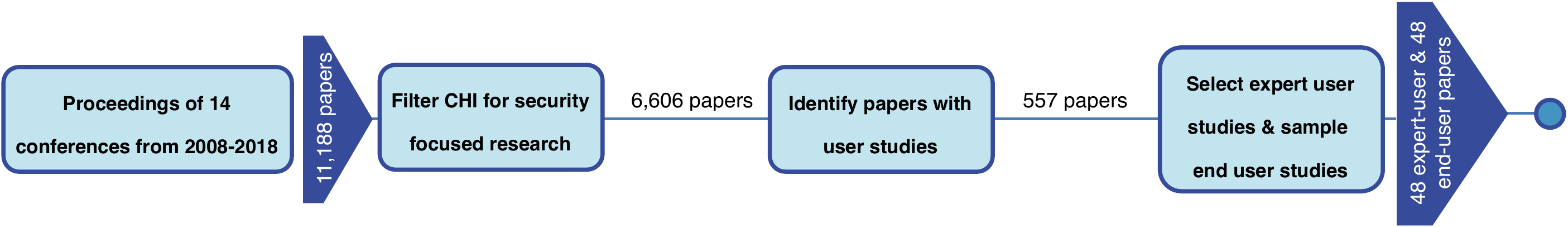}
\caption{Visual overview of the search process.}
\label{fig:analysis}
\end{figure*}

\section{Paper Search and Selection Process}
\label{sec:method}

In this section, we present our approach to building a representative corpus of human-factor studies and 
present the analysis criteria for our subsequent analysis.

\subsection{Search Process}
We perform a systematic literature review (SLR), inspired by Kitchenham et al.~\cite{Kitchenham2009}, to identify publications on human factors in computer security. \Fref{fig:analysis} presents an overview of the major steps of our selection and filtering process.

\subsubsection*{Inclusion and exclusion criteria}
First, we perform an comprehensive search across the most prominent computer security venues.
Specifically, we selected all top-tier (Tier 1 and 2) computer security and network operations venues, based on a common ranking.\footnote{See \url{http://faculty.cs.tamu.edu/guofei/sec_conf_stat.htm}. The list was updated after we had finished our search and now has 18 venues in tier 1 and 2, instead of 17. We acknowledge that this list does not constitute an `official' ranking, yet is commonly used within the community, even though it is critically acclaimed by some for its selection of venues.}
We consider venues that purely focus on cryptography, like Crypto or TCC, as out of scope.
Furthermore, we did not include workshops, as for example USEC, as our goal is comprehensiveness, not completeness, even though they also publish a sizeable number of human factors related security work.
We do, however, add the Symposium on Usable Privacy and Security (SOUPS, Tier 3) to this list, as it is a major venue for usable security.
We also add ACM CHI, the Conference on Human Factors in Computing Systems, which is the ``premier international conference of Human-Computer Interaction.''
Overall, we reviewed the proceedings of 14 conferences from 2008 to 2018, resulting in an initial set of 11,188 papers.
We specifically chose to limit our search scope to this period because we want to investigate the current development of the field.
Also, we do not present search keywords because all the papers were selected from these venues.
Next, we reduce the set of papers to 6,606 papers, by only including papers from ACM CHI that are presented in sessions related to security, privacy, passwords, and authentication.
We acknowledge that this might lead to individual papers within CHI being omitted.
However, to set a reasonable scope for the literature review, this limitation was necessary.
We read the title and abstracts of all 6,606 papers to identify those that investigate human factors.
The key criterion is the direct involvement of humans in the research, both online and offline, to study \emph{behavior} or \emph{actions}.
This means we exclude papers that only perform large-scale internet measurements to understand user behavior.
Furthermore, we also exclude papers that do not contain full-fledged user studies, for example
Czyz et al., who perform an unstructured inquiry via email to identify root causes of IPv4/IPv6 security misconfigurations~\cite{czyz2016}.
Ultimately, this selection process took over 1,300 working hours.
We identified 557 papers on human factors in security (see \Fref{tab:conf-sum} for an overview).
We will publish our annotated literature database along with this paper.
Finally, we discuss the limitations of our search process in Section \ref{sec:lim}.

\subsection{End Users and Expert Users: A Taxonomy}
\label{ssec:taxonomy}
Here, we developed a taxonomy of what users are being studied to explain how we arrive at the distinction between ``expert users'' and ``end users'' which we use for segmenting and sampling the literature in the next section.
This categorization is based on the task that is being studied, rather than on inherent properties of the user who participates in the study.
If the task is part of expert work, then we include the study as an expert study.
This means that even when a person which could be classified as a ``security professional'', if this person is participating in a study of an email user interface, this participation would not make the study an ``expert-user'' study.
Similarly, studies that subject ``non-experts'' to expert tasks, like vulnerability discovery in the case of Votipka et al.~\cite{Votipka2018}, do not become ``end user'' studies because of the utilized population.

\begin{itemize}
	\item \textbf{Expert Users (Building Systems):} Expert users are those that build and run systems. Contrary to end users, they directly influence the security of systems \emph{used by someone else}. Studies in this category deal with tools exclusively used in this context, that is, the process of providing a system for a third party (end users), and the processes and behaviors associated with the process of running these systems.
	\begin{itemize}
		\item \textbf{Developers:} Developers write the code for end user visible applications as well as the back-end systems that make these tools function. A common sub-distinction for developers is frontend vs. backend developers.
		\begin{itemize}
			\item \textbf{Frontend Developers: } Developers who work on the user interface of applications.
			\item \textbf{Backend Developers:} Developers who work on the backend, that is, they create application programming interfaces (APIs) that can be used by the frontend to handle database interactions and business logic.
			\item \textbf{Fullstack Developers:} Developers versed in frontend and backend tasks.
		\end{itemize}
		\item \textbf{Operators:} Operators are those running systems. They deploy and update software created by developers, configure network equipment, and provide \emph{services} to users. We note, that this distinction is difficult. On the one hand, we see that the community often utilizes ``developers'' as a covering term for everything that involves building and running a service or application, thereby covering operators. On the other hand, recent developments in how we run systems more and more merge the concept of operations and development, that is, DevOps~\cite{liabook}. Below, we provide a non-exhaustive set of examples of operators.
		\begin{itemize}
			\item \textbf{System Operators:} System operators operate systems in general, akin to fullstack developers, that is, they take care of systems from several of the following categories.
			\item \textbf{Network Operators:} Network operators deal with network infrastructure, that is, they configure network switches and routers, and are usually also in charge of designing the physical network.
			\item \textbf{Client Operators:} Client operators are among the most visible operators of an organization, as they deal with provisioning and providing patches to workstations, which are the most user-visible activities.
			\item \textbf{Help Desk Personnel:} Help desk personnel is commonly the first point of contact for users. Although help desk staff does not fall into the ``traditional'' operator categories, they often receive some operational permissions to handle common user requests.
		\end{itemize}
		\item \textbf{Security Experts:} While security professionals constitute their own class, they often overlap with other roles from development or operations. However, due to the context of our work, we detail them as a dedicated class.
		\begin{itemize}
			\item \textbf{CSIRT/SOC Workers:} Computer Security Incident Response Teams (CSIRTs) and Security Operations Center (SOC) workers handle threat intelligence feeds and incident reports received by an organization and follow up on potential threats.
			\item \textbf{Red/Blue Team Members:} Red and blue team members conduct assessments of an organization. While red teams attempt to gain access to systems as ``attackers,'' blue teams audit infrastructures to identify security issues and ``defend.''
			\item \textbf{Residential Security Experts:} Residential security experts often overlap with blue team work, and they are members of an organization who are in charge of assessing and reviewing security sensitive changes in code bases or concerning infrastructure.
		\end{itemize}
		\item \textbf{Researchers:} Some papers study computer-security researchers as their sample population. These studies account for the researchers' perspective in the computer security domain.
		\item \textbf{Computer Science (CS) students:} Many of the studies recruit computer science students as a proxy for expert users. These students have a technical background and are a convenient sample in academic research.
		\item \textbf{Others:} The remaining studies are categorized as 'other'. These include experts from various organizations such as those that develop cryptographic products~\cite{Haney2018_2}, studies that perform participant observation inside the organization~\cite{Sundaramurthy2016}, studies that include hackers~\cite{Mu2018} etc.
	\end{itemize}
		\item \textbf{End Users (Using Systems):} This group contains \emph{users} of systems. This means that this group is not involved with \emph{running} or \emph{changing} the systems they use, and they use these systems for personal---in a private and professional context---activities, such as reading or encrypting one's emails.
	\begin{itemize}
		\item \textbf{Applicable Subgroups:}
      For end users, various population slices are applicable.
      This ranges from studies of the elderly and their security behavior~\cite{frik2019privacy} to children~\cite{Lastdrager2017}, and it includes classifications of profession related subgroups, like journalists or aid workers.
      We identified the following sub-groups for our study, namely: the general public, university students/staff, specific users groups like journalists or air workers and children. 
	\end{itemize}
\end{itemize}
Improving human factors clearly requires different approaches and solutions for expert tasks compared to regular end-user tasks. 
One can design very different solutions given the stark contrast in training and competencies of experts compared to end users.
Furthermore, the stakes of individual human errors of experts are often higher. A simple error during the operation of a system of the development of software can easily affect hundreds to thousands to even millions of users.
This, in turn, may have a significant impact on how human factors need to be treated for these two different populations.

\subsection{Dataset Overview and Sampling}
The first observation we can make is that human factors research is on the rise, both in an absolute and a relative sense.
Starting at 21 papers in 2008 (3.4\% of all studies in the selected venues), the number of human factors papers rose to a total of 88 in 2018 (6.1\%).
Naturally, most of these papers appeared in SOUPS (198) and ACM CHI (133).
We do not consider all SOUPS papers because not all include user studies, that is, users were not directly involved in the research, for example, in the case of literature surveys and position papers.

From a human factors perspective, different user populations present different challenges, which also implies the need for different theories and methods.
The most important distinction we encountered across the corpus of papers is between  \emph{expert users} and end users, see \Fref{ssec:taxonomy}.
End user studies typically concern themselves with topics like interfaces used by the general population, or user behavior around widely-used technology.
In contrast to end users, expert users do have prior knowledge, training, or experience in software or hardware engineering,  networks, or systems operations, which they use to build systems.

Overall, we find that end user studies considerably outweigh expert user studies: 509 of 557 papers deal with end user (91.4\%), while only 48 papers (8.6\%) concern themselves with expert users.
The lack of human factor research on expert users is alarming. 
While numerically clearly a smaller group, the behavior of expert users typically affects more systems than just their own, thus having a comparatively larger impact on security than individual end users.
For example, system administrators making security misconfigurations can affect thousands or more regular users.
For our study, we review all the 48 expert user studied in depth.
To gain additional insights, we have also reviewed a group of end user papers. 
Since we cannot analyze all the 509 papers in depth, and the two groups are imbalanced, we have chosen a stratified random sample of 48 end-user publications.
Stratification was done by publication year, that is, we matched the distribution of expert user papers over time by randomly choosing papers from the end user group corresponding to the number of expert user papers per year.
To illustrate: since two papers on expert users appeared in 2008, we randomly selected 2 out of 19 end user publications in 2008.
We acknowledge that this might limit our view on the literature on end users.
However, given the vast body of existing literature, an exhaustive analysis is infeasible, and a stratified sample based on the temporal distribution of the expert user sample provides a reasonable trade-off between reliability and feasibility.

\begin{table*}[tb!]
\caption{Literature on human factors in security (HFS) vs. all papers, for major security venues between 2008 and 2018. For each year we list the number and share of HFS papers for that year, and how they are distributed over end users and expert users.}
\label{tab:conf-sum}
\resizebox{\textwidth}{!}{%
\begin{tabular}{lgwgwgwgwgwgwgwgwgwgwgw|gr}
\toprule
\rowcolor{white}\textbf{Conference} & \multicolumn{2}{c}{\textbf{2008}} & \multicolumn{2}{c}{\textbf{2009}} & \multicolumn{2}{c}{\textbf{2010}} & \multicolumn{2}{c}{\textbf{2011}} & \multicolumn{2}{c}{\textbf{2012}} & \multicolumn{2}{c}{\textbf{2013}} & \multicolumn{2}{c}{\textbf{2014}} & \multicolumn{2}{c}{\textbf{2015}} & \multicolumn{2}{c}{\textbf{2016}} & \multicolumn{2}{c}{\textbf{2017}} & \multicolumn{2}{c|}{\textbf{2018}} & \multicolumn{2}{c}{\textbf{Total}} \\
\rowcolor{white}&\rotatebox{90}{HFS}&\rotatebox{90}{All}&\rotatebox{90}{HFS}&\rotatebox{90}{All}&\rotatebox{90}{HFS}&\rotatebox{90}{All}&\rotatebox{90}{HFS}&\rotatebox{90}{All}&\rotatebox{90}{HFS}&\rotatebox{90}{All}&\rotatebox{90}{HFS}&\rotatebox{90}{All}&\rotatebox{90}{HFS}&\rotatebox{90}{All}&\rotatebox{90}{HFS}&\rotatebox{90}{All}&\rotatebox{90}{HFS}&\rotatebox{90}{All}&\rotatebox{90}{HFS}&\rotatebox{90}{All}&\rotatebox{90}{HFS}&\rotatebox{90}{All}& \rotatebox{90}{HFS}&\rotatebox{90}{All} \\\midrule

ACM AsiaCCS & - & 40 & 1 & 40 & 1 & 37 & 2 & 61 & 3 & 47 & 3 & 61 & 5 & 50 & - & 71 & 5 & 83 & 4 & 72 & 3 & 62 &  27 & 624 \\
ACM CCS & - & 52 & 1 & 58 & 1 & 55 & 1 & 61 & 3 & 81 & 6 & 96 & 5 & 138 & 4 & 131 & 6 & 146 & 9 & 159 & 9 & 140 &  45 & 1,117 \\
ACSAC & 1 & 45 & 1 & 48 & 1 & 42 & 2 & 41 & 3 & 45 & 1 & 40 & 2 & 47 & 4 & 47 & 3 & 47 & 4 & 47 & 8 & 60 &  30 & 509 \\
IEEE DSN & 1 & 58 & - & 64 & - & 65 & 1 & 52 & - & 51 & - & 68 & - & 57 & - & 50 & 1 & 58 & - & 55 & - & 62 &  3 & 640 \\
ESORICS & 1 & 37 & - & 42 & - & 42 & 2 & 36 & - & 50 & - & 43 & - & 58 & 1 & 57 & - & 59 & - & 56 & 1 & 55 &  5 & 535 \\
IEEE CSF & - & 22 & - & 22 & - & 23 & - & 21 & - & 25 & - & 19 & 2 & 29 & - & 35 & - & 33 & - & 34 & - & 27 &  2 & 290 \\
IEEE S\&P & - & 28 & 1 & 26 & 1 & 34 & 2 & 34 & 2 & 40 & - & 38 & 1 & 44 & 1 & 55 & 6 & 55 & 8 & 60 & 5 & 63 &  27 & 477 \\
ACM IMC & - & 31 & - & 41 & - & 47 & 1 & 42 & - & 45 & - & 42 & - & 42 & 1 & 44 & - & 46 & - & 42 & - & 43 &  2 & 465 \\
ISOC NDSS & - & 21 & 1 & 20 & - & 24 & 1 & 28 & 1 & 46 & 5 & 50 & 2 & 55 & 3 & 50 & - & 60 & 2 & 68 & 5 & 71 &  20 & 493 \\
PETS & - & 13 & 1 & 14 & - & 16 & 1 & 15 & 2 & 16 & 1 & 13 & 4 & 16 & - & 23 & 9 & 51 & 5 & 52 & 3 & 35 &  26 & 264 \\
RAID & - & 20 & 1 & 17 & - & 24 & - & 20 & - & 18 & - & 22 & - & 22 & 1 & 28 & 1 & 21 & - & 21 & - & 32 &  3 & 245 \\
SOUPS & 10 & 12 & 14 & 15 & 14 & 16 & 15 & 15 & 14 & 14 & 15 & 15 & 21 & 21 & 21 & 22 & 22 & 22 & 26 & 26 & 26 & 28 &  198 & 206 \\
USENIX Security & - & 27 & 1 & 26 & 2 & 30 & 1 & 35 & 4 & 43 & 2 & 45 & 4 & 67 & 5 & 67 & 1 & 72 & 5 & 85 & 11 & 100 &  36 & 597 \\
ACM CHI & 8 & 218 & 7 & 277 & 10 & 302 & 15 & 409 & 7 & 369 & 5 & 392 & 10 & 465 & 18 & 484 & 21 & 545 & 15 & 600 & 17 & 665 &  133 & 4,726 \\
\midrule
\rowcolor{white}\textbf{HFS papers (\%):} & \multicolumn{2}{r}{21  (3.4\%)}& \multicolumn{2}{r}{29  (4.1\%)}& \multicolumn{2}{r}{30  (4.0\%)}& \multicolumn{2}{r}{44  (5.1\%)}& \multicolumn{2}{r}{39  (4.4\%)}& \multicolumn{2}{r}{38  (4.0\%)}& \multicolumn{2}{r}{56  (5.0\%)}& \multicolumn{2}{r}{59  (5.1\%)}& \multicolumn{2}{r}{75  (5.8\%)}& \multicolumn{2}{r}{78  (5.7\%)}& \multicolumn{2}{r|}{88  (6.1\%)}&   557  &         (5.0\%) \\

\rowcolor{white}\emph{~End Users}   &              19 &                 &              26 &                 &              29 &                 &              42 &                 &              38 &                 &              36 &                 &              53 &                 &              54 &                 &              71 &                 &              70 &                 &              70 &                 &               509 &                 \\
\rowcolor{white}\emph{~Experts}     &               2 &                 &               3 &                 &               1 &                 &               2 &                 &               1 &                 &               2 &                 &               3 &                 &               5 &                 &               4 &                 &              8 &                 &              18 &                 &                48 &                 \\ \bottomrule
\end{tabular}%
}
\end{table*}

\subsection{Analysis Criteria}
\label{ssec:analysiscriteria}
We analyze the literature on six aspects: The general perspective on human factors, the sample used in the study, how this sample has been recruited, the research objective, how the authors utilized existing theory or methodology to inform their research design, and, ethical considerations.

\noindent\textbf{Perspective on Human Factors:}
In safety science, decades of research has fundamentally changed the understanding of human factors and human error.
The current perspective of safety science sees human error not as avoidable, but as a property of human work, which systems have to account for to ensure safe operations in the presence of error.
This evolution is summarized in five major stages, which we discuss in the next Section (see also \Fref{fig:hf-stages}).
We analyze how research on human factors in computer security compares to this understanding from safety science.

\noindent\textbf{Study Population:}
Naturally, we also investigate the samples used in contemporary research.
We identify the major types of populations based on how authors describe their samples.
This taxonomy is discussed in Subsection \ref{ssec:taxonomy}.
For end users, these groups are ``children'' (minors), the general public, university-affiliated users (like staff and students), and other specific user groups.
For example, some studies focus on users with social disorders~\cite{Neupane2018}, South Asian women~\cite{Sambasivan2018}, or users in relationships~\cite{Park2018, Lebeck2018}.
If no information about the sample population is available, then we mark the population as ``N/A.''

For expert users, we broadly differentiate between developers, operators, security professionals, researchers and computer science students.
Each of these categories is explained, along with the subdivisions, in Subsection \ref{ssec:taxonomy}.
When studies compare expert users to end users, a confusing edge case, we classify them as ``end users'' among the expert-user publications.
The remaining studies on expert users we categorize as ``other''.
This includes studies where a set of different experts from a specific organization or set of organizations are involved~\cite{Haney2018_2,Sundaramurthy2015,Sundaramurthy2016}, technical experts and end-users are recruited for a comparison study and their expertise is not specified~\cite{Skirpan2018}, or a study with hackers and testers~\cite{Votipka2018}.
As an additional point of reference, we also identify the geographic region from where samples are collected, and where the authors themselves are located.

For our analysis, we only consider the broad categories and not the subdivisions. For example, we talk about frontend, backend and fullstack developers in our taxonomy. During the analysis however, we broadly classify all these under the developer category.

\noindent\textbf{Recruitment:}
We analyze how researchers recruited participants.
For end users, we consider recruitment via crowd-sourcing platforms (like Amazon Mechanical Turk, or other crowdsourcing platforms like CrowdFlower~\cite{Chanchary2015, Angulo2015}), recruitment in the local city, at the local university, via personal contacts, a recruitment agency, social media, or ``other'' online channels.
For example, these online channels can be Craigslist~\cite{Gao2018, Zou2018, Ur2015}, Sampling Survey International~\cite{Redmiles2016},
or simply using other non-crowdsourcing platforms online, like message boards.

Similarly, for expert users, we distinguish between crowd-sourcing platforms, GitHub, the local university, personal contacts, industry contacts or industry organizations, social media, and ``other'' methods.
Other recruitment methods include recruitment at a conference~\cite{Haney2018_2, Krombholz2017}, public bug bounty data~\cite{Votipka2018}, or establishing an online brand~\cite{Dietrich2018}.
In case the authors fail to provide sufficient recruitment information, we mark it as ``N/A.''

\noindent\textbf{Research Objective:}
Concerning the research objective, we distinguish between studies that
\begin{enumerate*}[label=\emph{(\roman*)}, itemjoin={{, }}, itemjoin*={{, and }}, after={{. }}]
\item evaluate an artifact
  \item test hypotheses
  \item perform general exploratory research
  \item focus on gathering users' perspective on specific issues
\end{enumerate*}
Moreover, if authors evaluate an artifact, we check if they used an existing research framework for building and evaluating the artifacts, such as design science, and whether they include user feedback or evaluation results in the design of their artifact.

\noindent\textbf{Research Method:}
We systematize \emph{how} researchers conduct their studies by distinguishing between studies performed in a local laboratory, online, using interviews, surveys (including questionnaires), focus groups, or using observations. One study can have multiple research methods.

\noindent\textbf{Theory/Framework:}
Regarding the use of theories and frameworks, we scrutinize how authors use existing scientific theories.
Specifically, we investigate if they
\begin{enumerate*}[label=\emph{(\roman*)}, itemjoin={{, }}, itemjoin*={{, or }}, after={{. }}]
  \item use an existing theory to inform their research design or set out to validate and improve upon an existing theory
  \item mention an existing theory in the context of their results and observations
  \item neither use or mention a theory
\end{enumerate*}
In our analysis, we identified three major theories (Mental Models, Sensemaking, and the Theory of Reasoned Action). 
Furthermore, we closely study work that claims to use grounded theory, which is a methodology that creates theory through a systematic process of data gathering and interpretation.
Correspondingly, we do not mix it with the use of existing theories, but add an additional category, in which we explore whether authors
\begin{enumerate*}[label=\emph{(\roman*)}, itemjoin={{, }}, itemjoin*={{, or }}, after={{. }}]
  \item focus on the methodological parts of grounded theory to obtain observational results and generally inform their qualitative data analysis
  \item use a ``middle ground'' approach~\cite{sekaran2016research}, in which they contrast their findings with existing theories
  \item perform grounded theory to construct a new theory or model
\end{enumerate*}

\noindent\textbf{Ethics:}
Finally, we study whether the authors considered the ethical implications of their work.
We distinguish between
\begin{enumerate*}[label=\emph{(\roman*)}, itemjoin={{, }}, itemjoin*={{, and }}, after={{. }}]
  \item authors that obtained full clearance from their ethical review board
  \item those who discuss the ethical implications but did not or could not obtain a clearance from a review board, e.g., because their institution does not have one,
  \item authors who do not discuss the ethical implications of their work
\end{enumerate*}

\section{Perspective on Human Factors}
\label{sec:perspective}

\begin{figure*}[t!]
\centering
\includegraphics[width=1\textwidth]{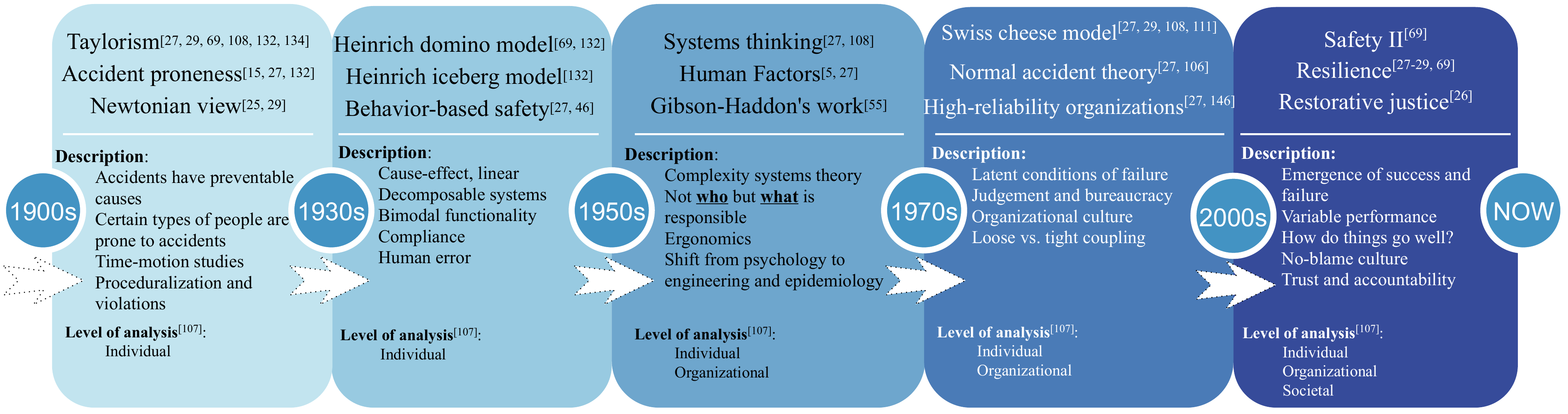}
\caption{Overview of the development of human factors research in safety science~\cite{dekker2019foundations}. Note how over time, the perspective shifted away from individual responsibility to, first, organizational factors, and finally to a societal perspective.
Ultimately, the focus is no longer on how to prevent human error, but instead, on how to facilitate proper resilient operations.}
\label{fig:hf-stages}
\end{figure*}

Research on human factors has emerged in safety science decades earlier than in computer security, which bears the question: What can we learn from safety science?
In this section, we first present the safety science perspective accompanied with a visual aid (\Fref{fig:hf-stages}).
We then present the computer security perspective on human factors research and what we observed in our literature review.
This is followed by a discussion on the synthesis of these two perspectives.
We discuss what we can learn from other domains and also insights that cannot be directly applied.
Finally, we present our key observations and recommendations at the end of the section.

\nocite{taylor1911, dekker2016, Addingtoolbox, dekker2019foundations, swuste2010safety, Hollnagel2013, dekker2019foundations, swuste2010safety, burnham2010accident, dekker2016, dekker2010system, van1999risks, swuste2010safety, Hollnagel2013, swuste2010safety, dekker2019foundations, frederick2000blame, Addingtoolbox, dekker2019foundations, dekker2019foundations, adams2006layman, guarnieri1992landmarks, dekker2016, Addingtoolbox, dekker2019foundations, reason1998achieving, dekker2019foundations, perrow1984normal, dekker2019foundations, weick1987organizational, Hollnagel2013, dekker2016, dekker2019foundations, Hollnagel2013, dekker2008resilience, dekker2018just}


\subsection{Safety Science Perspective}
\label{ssec:ssperspective}
Initially, the term \emph{human factors} described the application of scientific knowledge, concepts, models, and theories derived from social science disciplines, such as psychology, towards improving operational efficiency and reducing the human errors that led to accidents~\cite{adams2006layman}.
This early literature on human factors and human error has since gone through five major stages of development in the past century (\Fref{fig:hf-stages}).

For the first half of the 20\textsuperscript{th} century, the core ideas were that certain people are prone to accidents and  accidents are preventable by taking away the causes, for example, enforcing compliance with rules.
These ideas developed further and gave rise to the concepts of decomposable systems (a linear model where cause and effect is visible and wherein the system can be decomposed meaningfully into its parts and rearranged again into a whole) and bi-modal functionality (the components of the system can be in one of two modes of operation - either functioning correctly or not).
These two concepts led to the assumption that every failure has a root-cause and if we can find this root-cause, we can fix it and ensure safety.
In this case, the analysis is centered around the individual responsible for ``human error'' or failure.

During the second half of the century, the systems perspective emerged, as did the label of ``human-factors research.''
This changed the narrative from \emph{who} is responsible to \emph{what} is responsible, shifting the focus on the latent conditions behind failure.
The analysis now included both individual and organizational aspects.

Since the 2000s, the safety science domain has witnessed another shift in paradigm. 
The new perspective on safety is known as Safety-II.
The Safety-II approach takes into account that what is responsible for success.
Instead of creating the best way for people to comply, researchers take a step back to understand people and their variable performance in safety or security-critical operational environments.
The Safety-II approach does not replace the traditional approach to safety that has developed over the decades.
It is a complementary approach with a focus on proactive safety management.
In addition, we also see a shift in the way in which we deal with human error. 
Restorative Justice is an approach that focuses on repairing the harm through accountability and learning intead of responsibility and blame.
Also included now are the societal parameters in the analysis of human factors.

A key take-away from the contemporary perspective is that trying to \emph{eliminate} the human factor to build safe and secure systems is not the only way to improve safety.
It is also important to understand \emph{why systems do not fail}, in daily operations as well as in the presence of human error, and understand how the human factor contributes to \emph{success}.

\subsection{Computer Security Perspective}

We know that people are considered the weakest-link in computer security and many large-scale incidents/breaches are often blamed on human error.
This holds true for both end users and expert users.
Currently the most common solutions to this human problem are to eliminate the human-in-the-loop, training and education, compliance via policies and root-cause analysis (reactive security)~\cite{zimmermann2019}.
When mistakes occur, the route to security is to eliminate these mistakes by adding automation, protocols, or standards, thereby often even introducing new challenges for the secure operation of systems~\cite{Dietrich2018}.
For example, when an automated system behaves differently than expected by its operator.
This negative impact of automation is known as ``automation surprise''.

To be able to classify human factors in computer security literature more easily, we condense the perspective of safety science in the following way:
a) eliminating the human factor, that is, preventing errors, b) investigating the human factor to understand what makes things go the way they go and c) neither of these perspectives is identified.
We marked papers trying to ``eliminate'' the human factors in column ``HF Persp.'' with a \(\CIRCLE\) and those that are trying to understand the real-world phenomenon with a \(\Circle\) under ``Theory/Framework'' in \Fref{tab:exp-papers} and \Fref{tab:usr-papers}. If neither of this perspectives is identified, the paper is unmarked.

\begin{figure}[tb!]
    \centering
       \begin{subfigure}[t]{0.20\textwidth}
        \centering
        \includegraphics[width=\columnwidth,trim={0 0cm 0 0cm},clip,angle=0]{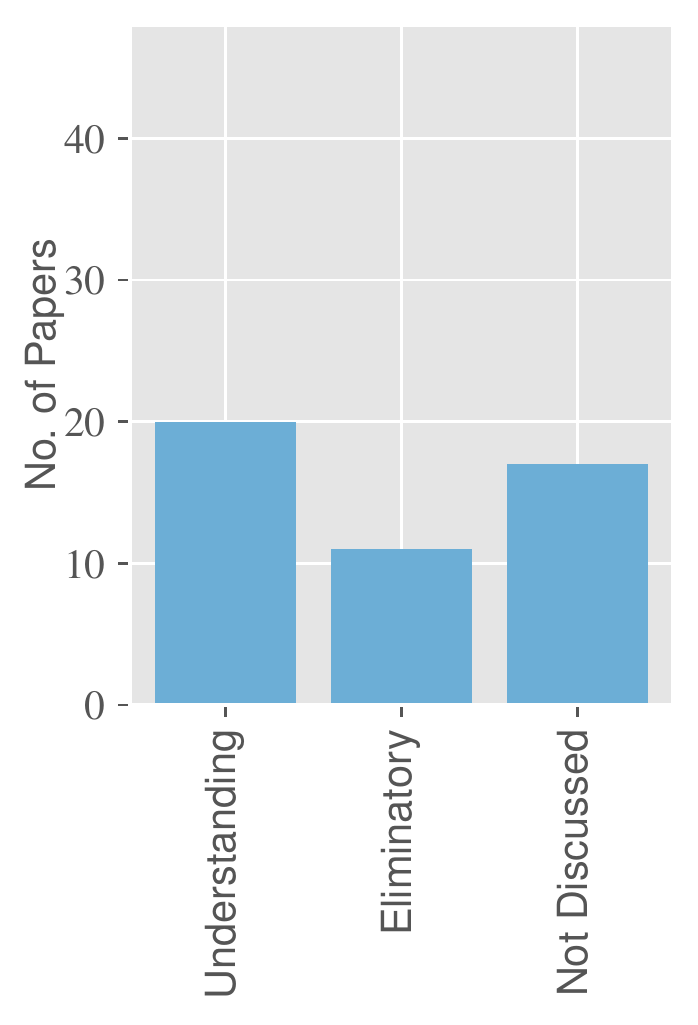}
        \caption{Expert Users}
        \label{fig:hf_perspective_exp}
    \end{subfigure}%
     ~
    \begin{subfigure}[t]{0.20\textwidth}
        \centering
        \includegraphics[width=\columnwidth,trim={0 0cm 0 0cm},clip,angle=0]{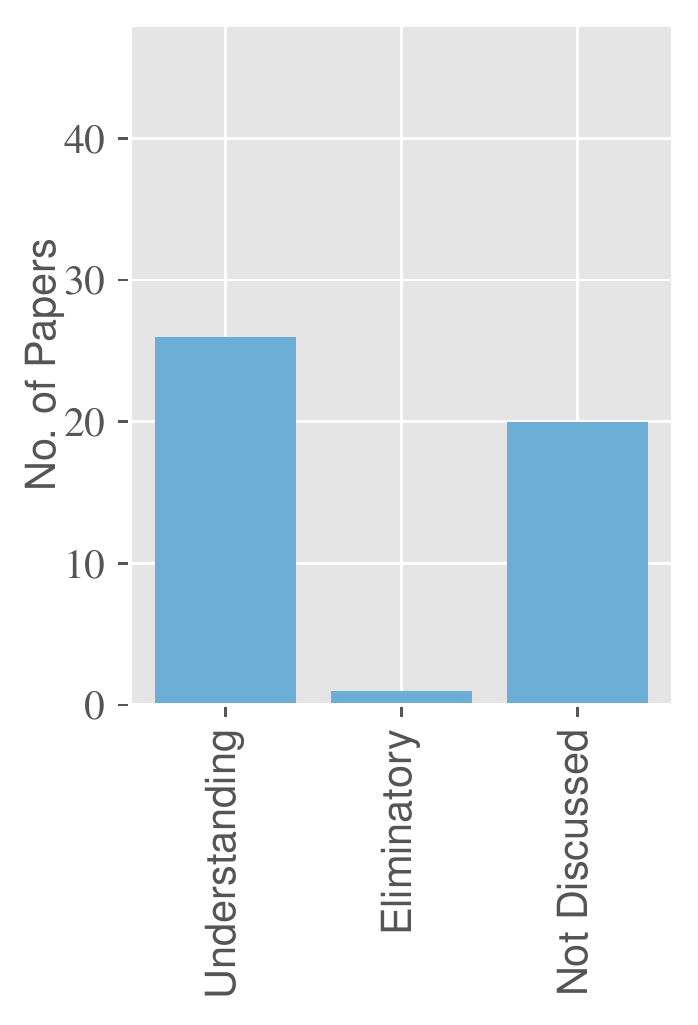}
        \caption{End Users}
        \label{fig:hf_perspective_end}
    \end{subfigure}
\caption{Overview of the perspective taken on the human factor in expert-user (\Fref{fig:hf_perspective_exp}) and end-user papers (\Fref{fig:hf_perspective_end}).
We see that the eliminatory perspective is still more prevalent in papers dealing with experts.
}
    \label{fig:hf_perspective}
\end{figure}

We find that \ExpTheoryandFrameworkHFEliminatoryTotal papers in the expert-user sample take the elimination perspective, as opposed to \UsrTheoryandFrameworkHFEliminatoryTotal paper from the end-user studies, see \Fref{fig:hf_perspective}.
These papers set the premise of error elimination by proposing complete or partial automation~\cite{Gorski2018, Nguyen2017, Krombholz2017, Falsina2015, Xie2012, Fahl2013}, emphasizing the role of policies, systems and frameworks~\cite{Kang2015, Gember-Jacobson2015} or focusing on ``human error'' as the root cause~\cite{Naiakshina2017, Fahl2014}.
We connect this observation to how we, in general, perceive experts and end users.
Professionals are expected to be knowledgeable and trained enough to not make mistakes.
Researchers, especially from the engineering field, implicitly assume experts should know better than to make certain mistakes in the operation and creation of systems.

In end-user studies, on the other hand, we often encountered the somewhat condescending notion that end users are ``the weakest link'' in systems' security.
While this notion initially points towards a similar elimination of the human factor, we observe that in practice that it is implicitly accepted that mistakes are unavoidable and that systems and environments should perform well in the face of these mistakes.
It seems that, here, the field has embraced another perspective.

We find that about half of the papers (\ExpTheoryandFrameworkHFUnderstandingTotal for experts and \UsrTheoryandFrameworkHFUnderstandingTotal for end users) take the perspective of understanding the human factors.
For expert users, this mostly consists of researching the perspectives of users~\cite{Thomas2018, Oliveira2018, Assal2018, Naiakshina2018, Wermke2018, Votipka2018, Dietrich2018, Skirpan2018, Haney2018, Gamero2017,Jaferian2014, Beckerle2013, Bauer2009}, organizational factors~\cite{Haney2018_2, Zanutto2017, Epstein2009} or both~\cite{Sundaramurthy2016, Sundaramurthy2015}.
They investigate the perceptions and attitudes of experts, such as operators, security experts, or developers, and they provide important insights into the real-life practices and processes of complex operations.
Studies on the organizational aspects are researching interactions among different stakeholders and the role of other factors such as culture.
These approaches are important because accounting for the sociotechnical factors creates a more realistic description and better equips us to deal with the operational uncertainties.

The end-user studies look at user perspectives around the security and privacy challenges of emerging AR technologies~\cite{Lebeck2018}, layman's understanding of privacy~\cite{Oates2018}, user behavior and opinions on the adoption of two-factor authentication~\cite{Colnago2018} and whether social disorders influence social engineering~\cite{Neupane2018}, to name some examples.
Just as expert-user studies, these papers provide important insights into users' perspective and real-world concerns.

\subsection{What can computer security learn from safety science?}
Despite all our efforts, serious security breaches and hacks continue to happen. 
Some contemporary research has emphasized the need to rethink the status quo and challenge the core assumptions underlying our current approach.
As discussed earlier, the traditional safety science approach sees people as the problem and assumed that systems are decomposable and bimodal. 
Zimmerman and Renaud argue that \textit{cybersecurity, currently} has similar assumptions underneath~\cite{zimmermann2019}.

Human factors research in the computer security domain is interdisciplinary. 
We are studying sociotechnical systems that are complex, unpredictable and emergent.
Due to this, the traditional assumptions (human as problem or decomposability) do not work well.

Zimmerman and Renaud propose the \textit{cybersecurity, differently} approach.
Drawing learning from other fields such as military, management and safety, they present some key principles of this new approach.
These are system emergence (vs. system decomposability), human as solution (vs. human as problem), deference to expertise (vs. policy compliance), encourage learning and communication (vs. constrain and control), focus on success (instead of solely preventing errors) and finally, balancing resistance and resilience~\cite{zimmermann2019}.

As explained before, the Safety-II approach does not replace the traditional approaches to safety but is complementary.
Similarly, the \textit{cybersecurity, differently} approach is not about radically changing the way in which we manage computer security.
It is about recognizing the sociotechnical aspects of computer security when addressing the human factors.
We must broaden our perspective on the management of human factors and explore modern principles along with traditional ones.

Finally, are there some learnings that cannot transcend from another domain to the computer security domain?
The human problem in the safety domain focuses on unintentional mistakes by well-intended humans. 
The case of intentional harm and sabotage is not addressed in this approach and is seen as a separate security concern. 
However, the computer security domain deals with both malicious actors and non-malicious human error.
Therefore, we need to remember this knowledge transfer does not address dealing with malicious actors.

\subsection{Observations and Recommendations}
\noindent\textbf{Key Observations:}
We find that past research on expert users 
mainly took an eliminatory stance.
That is, the research tried to remove the human factor from systems, for example, by introducing automation.
This approach is losing steam in safety science research, mostly based on the insight that the human factor cannot be ultimately eliminated, and, therefore, systems rooting their safety and security in this are ultimately never really safe.
Fortunately, the situation is better for end-user related work, which hardly takes the traditional perspective of eliminating the human factor and focuses on usability studies and learning the users' perspective.

\noindent\textbf{Key Recommendations:}
Given how often human error is considered to be the root cause of security vulnerabilities, we encourage the field to rethink the perspective that we take concerning human factors in computer security, especially when studying expert users.
One key takeaway is that in addition to preventing human error, we should also try to understand which behavior leads to secure outcomes, and how we can facilitate that behavior.
To accomplish this, we will have to investigate---especially expert users---in their daily interactions with the tools and issues we focus on, something that is hardly done at the moment (see \Fref{sec:ana-method}).

\begin{table*}[htb!]
\caption{Overview of expert related human factors in security research.}
\label{tab:exp-papers}
\resizebox{\textwidth}{!}{%
\scriptsize
\begin{tabular}{rrr|C{0.2em}C{0.2em}C{0.2em}C{0.2em}C{0.2em}C{0.2em}C{0.3em}|C{0.2em}C{0.2em}C{0.3em}|C{0.2em}C{0.2em}C{0.2em}C{0.2em}C{0.2em}C{0.2em}C{0.2em}C{0.3em}|C{0.2em}C{0.2em}C{0.2em}C{0.3em}|C{0.2em}C{0.2em}C{0.2em}C{0.2em}C{0.2em}C{0.3em}|C{0.2em}C{0.2em}C{0.2em}C{0.3em}|C{0.2em}C{0.2em}C{0.3em}|C{0.2em}}
\toprule

\textbf{Year} & \textbf{Idx.} & \textbf{Ref.} & \multicolumn{10}{c|}{\textbf{Sample}} & \multicolumn{8}{c|}{\textbf{Recruitment}} & \multicolumn{4}{c|}{\textbf{Res. Obj.}} & \multicolumn{6}{c|}{\textbf{Research Method}} & \multicolumn{7}{c|}{\textbf{Theory / Framework}} &  \\
 & & & \rotatebox{90}{Developers} & \rotatebox{90}{Operators} & \rotatebox{90}{Other} & \rotatebox{90}{Researchers} & \rotatebox{90}{Sec. Exp.} & \rotatebox{90}{CS Students} & \rotatebox{90}{End-Users} & \rotatebox{90}{Sample Loc.} &  \rotatebox{90}{Author Loc.} &  \rotatebox{90}{Ext. Validity} & \rotatebox{90}{MTurk} & \rotatebox{90}{GitHub} & \rotatebox{90}{Univ.} & \rotatebox{90}{Personal} & \rotatebox{90}{Industry} & \rotatebox{90}{Social M.} & \rotatebox{90}{Other} & \rotatebox{90}{N/A} & \rotatebox{90}{Eval.} & \rotatebox{90}{Hyp. Test} & \rotatebox{90}{Exploratory} & \rotatebox{90}{Perspective} & \rotatebox{90}{Lab} & \rotatebox{90}{Online} & \rotatebox{90}{Interview} & \rotatebox{90}{Survey} & \rotatebox{90}{Focus Groups} & \rotatebox{90}{Observations} & \rotatebox{90}{Mental Models} & \rotatebox{90}{Sensemaking} & \rotatebox{90}{TRA} & \rotatebox{90}{Other} & \rotatebox{90}{GT} & \rotatebox{90}{Design Eval.} & \rotatebox{90}{HF Persp.}& \rotatebox{90}{\textbf{Ethics}} \\

\textbf{2008} &~E1&\cite{Werlinger2008}&&&&&\(\CIRCLE\)&&&&\(\Circle\)&\(\LEFTcircle\)&&&&&&&\(\CIRCLE\)&&\(\CIRCLE\)&&&\(\CIRCLE\)&&&\(\RIGHTcircle\)&&&\(\RIGHTcircle\)&\(\Circle\)&&&\(\Circle\)&\(\Circle\)&\(\RIGHTcircle\)&\(\Circle\)&\(\Circle\)\\
\rowcolor{Gray}\cellcolor{white}  &~E2&\cite{Neto2008}&&\(\CIRCLE\)&&&&&&&\(\Circle\)&\(\LEFTcircle\)&&&&&&&&\(\CIRCLE\)&\(\CIRCLE\)&&&&\(\LEFTcircle\)&&&&&&&&&\(\Circle\)&&\(\RIGHTcircle\)&&\(\Circle\)\\
\midrule
\textbf{2009} &~E3&\cite{Epstein2009}&&&\(\CIRCLE\)&&&&&&\(\Circle\)&\(\Circle\)&&&&\(\CIRCLE\)&\(\CIRCLE\)&&&&&&&\(\CIRCLE\)&&&&\(\RIGHTcircle\)&&&&\(\Circle\)&&&&&\(\Circle\)&\(\Circle\)\\
\rowcolor{Gray}\cellcolor{white}  &~E4&\cite{Bauer2009}&&\(\CIRCLE\)&&&&&&&\(\Circle\)&\(\Circle\)&&&&\(\CIRCLE\)&&&&&&&&\(\CIRCLE\)&&&\(\RIGHTcircle\)&&&&\(\Circle\)&\(\Circle\)&&&&&\(\Circle\)&\(\Circle\)\\
  &~E5&\cite{Fenz2009}&&&&&\(\CIRCLE\)&&&\(\Circle\)&\(\Circle\)&\(\Circle\)&&&&&&&&\(\CIRCLE\)&\(\CIRCLE\)&&&&\(\RIGHTcircle\)&&\(\RIGHTcircle\)&&&&&&&&&\(\RIGHTcircle\)&&\(\Circle\)\\
\midrule
\rowcolor{Gray}\cellcolor{white}\textbf{2010} &~E6&\cite{Johnson2010}&&&\(\CIRCLE\)&&&&&&\(\Circle\)&\(\Circle\)&&&&&&&&\(\CIRCLE\)&\(\CIRCLE\)&&&&\(\RIGHTcircle\)&&\(\RIGHTcircle\)&\(\RIGHTcircle\)&&\(\RIGHTcircle\)&&&&&&\(\RIGHTcircle\)&&\(\Circle\)\\
\midrule
\textbf{2011} &~E7&\cite{Huang2011}&&\(\CIRCLE\)&&&&&&\(\Circle\)&\(\Circle\)&\(\Circle\)&&&\(\CIRCLE\)&&&&&&\(\CIRCLE\)&&&&\(\LEFTcircle\)&&&&&&&&&&&\(\RIGHTcircle\)&&\(\CIRCLE\)\\
\rowcolor{Gray}\cellcolor{white}  &~E8&\cite{Jaferian2011}&&&&\(\CIRCLE\)&\(\CIRCLE\)&&&\(\Circle\)&\(\Circle\)&\(\LEFTcircle\)&&&\(\CIRCLE\)&&&&\(\CIRCLE\)&&\(\CIRCLE\)&&&&\(\CIRCLE\)&&\(\CIRCLE\)&&\(\CIRCLE\)&&&&&\(\CIRCLE\)&\(\RIGHTcircle\)&\(\RIGHTcircle\)&&\(\CIRCLE\)\\
\midrule
\textbf{2012} &~E9&\cite{Xie2012}&\(\CIRCLE\)&&&&&\(\CIRCLE\)&&\(\Circle\)&\(\Circle\)&\(\LEFTcircle\)&&&\(\CIRCLE\)&\(\CIRCLE\)&&&&&\(\CIRCLE\)&&&&\(\CIRCLE\)&&\(\CIRCLE\)&&&&&\(\Circle\)&&&&\(\RIGHTcircle\)&\(\CIRCLE\)&\(\Circle\)\\
\midrule
\rowcolor{Gray}\cellcolor{white}\textbf{2013} &~E10&\cite{Fahl2013}&\(\CIRCLE\)&&&&&&&&\(\Circle\)&\(\Circle\)&&&&&&&\(\CIRCLE\)&&\(\CIRCLE\)&&&&&&\(\CIRCLE\)&&&&&&&\(\Circle\)&&\(\CIRCLE\)&\(\CIRCLE\)&\(\Circle\)\\
  &~E11&\cite{Beckerle2013}&&&\(\CIRCLE\)&&&\(\CIRCLE\)&&&\(\Circle\)&\(\Circle\)&&&\(\CIRCLE\)&&\(\CIRCLE\)&&&&\(\CIRCLE\)&\(\CIRCLE\)&&&\(\LEFTcircle\)&&&&&&&\(\Circle\)&&\(\Circle\)&&\(\CIRCLE\)&\(\Circle\)&\(\Circle\)\\
\midrule
\rowcolor{Gray}\cellcolor{white}\textbf{2014} &~E12&\cite{Jaferian2014}&&&&&\(\CIRCLE\)&&\(\CIRCLE\)&&\(\Circle\)&\(\LEFTcircle\)&\(\CIRCLE\)&&&&&&&\(\CIRCLE\)&\(\CIRCLE\)&&\(\CIRCLE\)&\(\CIRCLE\)&&\(\CIRCLE\)&\(\CIRCLE\)&&&&&&&\(\CIRCLE\)&\(\RIGHTcircle\)&\(\CIRCLE\)&\(\Circle\)&\(\Circle\)\\
  &~E13&\cite{Oliveira2014}&\(\CIRCLE\)&&&&&&&\(\Circle\)&\(\Circle\)&\(\Circle\)&&&\(\CIRCLE\)&&\(\CIRCLE\)&&&&&\(\CIRCLE\)&&&&&&\(\CIRCLE\)&&&&&&\(\CIRCLE\)&&&\(\Circle\)&\(\CIRCLE\)\\
\rowcolor{Gray}\cellcolor{white}  &~E14&\cite{Fahl2014}&&\(\CIRCLE\)&&&&&&&\(\Circle\)&\(\Circle\)&&&&&&&\(\CIRCLE\)&&&&&\(\CIRCLE\)&&&&\(\RIGHTcircle\)&&&&\(\Circle\)&&\(\Circle\)&&&\(\CIRCLE\)&\(\RIGHTcircle\)\\
\midrule
\textbf{2015} &~E15&\cite{Sundaramurthy2015}&&&\(\CIRCLE\)&&&&&\(\Circle\)&\(\Circle\)&\(\CIRCLE\)&&&&&&&&\(\CIRCLE\)&&&\(\CIRCLE\)&\(\CIRCLE\)&&&\(\RIGHTcircle\)&&&\(\RIGHTcircle\)&&&&\(\CIRCLE\)&\(\CIRCLE\)&&\(\Circle\)&\(\CIRCLE\)\\
\rowcolor{Gray}\cellcolor{white}  &~E16&\cite{Ion2015}&&&&&\(\CIRCLE\)&&\(\CIRCLE\)&\(\CIRCLE\)&\(\Circle\)&\(\LEFTcircle\)&\(\CIRCLE\)&&&\(\CIRCLE\)&&\(\CIRCLE\)&\(\CIRCLE\)&&&&&\(\CIRCLE\)&&&\(\CIRCLE\)&\(\CIRCLE\)&&&\(\Circle\)&\(\Circle\)&&&&&&\(\Circle\)\\
  &~E17&\cite{Kang2015}&&&\(\CIRCLE\)&&&\(\CIRCLE\)&\(\CIRCLE\)&\(\Circle\)&\(\Circle\)&\(\LEFTcircle\)&&&\(\CIRCLE\)&\(\CIRCLE\)&&&\(\CIRCLE\)&&&&&\(\CIRCLE\)&&&\(\RIGHTcircle\)&&&&\(\CIRCLE\)&&&&\(\Circle\)&&\(\CIRCLE\)&\(\Circle\)\\
\rowcolor{Gray}\cellcolor{white}  &~E18&\cite{Gember-Jacobson2015}&&\(\CIRCLE\)&&&&&&\(\Circle\)&\(\Circle\)&\(\LEFTcircle\)&&&\(\CIRCLE\)&&\(\CIRCLE\)&&\(\CIRCLE\)&&&&&\(\CIRCLE\)&&&&\(\CIRCLE\)&&&&&&\(\Circle\)&&&\(\CIRCLE\)&\(\Circle\)\\
  &~E19&\cite{Falsina2015}&\(\CIRCLE\)&&&&&&&&\(\Circle\)&\(\Circle\)&&&&\(\CIRCLE\)&&&\(\CIRCLE\)&&\(\CIRCLE\)&&&&&\(\RIGHTcircle\)&&&&&&&&\(\Circle\)&&\(\RIGHTcircle\)&\(\CIRCLE\)&\(\CIRCLE\)\\
\midrule
\rowcolor{Gray}\cellcolor{white}\textbf{2016} &~E20&\cite{DeLuca2016}&&&&&\(\CIRCLE\)&&\(\CIRCLE\)&\(\Circle\)&\(\Circle\)&\(\LEFTcircle\)&&&&&&&\(\CIRCLE\)&\(\CIRCLE\)&&&\(\CIRCLE\)&\(\CIRCLE\)&&&\(\CIRCLE\)&\(\CIRCLE\)&&&&&\(\Circle\)&\(\Circle\)&&&\(\Circle\)&\(\Circle\)\\
  &~E21&\cite{Sundaramurthy2016}&&&\(\CIRCLE\)&&&&&\(\Circle\)&\(\Circle\)&\(\LEFTcircle\)&&&&&&&&\(\CIRCLE\)&&&\(\CIRCLE\)&\(\CIRCLE\)&&&&&&\(\RIGHTcircle\)&&&&\(\CIRCLE\)&\(\RIGHTcircle\)&&\(\Circle\)&\(\CIRCLE\)\\
\rowcolor{Gray}\cellcolor{white}  &~E22&\cite{Yakdan2016}&&&&&\(\CIRCLE\)&\(\CIRCLE\)&&\(\Circle\)&\(\Circle\)&\(\Circle\)&&&\(\CIRCLE\)&&\(\CIRCLE\)&&&&\(\CIRCLE\)&&&&&\(\LEFTcircle\)&&\(\LEFTcircle\)&&&&\(\Circle\)&&&&\(\RIGHTcircle\)&&\(\Circle\)\\
  &~E23&\cite{Acar2016}&\(\CIRCLE\)&&&&&\(\CIRCLE\)&&\(\Circle\)&\(\Circle\)&\(\LEFTcircle\)&&&\(\CIRCLE\)&&&\(\CIRCLE\)&\(\CIRCLE\)&&&&\(\CIRCLE\)&\(\CIRCLE\)&\(\CIRCLE\)&&\(\CIRCLE\)&\(\CIRCLE\)&&&&\(\Circle\)&&\(\Circle\)&&&&\(\CIRCLE\)\\
\midrule
\rowcolor{Gray}\cellcolor{white}\textbf{2017} &~E24&\cite{Gallagher2017}&&&\(\CIRCLE\)&&&&\(\CIRCLE\)&\(\Circle\)&\(\Circle\)&\(\LEFTcircle\)&&&\(\CIRCLE\)&&&\(\CIRCLE\)&\(\CIRCLE\)&&&&&\(\CIRCLE\)&&&\(\CIRCLE\)&&&&\(\CIRCLE\)&&&&\(\Circle\)&&&\(\CIRCLE\)\\
  &~E25&\cite{Acar2017}&\(\CIRCLE\)&&&&&&&&\(\Circle\)&\(\LEFTcircle\)&&\(\CIRCLE\)&&&&&&&&\(\CIRCLE\)&\(\CIRCLE\)&&&\(\LEFTcircle\)&&\(\LEFTcircle\)&&&&\(\Circle\)&&\(\Circle\)&&&&\(\CIRCLE\)\\
\rowcolor{Gray}\cellcolor{white}  &~E26&\cite{Krombholz2017}&&&&&\(\CIRCLE\)&\(\CIRCLE\)&&\(\Circle\)&\(\Circle\)&\(\CIRCLE\)&&&\(\CIRCLE\)&&\(\CIRCLE\)&&\(\CIRCLE\)&&&&&\(\CIRCLE\)&\(\CIRCLE\)&&\(\CIRCLE\)&\(\CIRCLE\)&&&&\(\Circle\)&&&&&\(\CIRCLE\)&\(\RIGHTcircle\)\\
  &~E27&\cite{Naiakshina2017}&&&&&&\(\CIRCLE\)&&\(\Circle\)&\(\Circle\)&\(\LEFTcircle\)&&&\(\CIRCLE\)&&&&&&&&\(\CIRCLE\)&&\(\RIGHTcircle\)&&\(\RIGHTcircle\)&\(\RIGHTcircle\)&&&\(\Circle\)&&&&\(\Circle\)&&\(\CIRCLE\)&\(\RIGHTcircle\)\\
\rowcolor{Gray}\cellcolor{white}  &~E28&\cite{Nguyen2017}&\(\CIRCLE\)&&&&&\(\CIRCLE\)&&\(\Circle\)&\(\Circle\)&\(\CIRCLE\)&&&\(\CIRCLE\)&&\(\CIRCLE\)&&&&\(\CIRCLE\)&\(\CIRCLE\)&&&&\(\LEFTcircle\)&&\(\LEFTcircle\)&&&&&&&&\(\RIGHTcircle\)&\(\CIRCLE\)&\(\CIRCLE\)\\
  &~E29&\cite{Derr2017}&\(\CIRCLE\)&&&&&&&&\(\Circle\)&\(\LEFTcircle\)&&&&&\(\CIRCLE\)&&&&&&&\(\CIRCLE\)&&&&\(\LEFTcircle\)&&&\(\Circle\)&\(\Circle\)&&&&&\(\Circle\)&\(\CIRCLE\)\\
\rowcolor{Gray}\cellcolor{white}  &~E30&\cite{Gamero2017}&&&&\(\CIRCLE\)&&&&\(\Circle\)&\(\Circle\)&\(\Circle\)&&&&&&&\(\CIRCLE\)&\(\CIRCLE\)&&\(\CIRCLE\)&&\(\CIRCLE\)&&&&\(\LEFTcircle\)&&&&\(\Circle\)&&\(\Circle\)&&&\(\Circle\)&\(\CIRCLE\)\\
  &~E31&\cite{Acar2017_2}&\(\CIRCLE\)&&&&&&&&\(\Circle\)&\(\CIRCLE\)&&\(\CIRCLE\)&&&&&&&\(\CIRCLE\)&\(\CIRCLE\)&&&&\(\LEFTcircle\)&&\(\LEFTcircle\)&&&\(\Circle\)&&&&&\(\LEFTcircle\)&&\(\CIRCLE\)\\
\midrule
\rowcolor{Gray}\cellcolor{white}\textbf{2018} &~E32&\cite{Haney2018}&&&&&\(\CIRCLE\)&&&&\(\Circle\)&\(\Circle\)&&&&\(\CIRCLE\)&&&\(\CIRCLE\)&&&&\(\CIRCLE\)&\(\CIRCLE\)&&&\(\RIGHTcircle\)&&&&&&&\(\RIGHTcircle\)&\(\Circle\)&&\(\Circle\)&\(\CIRCLE\)\\
  &~E33&\cite{Mu2018}&&&\(\CIRCLE\)&\(\CIRCLE\)&\(\CIRCLE\)&\(\CIRCLE\)&&&\(\CIRCLE\)&\(\CIRCLE\)&&&\(\CIRCLE\)&&\(\CIRCLE\)&&&&\(\CIRCLE\)&&&\(\CIRCLE\)&\(\CIRCLE\)&&&\(\CIRCLE\)&\(\CIRCLE\)&&&\(\Circle\)&&\(\Circle\)&&\(\LEFTcircle\)&&\(\CIRCLE\)\\
\rowcolor{Gray}\cellcolor{white}  &~E34&\cite{Stock2018}&&\(\CIRCLE\)&&&&&&&\(\Circle\)&\(\LEFTcircle\)&&&&&&&\(\CIRCLE\)&&&&&\(\CIRCLE\)&&&&\(\LEFTcircle\)&&&\(\Circle\)&\(\Circle\)&&&&&&\(\RIGHTcircle\)\\
  &~E35&\cite{Dietrich2018}&&\(\CIRCLE\)&&&&&&\(\Circle\)&\(\Circle\)&\(\Circle\)&&&&&&&\(\CIRCLE\)&&&&&\(\CIRCLE\)&&&\(\CIRCLE\)&\(\CIRCLE\)&&&\(\Circle\)&\(\Circle\)&&&\(\Circle\)&&\(\Circle\)&\(\RIGHTcircle\)\\
\rowcolor{Gray}\cellcolor{white}  &~E36&\cite{Skirpan2018}&&&\(\CIRCLE\)&&&&\(\CIRCLE\)&&\(\Circle\)&\(\LEFTcircle\)&\(\CIRCLE\)&&&\(\CIRCLE\)&&\(\CIRCLE\)&\(\CIRCLE\)&&&&\(\CIRCLE\)&\(\CIRCLE\)&&&&\(\CIRCLE\)&&&&&&\(\RIGHTcircle\)&&&&\(\Circle\)\\
  &~E37&\cite{Votipka2018}&\(\CIRCLE\)&&\(\CIRCLE\)&&&&&&\(\Circle\)&\(\LEFTcircle\)&&&&\(\CIRCLE\)&\(\CIRCLE\)&&\(\CIRCLE\)&&&&&\(\CIRCLE\)&&&\(\RIGHTcircle\)&&&&&\(\Circle\)&&\(\Circle\)&\(\CIRCLE\)&&\(\Circle\)&\(\CIRCLE\)\\
\rowcolor{Gray}\cellcolor{white}  &~E38&\cite{Gorski2018}&\(\CIRCLE\)&&&&&&&&\(\Circle\)&\(\CIRCLE\)&&\(\CIRCLE\)&&\(\CIRCLE\)&&\(\CIRCLE\)&\(\CIRCLE\)&&\(\CIRCLE\)&&&&&\(\LEFTcircle\)&&&&&&&&\(\Circle\)&&\(\RIGHTcircle\)&\(\CIRCLE\)&\(\RIGHTcircle\)\\
  &~E39&\cite{Merrill2018}&\(\CIRCLE\)&&&&&&&&\(\Circle\)&\(\Circle\)&&&&\(\CIRCLE\)&&&&&&&\(\CIRCLE\)&\(\CIRCLE\)&&&\(\RIGHTcircle\)&&&&\(\Circle\)&&&\(\Circle\)&&&&\(\CIRCLE\)\\
\rowcolor{Gray}\cellcolor{white}  &~E40&\cite{Wermke2018}&\(\CIRCLE\)&&&&&&&&\(\Circle\)&\(\CIRCLE\)&&&&&\(\CIRCLE\)&&&&\(\CIRCLE\)&&&\(\CIRCLE\)&&\(\LEFTcircle\)&&\(\LEFTcircle\)&&&\(\Circle\)&&&\(\Circle\)&\(\Circle\)&\(\LEFTcircle\)&\(\Circle\)&\(\CIRCLE\)\\
  &~E41&\cite{Adams2018}&\(\CIRCLE\)&&&&&&\(\CIRCLE\)&&\(\Circle\)&\(\LEFTcircle\)&&&&&&\(\CIRCLE\)&&&&&\(\CIRCLE\)&\(\CIRCLE\)&&\(\RIGHTcircle\)&\(\RIGHTcircle\)&\(\RIGHTcircle\)&&&&&&\(\RIGHTcircle\)&&&&\(\CIRCLE\)\\
\rowcolor{Gray}\cellcolor{white}  &~E42&\cite{Naiakshina2018}&&&&&&\(\CIRCLE\)&&\(\Circle\)&\(\Circle\)&\(\LEFTcircle\)&&&\(\CIRCLE\)&&&&&&&\(\CIRCLE\)&&&\(\LEFTcircle\)&&&\(\LEFTcircle\)&&&&\(\Circle\)&\(\Circle\)&&\(\Circle\)&&\(\Circle\)&\(\RIGHTcircle\)\\
  &~E43&\cite{Assal2018}&\(\CIRCLE\)&&&&&&&\(\Circle\)&\(\Circle\)&\(\LEFTcircle\)&&&&\(\CIRCLE\)&&\(\CIRCLE\)&&&&&\(\CIRCLE\)&\(\CIRCLE\)&&&\(\RIGHTcircle\)&&&&&&&\(\Circle\)&\(\Circle\)&&\(\Circle\)&\(\CIRCLE\)\\
\rowcolor{Gray}\cellcolor{white}  &~E44&\cite{Oliveira2018}&\(\CIRCLE\)&&&&&\(\CIRCLE\)&&\(\CIRCLE\)&\(\Circle\)&\(\Circle\)&&&\(\CIRCLE\)&\(\CIRCLE\)&&\(\CIRCLE\)&&&&&&\(\CIRCLE\)&&\(\LEFTcircle\)&&\(\LEFTcircle\)&&&&\(\Circle\)&&\(\Circle\)&&&\(\Circle\)&\(\CIRCLE\)\\
  &~E45&\cite{Thomas2018}&&&&&\(\CIRCLE\)&&&&\(\Circle\)&\(\LEFTcircle\)&&&&\(\CIRCLE\)&&&&&&&&\(\CIRCLE\)&&&\(\RIGHTcircle\)&&&&&\(\Circle\)&&\(\Circle\)&&&\(\Circle\)&\(\CIRCLE\)\\
\rowcolor{Gray}\cellcolor{white}  &~E46&\cite{Simko2018}&\(\CIRCLE\)&&\(\CIRCLE\)&&&\(\CIRCLE\)&&&\(\Circle\)&\(\Circle\)&&&&&&&&\(\CIRCLE\)&\(\CIRCLE\)&&&&\(\CIRCLE\)&&&&&&&&&&&\(\LEFTcircle\)&\(\CIRCLE\)&\(\CIRCLE\)\\
  &~E47&\cite{Hansch2018}&&&&&&\(\CIRCLE\)&&\(\Circle\)&\(\Circle\)&\(\LEFTcircle\)&&&\(\CIRCLE\)&&&&&&&\(\CIRCLE\)&&&\(\LEFTcircle\)&&&\(\LEFTcircle\)&&&&&&\(\Circle\)&&&&\(\CIRCLE\)\\
\rowcolor{Gray}\cellcolor{white}  &~E48&\cite{Haney2018_2}&&&\(\CIRCLE\)&&&&&\(\CIRCLE\)&\(\Circle\)&\(\LEFTcircle\)&&&&\(\CIRCLE\)&&&\(\CIRCLE\)&&&&&\(\CIRCLE\)&&&\(\RIGHTcircle\)&&&&&&&\(\Circle\)&\(\Circle\)&&\(\Circle\)&\(\CIRCLE\)\\
\midrule
\(\sum\) &&&17&7&12&3&11&13&7&&&&3&3&17&15&11&8&20&9&18&8&12&29&14&10&24&25&2&4&12&19&2&29&&&31\\\bottomrule

& \multicolumn{2}{r}{Legend:} & \multicolumn{36}{l}{\emph{Location:} \(\Circle\): Western (Europe, North America); \(\LEFTcircle\): Non-Western; \(\CIRCLE\): International (Multiple Regions); No Marker: Unknown;} \\
& \multicolumn{2}{r}{} & \multicolumn{36}{l}{\emph{External Validity:} \(\CIRCLE\): Considered and addressed; \(\LEFTcircle\): Mentioned as a limitation; \(\Circle\): Not discussed;} \\
& \multicolumn{2}{r}{} & \multicolumn{36}{l}{\emph{Methods:} \(\CIRCLE\): Mixed Methods; \(\LEFTcircle\): Quantiative; \(\RIGHTcircle\): Qualitative;} \\
& \multicolumn{2}{r}{} & \multicolumn{36}{l}{\emph{Theories:} \(\CIRCLE\): Used; \(\RIGHTcircle\): Mentioned; \(\Circle\): Suggested;} \\
& \multicolumn{2}{r}{} & \multicolumn{36}{l}{\emph{Grounded Theory:} \(\CIRCLE\): Full; \(\RIGHTcircle\): Middleground; \(\Circle\): Analytical; } \\
& \multicolumn{2}{r}{} & \multicolumn{36}{l}{\emph{Evaluation of Artifact}: \(\CIRCLE\): Before and After; \(\LEFTcircle\): Before; \(\RIGHTcircle\): After; } \\
& \multicolumn{2}{r}{} & \multicolumn{36}{l}{\emph{HF Perspective}: \(\CIRCLE\): Eliminatory; \(\Circle\): Understanding; } \\
& \multicolumn{2}{r}{} & \multicolumn{36}{l}{\emph{Ethics:} \(\CIRCLE\): Review with HREC; \(\RIGHTcircle\): Review without HREC; \(\Circle\): Not discussed;}

\end{tabular}%
}

\vspace{-1em}
\end{table*}

\begin{table*}[htb!]
\caption{Overview of end user related human factors in security research.}
\label{tab:usr-papers}
\resizebox{\textwidth}{!}{%
\scriptsize
\begin{tabular}{rC{3em}r|C{0.75em}C{0.75em}C{0.75em}C{0.75em}C{0.75em}|C{0.2em}C{0.2em}C{0.3em}|C{0.2em}C{0.2em}C{0.2em}C{0.2em}C{0.2em}C{0.2em}C{0.2em}C{0.3em}|C{0.2em}C{0.2em}C{0.2em}C{0.3em}|C{0.2em}C{0.2em}C{0.2em}C{0.2em}C{0.2em}C{0.3em}|C{0.2em}C{0.2em}C{0.2em}C{0.3em}|C{0.2em}C{0.2em}C{0.3em}|C{0.2em}}
\toprule

\textbf{Year} & \textbf{Idx.} & \textbf{Ref.} & \multicolumn{8}{c|}{\textbf{Sample}} & \multicolumn{8}{c|}{\textbf{Recruitment}} & \multicolumn{4}{c|}{\textbf{Res. Obj.}} & \multicolumn{6}{c|}{\textbf{Research Method}} & \multicolumn{7}{c|}{\textbf{Theory / Framework}} &  \\
 & & & \rotatebox{90}{Children} & \rotatebox{90}{N/A} & \rotatebox{90}{Gen. Pub.} & \rotatebox{90}{Univ.} & \rotatebox{90}{Spec. Users} &\rotatebox{90}{Sample Loc.} &  \rotatebox{90}{Author Loc.} &  \rotatebox{90}{Ext. Validity}  & \rotatebox{90}{MTurk} & \rotatebox{90}{City} & \rotatebox{90}{Univ.} & \rotatebox{90}{Personal} & \rotatebox{90}{Soc. Media} & \rotatebox{90}{Oth. Online} & \rotatebox{90}{Agency} & \rotatebox{90}{N/A} & \rotatebox{90}{Eval.} & \rotatebox{90}{Hyp. Test} & \rotatebox{90}{Exploratory} & \rotatebox{90}{Perspective} & \rotatebox{90}{Lab} & \rotatebox{90}{Online} & \rotatebox{90}{Interview} & \rotatebox{90}{Survey} & \rotatebox{90}{Focus Groups} & \rotatebox{90}{Observations} & \rotatebox{90}{Mental Models} & \rotatebox{90}{Sensemaking} & \rotatebox{90}{TRA} & \rotatebox{90}{Other} & \rotatebox{90}{GT} & \rotatebox{90}{Design Eval.} & \rotatebox{90}{HF Persp.}& \rotatebox{90}{\textbf{Ethics}} \\

\textbf{2008} &NE1&\cite{Forget2008}&&&&\(\CIRCLE\)&&\(\Circle\)&\(\Circle\)&\(\LEFTcircle\)&&&\(\CIRCLE\)&&&&&&\(\CIRCLE\)&&&&\(\LEFTcircle\)&&&\(\LEFTcircle\)&&&&&&\(\Circle\)&&\(\RIGHTcircle\)&&\(\CIRCLE\)\\
\rowcolor{Gray}\cellcolor{white}  &NE2&\cite{Egelman2008}&&&\(\CIRCLE\)&&&\(\Circle\)&\(\Circle\)&\(\CIRCLE\)&&\(\CIRCLE\)&&&&\(\CIRCLE\)&&&&&&&\(\CIRCLE\)&&&&&&\(\RIGHTcircle\)&&&&&&\(\Circle\)&\(\Circle\)\\
\midrule
\textbf{2009} &NE3&\cite{Mcdonald2009}&&&\(\CIRCLE\)&&&&\(\Circle\)&\(\Circle\)&&&&\(\CIRCLE\)&&\(\CIRCLE\)&&&&\(\CIRCLE\)&\(\CIRCLE\)&&&\(\LEFTcircle\)&&&&&&&&\(\Circle\)&&&&\(\Circle\)\\
\rowcolor{Gray}\cellcolor{white}  &NE4&\cite{Klasnja2009}&&&\(\CIRCLE\)&&\(\CIRCLE\)&\(\Circle\)&\(\Circle\)&\(\Circle\)&&&&&&&\(\CIRCLE\)&&&&\(\CIRCLE\)&\(\CIRCLE\)&&\(\CIRCLE\)&\(\CIRCLE\)&&&&&&&\(\RIGHTcircle\)&\(\Circle\)&&\(\Circle\)&\(\Circle\)\\
  &NE5&\cite{Karlof2009}&&&&\(\CIRCLE\)&&\(\Circle\)&\(\Circle\)&\(\CIRCLE\)&&&\(\CIRCLE\)&&&&&&\(\CIRCLE\)&\(\CIRCLE\)&&&&\(\CIRCLE\)&&&&&&&&\(\CIRCLE\)&&\(\RIGHTcircle\)&&\(\CIRCLE\)\\
\midrule
\rowcolor{Gray}\cellcolor{white}\textbf{2010} &NE6&\cite{Ion2010}&&&\(\CIRCLE\)&&\(\CIRCLE\)&\(\Circle\)&\(\CIRCLE\)&\(\LEFTcircle\)&&&\(\CIRCLE\)&&&\(\CIRCLE\)&&&\(\CIRCLE\)&&&&\(\RIGHTcircle\)&&&&&&\(\RIGHTcircle\)&&&&&\(\RIGHTcircle\)&\(\Circle\)&\(\Circle\)\\
\midrule
\textbf{2011} &NE7&\cite{Zhu2011}&&&&\(\CIRCLE\)&&\(\Circle\)&\(\Circle\)&\(\CIRCLE\)&&&\(\CIRCLE\)&&&&&&&\(\CIRCLE\)&&&\(\LEFTcircle\)&&&\(\LEFTcircle\)&&&&&&\(\CIRCLE\)&&&&\(\CIRCLE\)\\
\rowcolor{Gray}\cellcolor{white}  &NE8&\cite{Shin2011}&&&&\(\CIRCLE\)&&\(\Circle\)&\(\Circle\)&\(\LEFTcircle\)&&&\(\CIRCLE\)&&&&&&\(\CIRCLE\)&\(\CIRCLE\)&&&\(\LEFTcircle\)&&&\(\LEFTcircle\)&&&&&&&&\(\RIGHTcircle\)&&\(\CIRCLE\)\\
\midrule
\textbf{2012} &NE9&\cite{Sae_Bae2012}&&\(\CIRCLE\)&&&&&\(\Circle\)&\(\Circle\)&&&&&&&&\(\CIRCLE\)&\(\CIRCLE\)&&&&\(\LEFTcircle\)&&&\(\LEFTcircle\)&&&&&&\(\RIGHTcircle\)&&\(\RIGHTcircle\)&&\(\Circle\)\\
\midrule
\rowcolor{Gray}\cellcolor{white}\textbf{2013} &NE10&\cite{Ruoti2013}&&&&\(\CIRCLE\)&&\(\Circle\)&\(\Circle\)&\(\LEFTcircle\)&&&\(\CIRCLE\)&&&&&&\(\CIRCLE\)&&&&\(\LEFTcircle\)&&&&&&&&&\(\Circle\)&&\(\RIGHTcircle\)&&\(\CIRCLE\)\\
  &NE11&\cite{Denning2013}&&&\(\CIRCLE\)&&\(\CIRCLE\)&\(\CIRCLE\)&\(\Circle\)&\(\Circle\)&&&\(\CIRCLE\)&&&\(\CIRCLE\)&&&\(\CIRCLE\)&&&\(\CIRCLE\)&\(\RIGHTcircle\)&&&\(\RIGHTcircle\)&&&&&&\(\Circle\)&&\(\RIGHTcircle\)&&\(\CIRCLE\)\\
\midrule
\rowcolor{Gray}\cellcolor{white}\textbf{2014} &NE12&\cite{Wash2014}&&&&\(\CIRCLE\)&&\(\Circle\)&\(\Circle\)&\(\LEFTcircle\)&&&\(\CIRCLE\)&&&&&&&&&\(\CIRCLE\)&&&\(\CIRCLE\)&\(\CIRCLE\)&&&\(\Circle\)&&&&&&\(\Circle\)&\(\CIRCLE\)\\
  &NE13&\cite{Aviv2014}&&&&&\(\CIRCLE\)&&\(\Circle\)&\(\Circle\)&\(\CIRCLE\)&&&&&&&&&&\(\CIRCLE\)&\(\CIRCLE\)&&&&\(\LEFTcircle\)&&&&&&\(\CIRCLE\)&&&&\(\CIRCLE\)\\
\rowcolor{Gray}\cellcolor{white}  &NE14&\cite{Mohamed2014}&&&&\(\CIRCLE\)&&\(\Circle\)&\(\CIRCLE\)&\(\Circle\)&&&&&&&&\(\CIRCLE\)&\(\CIRCLE\)&\(\CIRCLE\)&&&\(\LEFTcircle\)&&&\(\LEFTcircle\)&&&&&&\(\CIRCLE\)&&\(\RIGHTcircle\)&&\(\RIGHTcircle\)\\
\midrule
\textbf{2015} &NE15&\cite{Ur2015}&&&\(\CIRCLE\)&\(\CIRCLE\)&&\(\Circle\)&\(\Circle\)&\(\LEFTcircle\)&&\(\CIRCLE\)&\(\CIRCLE\)&&&\(\CIRCLE\)&&&&&\(\CIRCLE\)&\(\CIRCLE\)&\(\RIGHTcircle\)&&\(\RIGHTcircle\)&&&&\(\Circle\)&&\(\Circle\)&&&&\(\Circle\)&\(\CIRCLE\)\\
\rowcolor{Gray}\cellcolor{white}  &NE16&\cite{Angulo2015}&&&&&\(\CIRCLE\)&\(\CIRCLE\)&\(\LEFTcircle\)&\(\CIRCLE\)&&&&\(\CIRCLE\)&&\(\CIRCLE\)&&&&&\(\CIRCLE\)&\(\CIRCLE\)&&&\(\CIRCLE\)&\(\CIRCLE\)&&&&&&\(\RIGHTcircle\)&&&\(\Circle\)&\(\Circle\)\\
  &NE17&\cite{Chanchary2015}&&&&&\(\CIRCLE\)&&\(\Circle\)&\(\Circle\)&&&&&&\(\CIRCLE\)&&&&&\(\CIRCLE\)&\(\CIRCLE\)&&\(\LEFTcircle\)&&&&&\(\Circle\)&\(\Circle\)&&&&&\(\Circle\)&\(\CIRCLE\)\\
\rowcolor{Gray}\cellcolor{white}  &NE18&\cite{Bianchi2015}&&&&&\(\CIRCLE\)&&\(\Circle\)&\(\Circle\)&\(\CIRCLE\)&&&&&&&&\(\CIRCLE\)&&&&&\(\LEFTcircle\)&&&&&&&&&&\(\RIGHTcircle\)&&\(\CIRCLE\)\\
  &NE19&\cite{Hupperich2015}&&&&\(\CIRCLE\)&\(\CIRCLE\)&&\(\Circle\)&\(\LEFTcircle\)&&&\(\CIRCLE\)&\(\CIRCLE\)&&\(\CIRCLE\)&&&\(\CIRCLE\)&&&&&&&\(\LEFTcircle\)&&&&&&&&\(\RIGHTcircle\)&&\(\RIGHTcircle\)\\
\midrule
\rowcolor{Gray}\cellcolor{white}\textbf{2016} &NE20&\cite{Fagan2016}&&&&&\(\CIRCLE\)&&\(\Circle\)&\(\LEFTcircle\)&\(\CIRCLE\)&&&&&&&&&&&\(\CIRCLE\)&&&&\(\CIRCLE\)&&&&&&\(\CIRCLE\)&\(\RIGHTcircle\)&&\(\Circle\)&\(\Circle\)\\
  &NE21&\cite{Mathur2016}&&&&\(\CIRCLE\)&\(\CIRCLE\)&\(\Circle\)&\(\Circle\)&\(\LEFTcircle\)&&&\(\CIRCLE\)&&\(\CIRCLE\)&&&&\(\CIRCLE\)&&&\(\CIRCLE\)&\(\RIGHTcircle\)&&\(\RIGHTcircle\)&&&&&&&\(\CIRCLE\)&&\(\CIRCLE\)&\(\Circle\)&\(\CIRCLE\)\\
\rowcolor{Gray}\cellcolor{white}  &NE22&\cite{Redmiles2016}&&&\(\CIRCLE\)&&&\(\Circle\)&\(\Circle\)&\(\CIRCLE\)&&&\(\CIRCLE\)&\(\CIRCLE\)&\(\CIRCLE\)&&&&&&&\(\CIRCLE\)&&&\(\RIGHTcircle\)&&&&\(\Circle\)&&\(\Circle\)&&\(\Circle\)&&\(\Circle\)&\(\CIRCLE\)\\
  &NE23&\cite{Tischer2016}&&&&\(\CIRCLE\)&&\(\Circle\)&\(\Circle\)&\(\Circle\)&&&\(\CIRCLE\)&&&&&&&&\(\CIRCLE\)&\(\CIRCLE\)&&&&\(\CIRCLE\)&&&&&\(\Circle\)&\(\Circle\)&&&&\(\CIRCLE\)\\
\midrule
\rowcolor{Gray}\cellcolor{white}\textbf{2017} &NE24&\cite{Lastdrager2017}&\(\CIRCLE\)&&&&&\(\Circle\)&\(\Circle\)&\(\LEFTcircle\)&&&&&&&&\(\CIRCLE\)&\(\CIRCLE\)&&\(\CIRCLE\)&&\(\LEFTcircle\)&&&&&&&&&\(\Circle\)&&\(\RIGHTcircle\)&&\(\CIRCLE\)\\
  &NE25&\cite{Ruoti2017}&&&\(\CIRCLE\)&&&\(\Circle\)&\(\Circle\)&\(\LEFTcircle\)&&\(\CIRCLE\)&&&&&&&&&&\(\CIRCLE\)&&&\(\RIGHTcircle\)&&&&&&&\(\RIGHTcircle\)&\(\CIRCLE\)&&\(\Circle\)&\(\CIRCLE\)\\
\rowcolor{Gray}\cellcolor{white}  &NE26&\cite{Tian2017}&&&&&\(\CIRCLE\)&\(\Circle\)&\(\Circle\)&\(\LEFTcircle\)&\(\CIRCLE\)&&\(\CIRCLE\)&&&&&&\(\CIRCLE\)&&&\(\CIRCLE\)&\(\CIRCLE\)&&&\(\CIRCLE\)&&&\(\CIRCLE\)&&&&&\(\RIGHTcircle\)&&\(\CIRCLE\)\\
  &NE27&\cite{Liu2017}&&\(\CIRCLE\)&&&&&\(\Circle\)&\(\Circle\)&&&&&&&&\(\CIRCLE\)&\(\CIRCLE\)&&&&\(\LEFTcircle\)&&&&&&&&&&&\(\RIGHTcircle\)&&\(\CIRCLE\)\\
\rowcolor{Gray}\cellcolor{white}  &NE28&\cite{Shirvanian2017}&&&&&\(\CIRCLE\)&&\(\Circle\)&\(\LEFTcircle\)&\(\CIRCLE\)&&&&&&&&\(\CIRCLE\)&&&&&\(\LEFTcircle\)&&&&&&&&&&\(\RIGHTcircle\)&\(\CIRCLE\)&\(\CIRCLE\)\\
  &NE29&\cite{Zhang2017}&&&&\(\CIRCLE\)&&\(\Circle\)&\(\Circle\)&\(\LEFTcircle\)&&&\(\CIRCLE\)&&&&&&\(\CIRCLE\)&&&&\(\LEFTcircle\)&&&&&&&&&&&\(\RIGHTcircle\)&&\(\CIRCLE\)\\
\rowcolor{Gray}\cellcolor{white}  &NE30&\cite{Chatterjee2017}&&&&&\(\CIRCLE\)&&\(\LEFTcircle\)&\(\LEFTcircle\)&\(\CIRCLE\)&&&\(\CIRCLE\)&&&&&\(\CIRCLE\)&&&&&\(\LEFTcircle\)&&&&&&&&&&\(\CIRCLE\)&&\(\CIRCLE\)\\
  &NE31&\cite{Abu_salma2017}&&&&\(\CIRCLE\)&\(\CIRCLE\)&\(\Circle\)&\(\Circle\)&\(\LEFTcircle\)&&&\(\CIRCLE\)&&&\(\CIRCLE\)&&&&&&\(\CIRCLE\)&&&\(\RIGHTcircle\)&&&&\(\CIRCLE\)&&&&\(\Circle\)&&\(\Circle\)&\(\CIRCLE\)\\
\midrule
\rowcolor{Gray}\cellcolor{white}\textbf{2018} &NE32&\cite{Murillo2018}&&&\(\CIRCLE\)&&\(\CIRCLE\)&\(\Circle\)&\(\Circle\)&\(\LEFTcircle\)&&&&\(\CIRCLE\)&&&\(\CIRCLE\)&&&&\(\CIRCLE\)&\(\CIRCLE\)&&&\(\RIGHTcircle\)&&\(\RIGHTcircle\)&&\(\RIGHTcircle\)&&&&&&\(\Circle\)&\(\Circle\)\\
  &NE33&\cite{Rashidi2018}&&&&\(\CIRCLE\)&&\(\Circle\)&\(\Circle\)&\(\LEFTcircle\)&&&\(\CIRCLE\)&&&&&&&&&\(\CIRCLE\)&&&\(\RIGHTcircle\)&&&&&&&\(\CIRCLE\)&\(\Circle\)&&\(\Circle\)&\(\CIRCLE\)\\
\rowcolor{Gray}\cellcolor{white}  &NE34&\cite{Sambasivan2018}&&&&&\(\CIRCLE\)&\(\CIRCLE\)&\(\CIRCLE\)&\(\LEFTcircle\)&&&&\(\CIRCLE\)&&&\(\CIRCLE\)&&&&\(\CIRCLE\)&\(\CIRCLE\)&&&&&\(\RIGHTcircle\)&&\(\Circle\)&&&\(\Circle\)&&&\(\Circle\)&\(\RIGHTcircle\)\\
  &NE35&\cite{Habib2018}&&&&&\(\CIRCLE\)&\(\Circle\)&\(\Circle\)&\(\LEFTcircle\)&\(\CIRCLE\)&&&&&\(\CIRCLE\)&&&&&\(\CIRCLE\)&\(\CIRCLE\)&&&&\(\CIRCLE\)&&&\(\Circle\)&&\(\Circle\)&&&&\(\Circle\)&\(\CIRCLE\)\\
\rowcolor{Gray}\cellcolor{white}  &NE36&\cite{Habib2018_2}&&&&&\(\CIRCLE\)&\(\Circle\)&\(\Circle\)&\(\LEFTcircle\)&\(\CIRCLE\)&&&&&&&&&&&\(\CIRCLE\)&&&&\(\CIRCLE\)&&&\(\Circle\)&\(\Circle\)&&&&&\(\Circle\)&\(\CIRCLE\)\\
  &NE37&\cite{Zou2018}&&&\(\CIRCLE\)&\(\CIRCLE\)&&\(\Circle\)&\(\Circle\)&\(\LEFTcircle\)&&&\(\CIRCLE\)&&\(\CIRCLE\)&\(\CIRCLE\)&&&&&&\(\CIRCLE\)&&&\(\RIGHTcircle\)&&&&\(\CIRCLE\)&&&&&&\(\Circle\)&\(\CIRCLE\)\\
\rowcolor{Gray}\cellcolor{white}  &NE38&\cite{Karunakaran2018}&&&&&\(\CIRCLE\)&\(\CIRCLE\)&\(\Circle\)&\(\LEFTcircle\)&\(\CIRCLE\)&&&&&\(\CIRCLE\)&&&&&&\(\CIRCLE\)&&&&\(\CIRCLE\)&&&\(\Circle\)&&&\(\Circle\)&&&\(\Circle\)&\(\Circle\)\\
  &NE39&\cite{Park2018}&&&&&\(\CIRCLE\)&\(\Circle\)&\(\Circle\)&\(\LEFTcircle\)&\(\CIRCLE\)&&&&&&&&&&\(\CIRCLE\)&\(\CIRCLE\)&&&&\(\CIRCLE\)&&&\(\Circle\)&&\(\Circle\)&&&&\(\Circle\)&\(\Circle\)\\
\rowcolor{Gray}\cellcolor{white}  &NE40&\cite{Neupane2018}&&&&&\(\CIRCLE\)&\(\Circle\)&\(\Circle\)&\(\LEFTcircle\)&&&\(\CIRCLE\)&&\(\CIRCLE\)&&&&&\(\CIRCLE\)&&&\(\CIRCLE\)&&&\(\CIRCLE\)&&&&&&\(\RIGHTcircle\)&&&\(\Circle\)&\(\CIRCLE\)\\
  &NE41&\cite{Gao2018}&&&\(\CIRCLE\)&\(\CIRCLE\)&&\(\Circle\)&\(\Circle\)&\(\LEFTcircle\)&&&\(\CIRCLE\)&&\(\CIRCLE\)&\(\CIRCLE\)&&&&\(\CIRCLE\)&\(\CIRCLE\)&&&\(\CIRCLE\)&&\(\CIRCLE\)&&&&&&\(\CIRCLE\)&&&\(\Circle\)&\(\CIRCLE\)\\
\rowcolor{Gray}\cellcolor{white}  &NE42&\cite{Schwarz2018}&&\(\CIRCLE\)&&\(\CIRCLE\)&&\(\Circle\)&\(\Circle\)&\(\Circle\)&&&&\(\CIRCLE\)&&&&&\(\CIRCLE\)&&&&\(\LEFTcircle\)&&&&&&&&&&&\(\RIGHTcircle\)&&\(\Circle\)\\
  &NE43&\cite{Colnago2018}&&&&\(\CIRCLE\)&&\(\Circle\)&\(\Circle\)&\(\LEFTcircle\)&&&\(\CIRCLE\)&&&&&&&&\(\CIRCLE\)&&&&&\(\CIRCLE\)&&&&&&\(\RIGHTcircle\)&&&\(\Circle\)&\(\CIRCLE\)\\
\rowcolor{Gray}\cellcolor{white}  &NE44&\cite{Das2018}&&&&&\(\CIRCLE\)&\(\Circle\)&\(\Circle\)&\(\LEFTcircle\)&\(\CIRCLE\)&&&&&&&&&&&\(\CIRCLE\)&&&&\(\CIRCLE\)&&&&&&\(\CIRCLE\)&&&&\(\CIRCLE\)\\
  &NE45&\cite{Reeder2018}&&&&&\(\CIRCLE\)&&\(\Circle\)&\(\CIRCLE\)&\(\CIRCLE\)&&&&&\(\CIRCLE\)&&&&&&\(\CIRCLE\)&&&&\(\CIRCLE\)&&&&&&\(\Circle\)&&&\(\Circle\)&\(\CIRCLE\)\\
\rowcolor{Gray}\cellcolor{white}  &NE46&\cite{Oates2018}&\(\CIRCLE\)&&\(\CIRCLE\)&&&\(\Circle\)&\(\Circle\)&\(\LEFTcircle\)&&&&&&\(\CIRCLE\)&&&&&\(\CIRCLE\)&&&\(\RIGHTcircle\)&&&&&\(\CIRCLE\)&&&\(\CIRCLE\)&&&\(\Circle\)&\(\CIRCLE\)\\
  &NE47&\cite{Redmiles2018}&&&&&\(\CIRCLE\)&\(\Circle\)&\(\CIRCLE\)&\(\CIRCLE\)&\(\CIRCLE\)&&&&&\(\CIRCLE\)&&&\(\CIRCLE\)&&&&&&&\(\LEFTcircle\)&&&\(\Circle\)&&\(\Circle\)&&&\(\LEFTcircle\)&&\(\Circle\)\\
\rowcolor{Gray}\cellcolor{white}  &NE48&\cite{Lebeck2018}&&&&&\(\CIRCLE\)&\(\Circle\)&\(\Circle\)&\(\LEFTcircle\)&&&&\(\CIRCLE\)&\(\CIRCLE\)&&&&&&\(\CIRCLE\)&\(\CIRCLE\)&\(\RIGHTcircle\)&&\(\RIGHTcircle\)&&&&\(\Circle\)&&&\(\Circle\)&\(\Circle\)&&\(\Circle\)&\(\CIRCLE\)\\
\midrule
\(\sum\) &&&2&3&12&18&25&&&&13&3&21&9&6&16&3&4&19&7&16&25&18&9&12&23&2&0&18&2&6&26&&&27\\\bottomrule

& \multicolumn{2}{r}{Legend:} & \multicolumn{34}{l}{\emph{Location:} \(\Circle\): Western (Europe, North America); \(\LEFTcircle\): Non-Western; \(\CIRCLE\): International (Multiple Regions); No Marker: Unknown;} \\
& \multicolumn{2}{r}{} & \multicolumn{34}{l}{\emph{External Validity:} \(\CIRCLE\): Considered and addressed; \(\LEFTcircle\): Mentioned as a limitation; \(\Circle\): Not discussed;} \\
& \multicolumn{2}{r}{} & \multicolumn{34}{l}{\emph{Methods:} \(\CIRCLE\): Mixed Methods; \(\LEFTcircle\): Quantiative; \(\RIGHTcircle\): Qualitative;} \\
& \multicolumn{2}{r}{} & \multicolumn{34}{l}{\emph{Theories:} \(\CIRCLE\): Used; \(\RIGHTcircle\): Mentioned; \(\Circle\): Suggested;} \\
& \multicolumn{2}{r}{} & \multicolumn{34}{l}{\emph{Grounded Theory:} \(\CIRCLE\): Full; \(\RIGHTcircle\): Middleground; \(\Circle\): Analytical; } \\
& \multicolumn{2}{r}{} & \multicolumn{34}{l}{\emph{Evaluation of Artifact}: \(\CIRCLE\): Before and After; \(\LEFTcircle\): Before; \(\RIGHTcircle\): After; } \\
& \multicolumn{2}{r}{} & \multicolumn{34}{l}{\emph{HF Perspective}: \(\CIRCLE\): Eliminatory; \(\Circle\): Understanding; } \\
& \multicolumn{2}{r}{} & \multicolumn{34}{l}{\emph{Ethics:} \(\CIRCLE\): Review with HREC; \(\RIGHTcircle\): Review without HREC; \(\Circle\): Not discussed;}

\end{tabular}%
}

\vspace{-1em}
\end{table*}

\section{Sample Population and Recruitment}
\label{sec:populationsample}
Next, we look at the population samples, that is, \emph{who} researchers investigate and how they recruit the participants.

\begin{figure*}[!t]
    \centering
       \begin{subfigure}[t]{0.5\textwidth}
        \centering
        \includegraphics[width=\columnwidth,trim={0 4cm 0 9.5cm},clip]{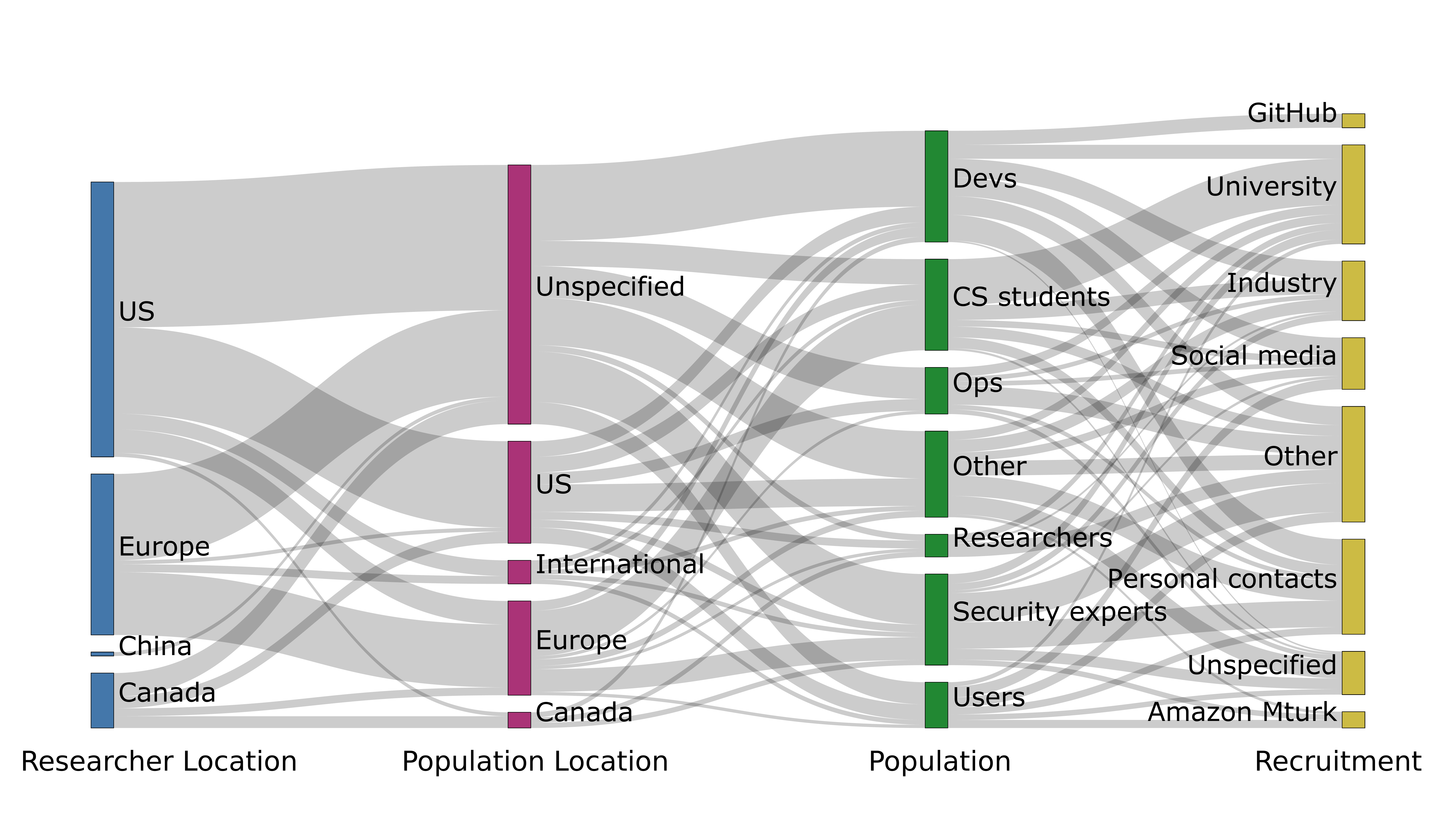}
        \caption{Expert Users}
        \label{fig:exp}
    \end{subfigure}%
     ~
    \begin{subfigure}[t]{0.5\textwidth}
        \centering
        \includegraphics[width=\columnwidth,trim={0 4cm 0 9.5cm},clip]{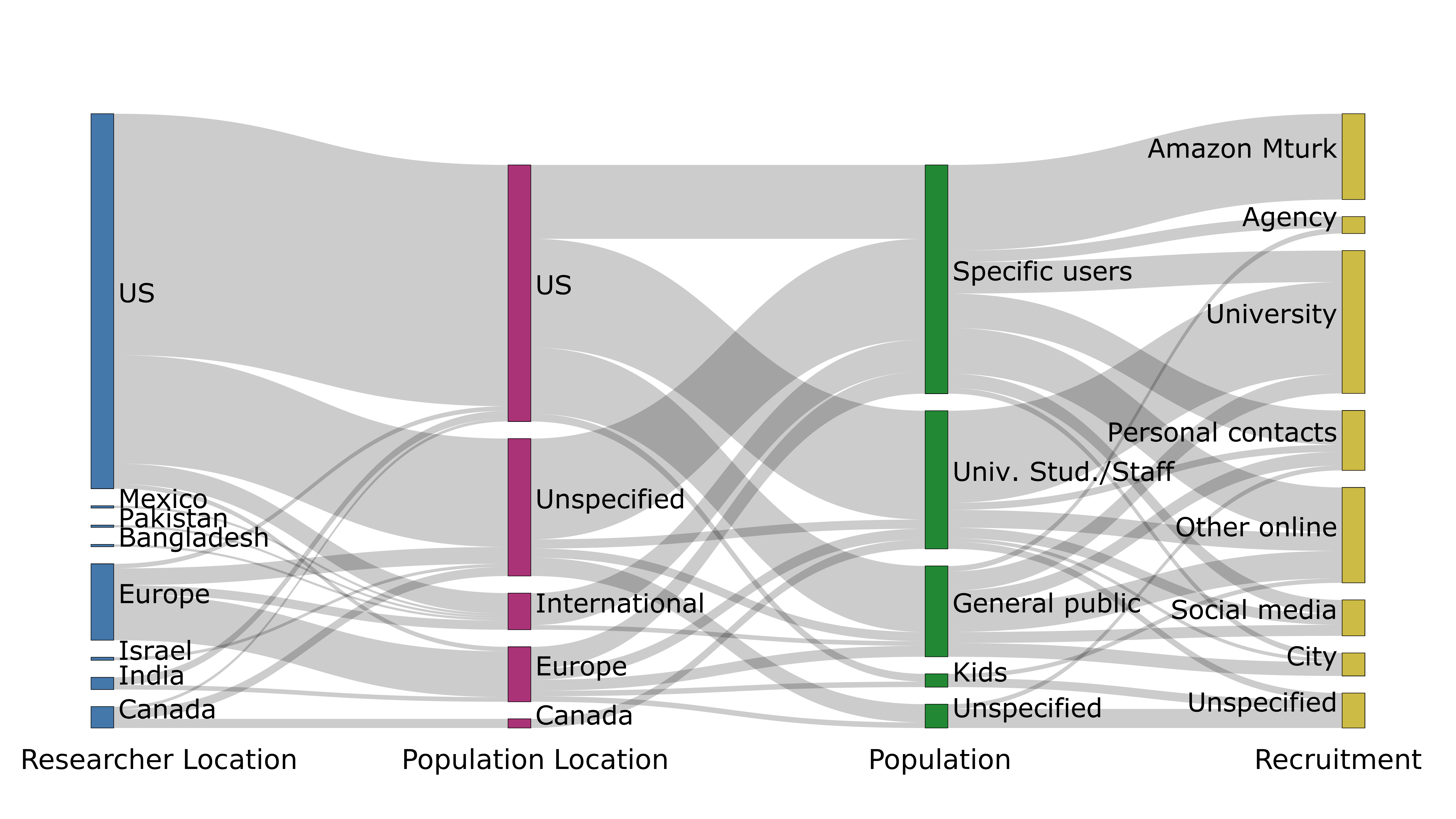}
        \caption{End Users}
        \label{fig:end}
    \end{subfigure}
    \caption{Overview of Authors' location in comparison to the population types, recruitment methods, and population locations.
The figure depicts how the share of publications across properties, for example, US, Europe, etc., for each category (Researcher Location, Population Location, Population, and Recruitment) connects to the other categories.
For example, in \Fref{fig:exp}, we see that the majority of studies on US populations is contributed by researchers located in the United States.
Similarly, the authors' location in \Fref{fig:end} predetermines the populations' location, apart from studies using a population of \emph{specific users} where the population's location usually is not disclosed.
This is similar to \Fref{fig:exp} in so far, that Expert Users are predominantly recruited as a population of \emph{specific users}.
}
    \label{fig:pop}
\vspace{-1em}
\end{figure*}

Our results are summarized in columns ``Sample'' and ``Recruitment'' in \Fref{tab:exp-papers} and \Fref{tab:usr-papers}, and we visualize the geographic distribution of authors in \Fref{fig:pop}.

\subsection{Population Selection and Recruitment}
In the expert-user studies sample, we discover that Computer Science students and security experts are the most utilized populations.
This holds true even for end-user studies.
In other words, university students are the most popular population sample being studied for both expert and end-user studies.
This is to be expected: members of the (local) university are easily accessible for university researchers, that is, they constitute a convenience sample.
Interestingly, only one of the papers is specifically studying college students as their intended research subject~\cite{Rashidi2018}, while the remainder used them as a convenient proxy for end-user and expert-user populations.

Regarding recruiting participants from these populations, we identified eight categories for both expert and end-user samples, though not exactly the same categories. 
For expert-user research, the most popular recruitment method is via personal contacts and university channels.
We note that it seems to be convenient to find experts through one's personal networks, specially for researchers working in the same field of expertise.
For end users, university channels, like local (physical) message boards and on-campus recruitment, are the most popular recruitment method, followed by Amazon MTurk.
Similar to the reason why university students are most studied, this is probably due to the fact that university channels are a convenient recruitment method.

\begin{figure*}[!t]
\centering
\begin{minipage}[t]{0.49\textwidth}
	\centering
       \begin{subfigure}[t]{0.40\textwidth}
        \centering
        \includegraphics[width=\columnwidth,trim={0 0cm 0 0cm},clip,angle=0]{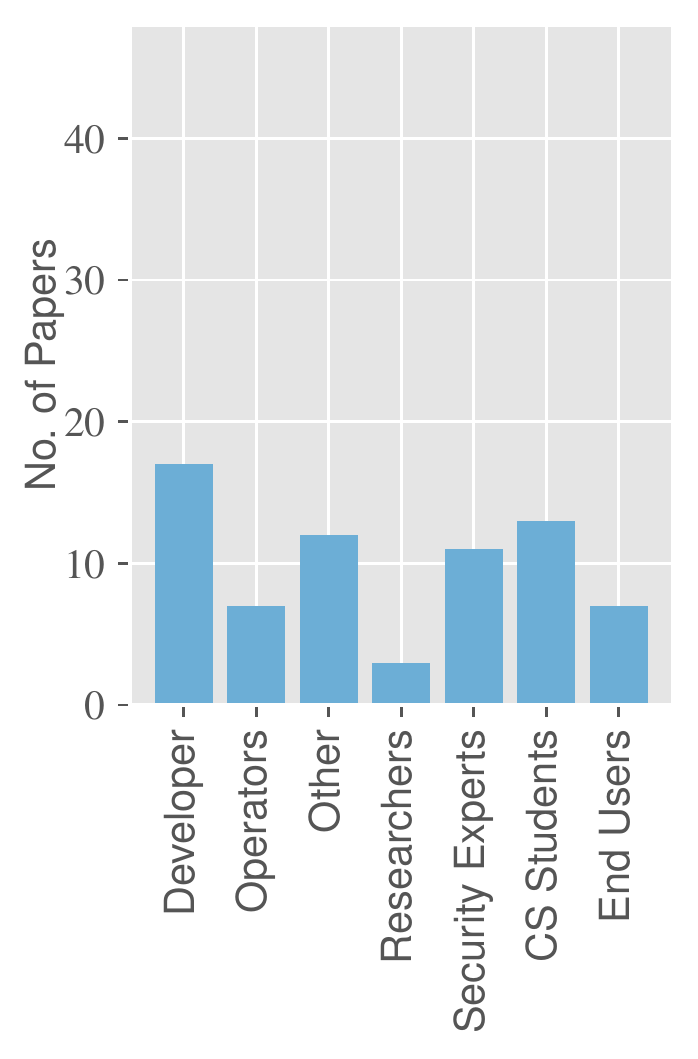}
        \caption{Expert Users}
        \label{fig:sample_exp}
    \end{subfigure}%
     ~
    \begin{subfigure}[t]{0.40\textwidth}
        \centering
        \includegraphics[width=\columnwidth,trim={0 0cm 0 0cm},clip,angle=0]{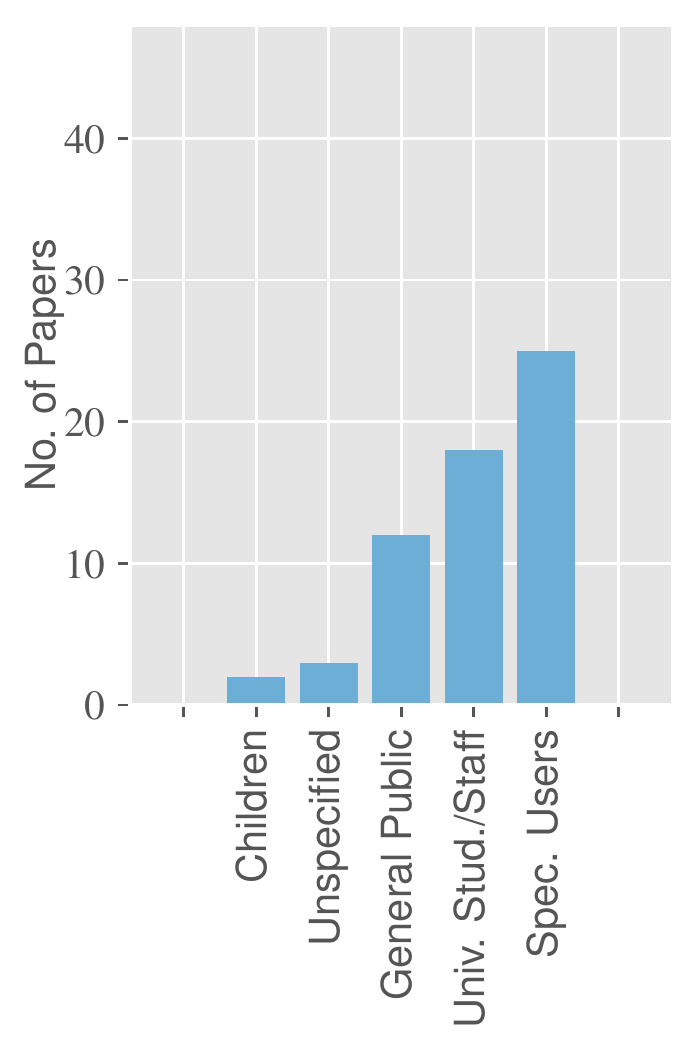}
        \caption{End Users}
        \label{fig:sample_end}
    \end{subfigure}
\caption{Overview of the studied populations for expert users (\Fref{fig:sample_exp}) and end users (\Fref{fig:sample_end}). Note that for end users the focus is on specific users, such as users using a \emph{specific} software, while for expert users the perspective is more on general observations tied to the function of the study participants (developers, operators, etc.).}
    \label{fig:sample}
\end{minipage}
\hfill
\begin{minipage}[t]{0.49\textwidth}
    \centering
       \begin{subfigure}[t]{0.40\textwidth}
        \centering
        \includegraphics[width=\columnwidth,trim={0cm 0cm 0.0cm 0.6cm},clip,angle=0]{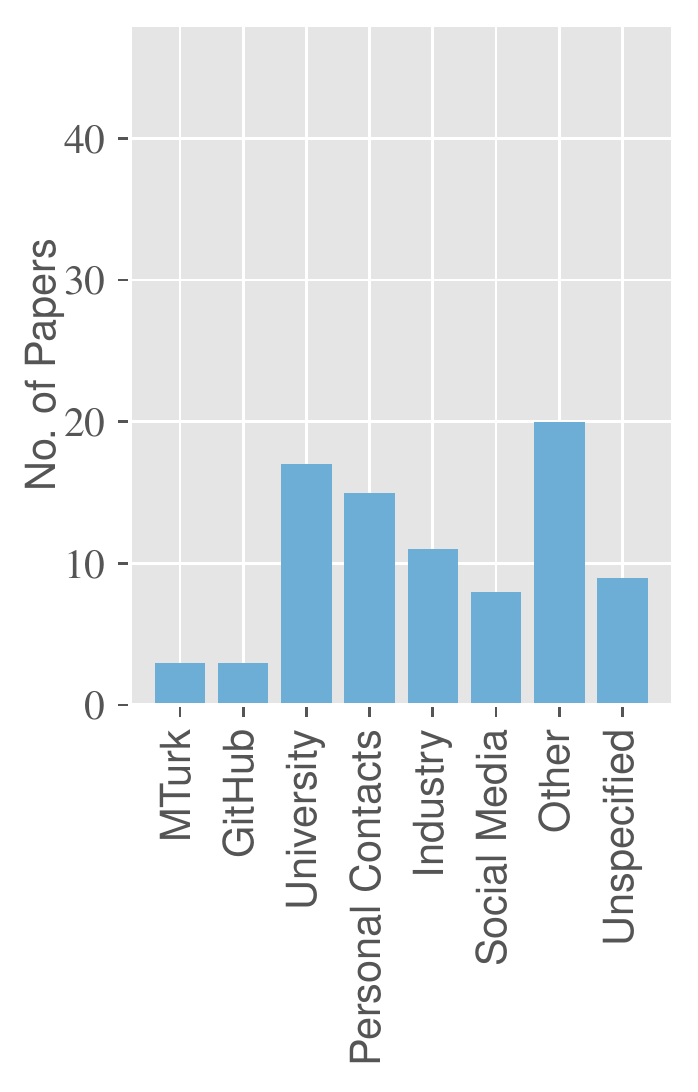}
        \caption{Expert Users}
        \label{fig:rec_exp}
    \end{subfigure}%
     ~
    \begin{subfigure}[t]{0.40\textwidth}
        \centering
        \includegraphics[width=\columnwidth,trim={0 0cm 0 0.6cm},clip,angle=0]{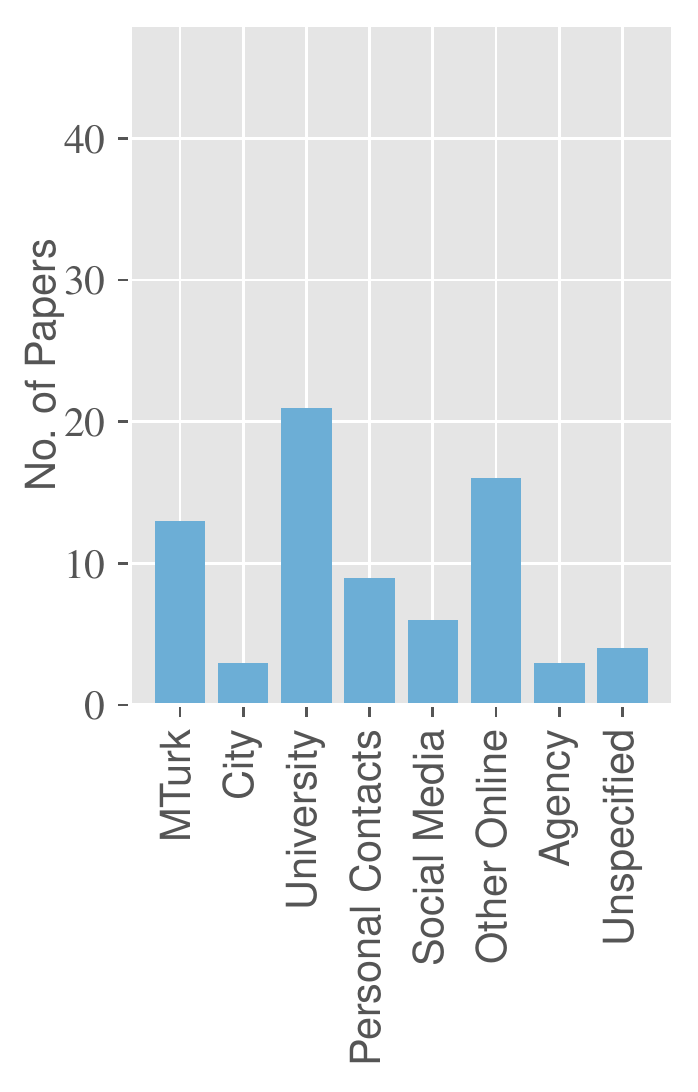}
        \caption{End Users}
        \label{fig:rec_end}
    \end{subfigure}
\caption{Overview of recruitment channels for expert users (\Fref{fig:rec_exp}) and end users (\Fref{fig:rec_end}). For both samples, we see convenience samples being prevalent, that is, recruitment at the local university, or via personal contacts. Naturally, university sampling is more common for end-user studies, as the local university corresponds closer to the target population. 
}
    \label{fig:rec}
\end{minipage}
\vspace{-1.5em}
\end{figure*}

\subsection{Population Location}
We find that in a large number of the studies, the population sample is based in North America or Europe.
Only four end-user and expert-user studies each report an international population sample.
Hence, overall, the western user population is the most represented.
This follows from our observation on convenience sampling, as we also see that most research itself is contributed by authors from the U.S. and, to a lesser degree, Europe.
In \Fref{tab:exp-papers} and \Fref{tab:usr-papers}, we mark western authors and populations with \(\Circle\), authors and populations from other regions with \(\LEFTcircle\), and international collaborations and populations with \(\CIRCLE\).

In our analysis of the end-user studies, we find that the majority of the \UsrPopulationpoplocUnarkedTotal papers where the location of the population is not reported, are studying ``specific users'' (see \Fref{fig:pop}).
Specific users, as explained earlier, refer to users of specific online channels, such as MTurk or the Security Behaviour Observatory, or other specific groups, like users with social disorders.
For expert users, the group of papers not specifying the location of the studied population is even larger (\ExpPopulationpoplocUnarkedTotal/\ExpPopulationpoplocItemsTotal) (see \Fref{fig:exp}).

A likely explanation for this imbalance is that ``expert users'' are a form of ``specific users.''
We conjecture that it is difficult to report the location of the population when people are recruited through online channels, which is the case in a large number of the expert-user studies.
Similarly, when selecting for a specific type of users, expert or end-user alike, it may seem reasonable to not focus on the users' location.
However, even when investigating specific users using an online service, the authors' location may predetermine the recruited population's location, for example, due to the language used for recruitment, or due to the service used being biased towards a population, like Amazon MTurk~\cite{redmiles2019well}.

\subsection{Challenges in Recruitment}
\label{ssec:challengesinrec}

In end-user studies, recruiting a representative sample is difficult, as the use of technology is inherently global and cultural differences may influence the effectiveness of security measures~\cite{golla2018site}.
In this case, it is better to acknowledge the limitations of one's population sample and report on the resulting restrictions on the generalizability of the results.

For expert-user studies, representativeness is even more challenging. Recruitment channels are more limited and willingness to participate is often reduced due to the high workload of experts~\cite{Dietrich2018}.

In their work on exploring a convenience sample, Acar et al.~\cite{Acar2017} further discuss the challenges in recruiting participants for expert-user studies.
Different from end-user studies, where recruitment is fairly straightforward (MTurk, posting flyers, classifieds etc.), no well established recruiting processes exist for expert-user studies~\cite{Acar2017}.
This is because it can be difficult to contact and invite professionals for in-lab studies, to find professionals locally, find free time in the experts' schedule or simply to provide enough incentives~\cite{Dietrich2018}.
These observations close the loop to our earlier remarks on convenience samples, such as from a local university or via personal contacts: It is simply easier.
However, when following this path, it is imperative to account for the limitations this introduces for the external validity of the obtained results.

\subsection{External Validity}

The limitations in study populations connect to the matter of external or rather global validity.
External validity is an important parameter to be evaluated to understand the generalizability of results.
To ensure external validity in quantitative studies, the researchers must restrict claims which cannot be generalized to all end or expert users.
This can be due to the interaction of several factors, like participant selection, experimental setting or temporal factors~\cite{creswell2017}.
For qualitative research, generalization has a different meaning.
This is because the intent of qualitative inquiry is not to generalize the findings but to understand a phenomenon in its specific context.
To ensure replicability in such cases, it is crucial to properly document the data collection and interpretation procedures used.

During our evaluation, we find that a majority of studies in both our samples do mention or discuss the generalizability of their findings (\UsrPopulationextvalMenLimitTotal for end-user studies and \ExpPopulationextvalMenLimitTotal for expert-user studies), usually in the form of stated limitations (marked \(\LEFTcircle\)).
However, only \UsrPopulationextvalConsideredAddrTotal end-user studies and \ExpPopulationextvalConsideredAddrTotal expert-user studies take steps to address threats to external validity (marked \(\CIRCLE\)).
Examples of the steps taken include not using a laboratory setting and employing deception~\cite{Karlof2009}, assuring theoretical saturation of the sample~\cite{Redmiles2016}, experience sampling in a wider population~\cite{Reeder2018}, and the global recruitment of specific developer groups (e.g., Google Play or Python developers)~\cite{Wermke2018, Gorski2018}.
However, this leaves \UsrPopulationextvalNotDiscussedTotal end-user and \UsrPopulationextvalNotDiscussedTotal expert-user studies that do not address the generalizability of their findings or mention the limitations thereof, which we mark with a \(\Circle\).
In general, there seems to be a trend to acknowledge limitations, as we find an increasing number of recent papers discuss their generalizability limitations compared to older work.
This still leaves the issue that generalization often means generalization to a U.S. or western population instead of a global population, see, for example, Redmiles et al. from 2019~\cite{redmiles2019well} without explicitly stating this limitation.
Given that \textit{``Most People are not WEIRD \(\lbrack\)(Western, educated, industrialized, rich and democratic)\(\rbrack\)''}~\cite{henrich2010most}, this means that human factors work for expert and end users alike in our community has so far neglected the concerns of the majority of earth's population.
It is imperative to fill this gap in the future.

\subsection{Observations and Recommendations}

\noindent\textbf{Key Observations:}

We find that population samples are dominated by convenience sampling, that is, in the local environment of the researchers or via their personal contacts.
In some cases, we observe Computer Science students being substituted for operators with operational experience~\cite{Krombholz2017}.
Such limitations are regularly not discussed, or only mentioned as a limitation, while general conclusions are drawn.
We tried to be representative by surveying the top security research venues on a global stage.

We found that samples are nearly exclusively sourced from western countries (the U.S., Europe, Australia), without researchers acknowledging that the specific socio-economic background of their population might influence their results.

\noindent\textbf{Key Recommendations:}
In future research, we, the community, must investigate more diverse population samples in terms of where the sample is located in the world to avoid selection bias.
We acknowledge, that this is a hard problem.

However, it is important to have a varied population represented in the top-tier computer security venues. 
Removing systemic bias within the field is a lengthy process, which cannot be paraphrased in a paragraph.
As a point of reference, we recommend a paper by Guillory~\cite{guillory2020}, who takes a stance on systemic racism in AI.
Addressing this problem entails a cultural change in hiring researchers, mentoring early career researchers, and international collaboration.
Indeed, looking at the surveyed papers, we find that international collaboration with researchers from non-western regions, for example, Sambasivan et al.~\cite{Sambasivan2018}, holds promise for research which allows us to explore and understand the impact of one's socio-economic background on security behavior.
The main point here is not ``utilizing'' researchers from the global south in the classical post-colonial western modus operandi to ``get access to samples otherwise inaccessible,'' but instead collaborating with researchers as the peers they are to allow the wider community a better understanding of differences, and shaping technology in a way that enables secure behavior for humans taking their diverse backgrounds into account.
This equally pertains to the perspective of hiring and mentoring, or as Guillory phrased it: \textit{``While substantial research has shown that diverse teams achieve better
performance \(\lbrack\)...\(\rbrack\), we reject this predatory view of diversity in which the worth of underrepresented people is tied to their value add to in-group members''}~\cite{guillory2020}.
Especially given the dominance of western economies not only in research, but also the development of tools and technologies, these steps are imperative to build a securely usable digital and global world.

Nevertheless, research on a population from a specific region has independent scientific value.
However, if we focus our research on a specific region or socio-economic background, we must report the location of the population along with recruitment method, sample size, demographics and discuss the generalizability of the findings to a specific population.
While we see more work acknowledging limitations with regard to their sample population, simply acknowledging the current U.S./western bias is a limitation which we, as a community, must overcome.
Furthermore, convenience sampling, which is currently common, must receive more scrutiny to ensure that results generalize outside its narrow scope, for example, beyond the university-attending population (see WEIRD~\cite{henrich2010most}).
It is important to place the research in the global context and work towards reducing biased data which can have serious real-world consequences~\cite{silva2019}.
If this is not feasible due to the constraints of the research project, the researchers must strive to discuss these limitations in terms of the cultural context and generalizability.

\begin{figure}[tb!]
    \centering
       \begin{subfigure}[t]{0.20\textwidth}
        \centering
        \includegraphics[width=\columnwidth,trim={0 0cm 0 0cm},clip,angle=0]{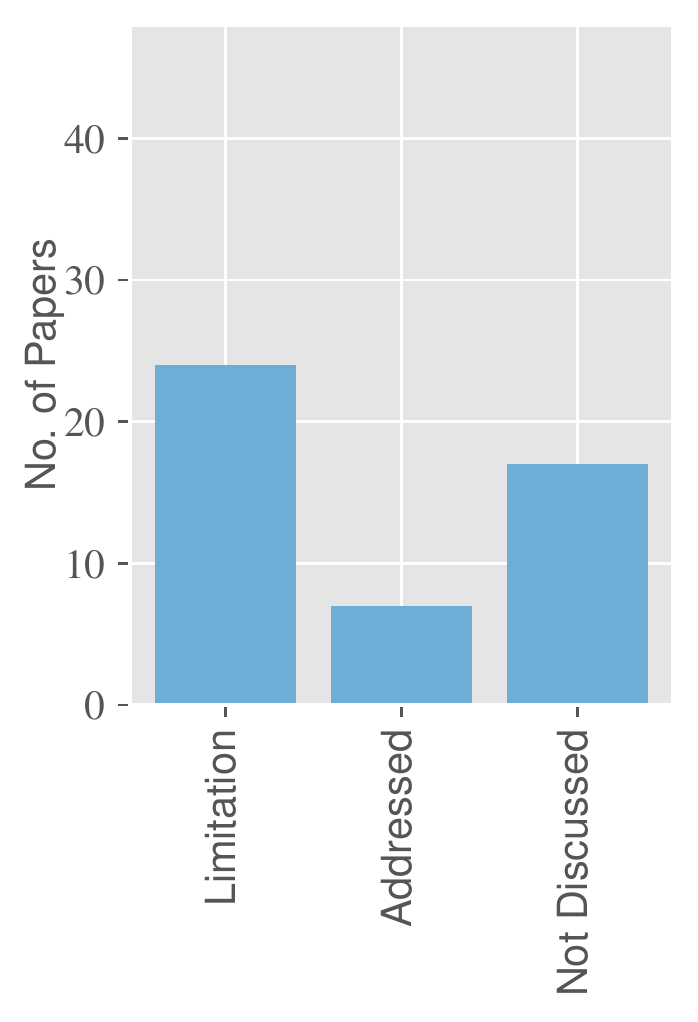}
        \caption{Expert Users}
        \label{fig:extval_exp}
    \end{subfigure}%
     ~
    \begin{subfigure}[t]{0.20\textwidth}
        \centering
        \includegraphics[width=\columnwidth,trim={0 0cm 0 0cm},clip,angle=0]{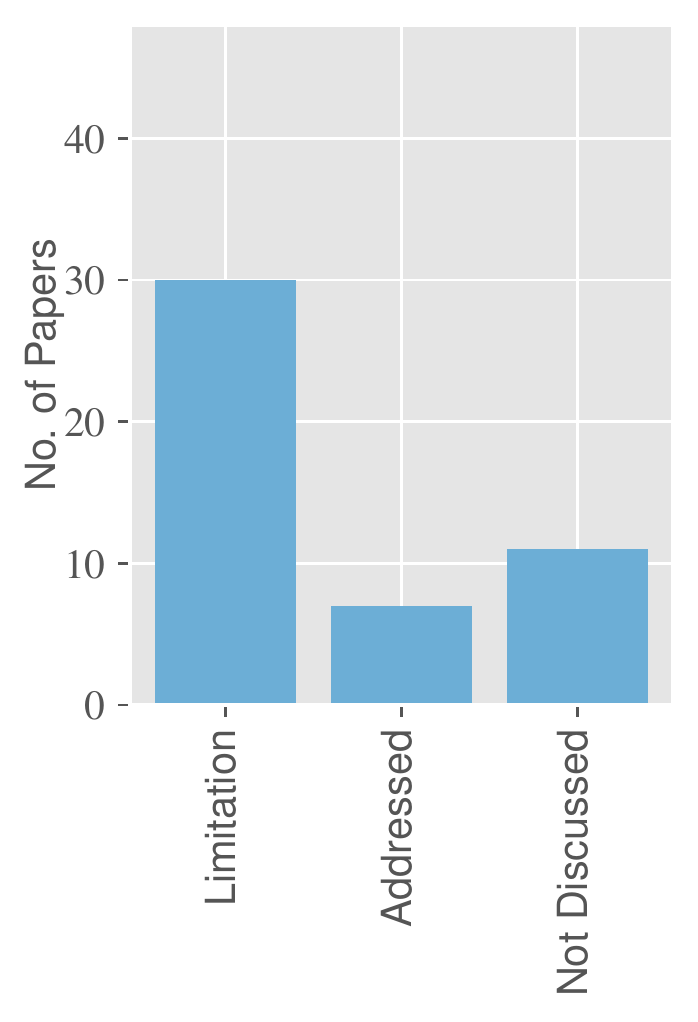}
        \caption{End Users}
        \label{fig:extval_end}
    \end{subfigure}
\caption{Overview of how studies address the external validity of their results for expert users (\Fref{fig:extval_exp}) and end users (\Fref{fig:extval_end}).
Note that only a fraction of papers tries to actively address this limitation, instead of simply stating it.
Furthermore, in expert-user studies, this limitation is more frequently not even mentioned or discussed.
}
    \label{fig:extval}
\end{figure}
Finally, to help the generalizability of the results, we suggest the use of theoretical frameworks.
These can be used to inform the research design as well as aid the external validity of the findings.
We discuss the use of theories in detail in Section~\ref{sec:ana-theo}.

\section{Research Objective}
\label{sec:ana-obj}

Following, we investigate the research objective of human factors in security research, that is, \emph{what} researchers are investigating.
For an overview of our findings, please see column ``Res. Obj.'' in \Fref{tab:exp-papers} and \Fref{tab:usr-papers}.

\subsection{User Perspective and Exploration}
\begin{figure*}[!t]
\begin{minipage}[t]{0.49\textwidth}
    \centering
       \begin{subfigure}[t]{0.40\textwidth}
        \centering
        \includegraphics[width=\columnwidth,trim={0 0cm 0 0cm},clip,angle=0]{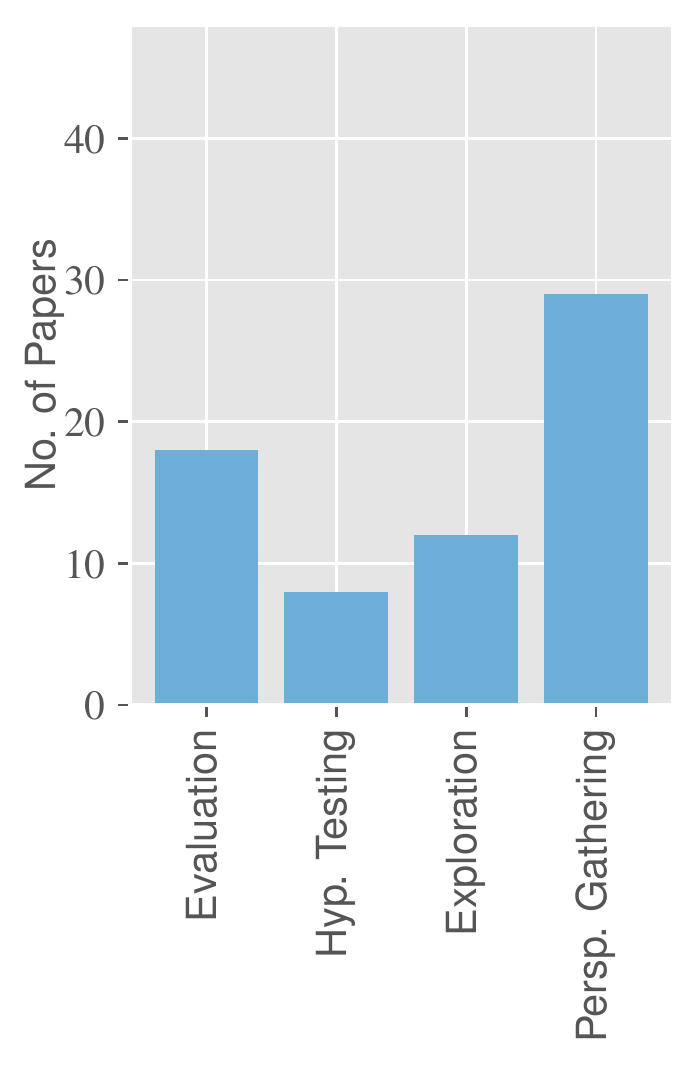}
        \caption{Expert Users}
        \label{fig:obj_exp}
    \end{subfigure}%
     ~
    \begin{subfigure}[t]{0.40\textwidth}
        \centering
        \includegraphics[width=\columnwidth,trim={0 0cm 0 0cm},clip,angle=0]{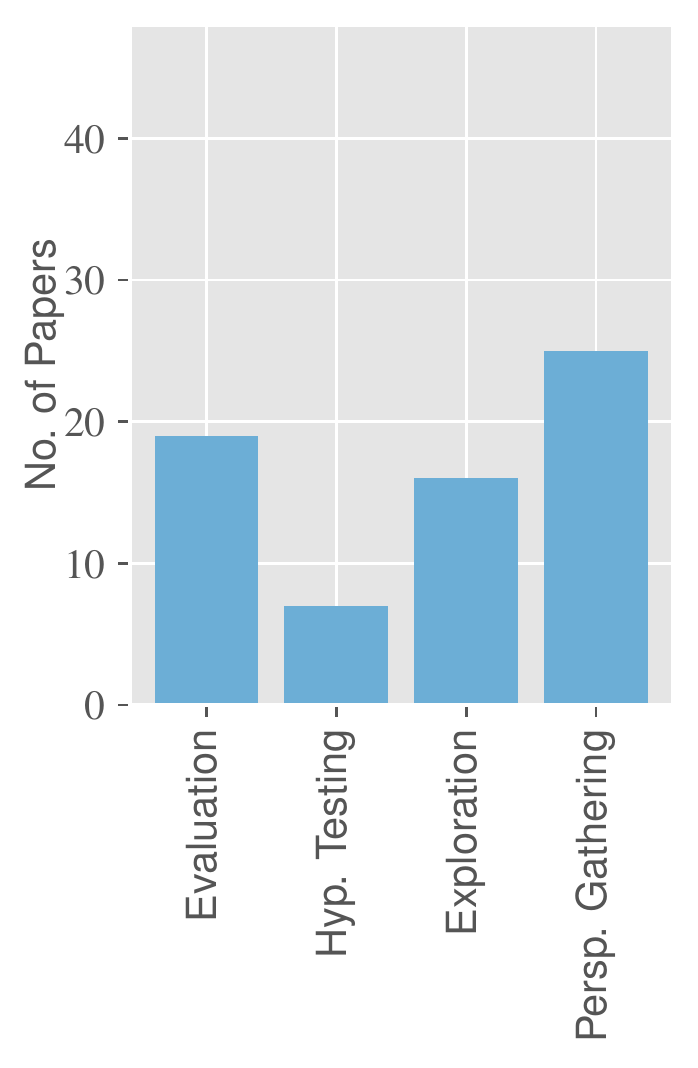}
        \caption{End Users}
        \label{fig:obj_end}
    \end{subfigure}
\caption{Overview of research objectives in expert-user (\Fref{fig:obj_exp}) and end-user papers (\Fref{fig:obj_end}).
We find no fundamental differences in this parameter, apart from a slightly higher number of perspective gathering work in the expert-user sample.
We conjecture that this is due to work on expert users only now becoming more prevalent.
}
    \label{fig:obj}
\end{minipage}
\hfill
\begin{minipage}[t]{0.49\textwidth}
    \centering
       \begin{subfigure}[t]{0.40\textwidth}
        \centering
        \includegraphics[width=\columnwidth,trim={-0cm -2cm -0cm -0cm},clip,angle=0]{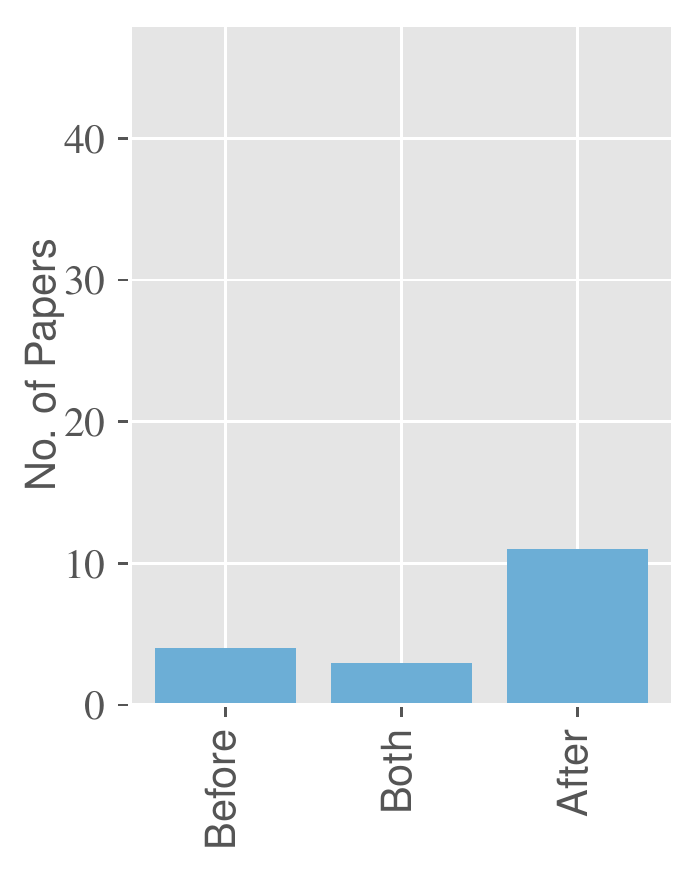}
        \caption{Expert Users}
        \label{fig:ds_exp}
    \end{subfigure}%
     ~
    \begin{subfigure}[t]{0.40\textwidth}
        \centering
        \includegraphics[width=\columnwidth,trim={0 -2cm 0 0cm},clip,angle=0]{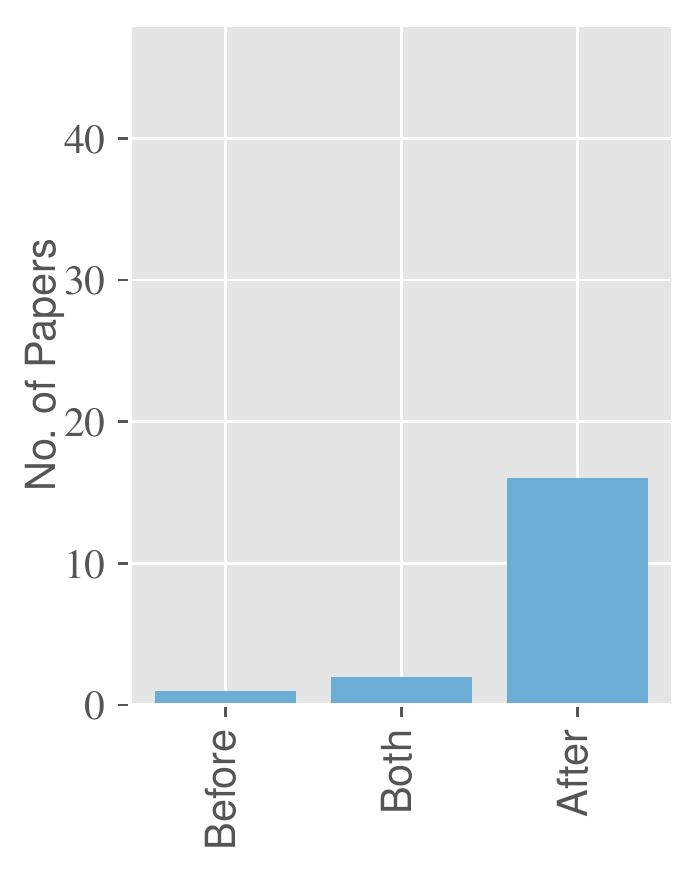}
        \caption{End Users}
        \label{fig:ds_end}
    \end{subfigure}
\caption{Overview when artifacts are evaluated and whether user perspectives/requirements are collected before the artifact is being developed in expert-user (\Fref{fig:ds_exp}) and end-user papers (\Fref{fig:ds_end}).
We find a very classical approach of \emph{first} building a system and \emph{then} evaluating it, instead of first collecting users' requirements and perspectives.
}
    \label{fig:ds}
\end{minipage}
\end{figure*}

Investigating the perspective of the user is the most common research goal across both expert and end users (see \Fref{fig:obj}).
For example, Dietrich et al. investigate system operators' perspective on security misconfigurations~\cite{Dietrich2018}.
However, exploratory research is more prevalent for end-user studies, while a stronger emphasis is put on perspective gathering in expert related studies.
Note the distinction between exploratory research and research trying to understand users' perspective:
While the former tries explore a new area from an external point of view, the latter strives to describe how a specific user group \emph{perceives} an issue.
Interestingly, earlier work on expert users is dominated by work that evaluates artifacts, while more recent work shifted towards looking at their perspective on specific issues.
This is in line with a mechanic in very early research focusing on end users, for example Whitten and Tygar~\cite{whitten1999johnny}, which also started out by evaluating artifacts, and then matured into considering users' perspectives.

For expert-user literature, a majority of it is concerned with gathering the user perspective and \ExpResearchfocusExploreMarkedTotal publications are exploratory research.
For end-user publications, there is a similar distribution between papers that are gathering the users' perspective and those that are exploratory.
Gathering users' perspective is common for issues that are prevalent and understudied.
Hence, in these cases, perspective gathering research is exploratory by nature.

Compared to expert-user research, slightly more end-user studies are exploratory.
This might be the case because end-user research has been more prevalent and expert-user research is only slowly getting traction in the last few years.
For both user categories, however, exploration itself is not the sole aim of most research.

\subsection{Evaluation and Rigorous Design}
Artifact evaluation is similarly common between end-user and expert-user studies, including the overlap with other research objectives.
In both cases, about half of the existing research is solely performing an evaluation study and the remainder overlaps with the other aims.

Most evaluation studies evaluate an existing or new artifact, but not all of them directly evaluate the \emph{usability} of an artifact.
For example, Wermke et al. performed a (non-user) evaluation of a tool to study obfuscation in Android applications~\cite{Wermke2018}.
For all evaluation studies, we identify under ``Design Eval.'' as part of the ``Theory/Framework'' columns whether the evaluation was purely done to test something \emph{after} (\(\RIGHTcircle\)) it has been built, if they first collect users' input to then design an artifact (\(\LEFTcircle\)), or if they combine both approaches (\(\CIRCLE\)).
Only \UsrTheoryandFrameworkDSBeforeAfterTotal end-user studies and \ExpTheoryandFrameworkDSBeforeAfterTotal expert-user studies gather requirements and input before designing an artifact, and later evaluate their artifact against the users again, see \Fref{fig:ds}.
A further \UsrTheoryandFrameworkDSBeforeTotal end-user study and \ExpTheoryandFrameworkDSBeforeTotal expert-user studies gather input from users before designing the artifact without validating the created artifact afterwards, again, see \Fref{fig:ds}.
This approach has the disadvantage that users' requirements are not incorporated in the design process of the artifact, which is problematic because the users' actual requirements may be different from the imagined user requirements, thus leading to poor artifacts.
In industry, most development processes incorporate a user-driven design component, hoping to prevent a requirements mismatch~\cite{vredenburg2002survey}.

The information systems community has already recognized the missing rigor in their artifact design and evaluation.
To counteract this limitation, they formalized a processes known as ``Design Science Research'' (e.g., see March and Smith~\cite{march1995design} or Hevner et al.\@~\cite{von2004design,hevner2007three}).
We suggest that studies in computer security that are in fact designing and evaluating an artifact also leverage the Design Science framework~\cite{von2004design,hevner2007three}.
Unfortunately, we could not identify any paper in our sample that explicitly uses the Design Science framework to inform their research.

\notext{
\subsubsection{Design Science: A Summary}
\label{ssec:ds}
Hevner summarizes the methodology of design science based research as three conjoined cycles: the relevance cycle, the rigor cycle, and the design cycle~\cite{hevner2007three}.
The relevance cycle includes the initial requirements engineering, and definition of acceptance criteria for the final evaluation of the artifact.
In this stage, the researchers should explore and include stakeholders' requirements for a specific artifact.
In the next cycle, the rigor cycle, researchers explore the related work and bodies of past research.
This serves to ensure that their solution is innovative, yet grounded in prior knowledge, and does not repeat past mistakes~\cite{fiebig2018learning}.
Last, the design cycle oscillates between implementing and evaluating the artifact, in line with ideas from the user-driven design world~\cite{vredenburg2002survey}.
Of course, one might naturally claim that this is how computer scientists and engineers have been developing and researching IT solutions since the beginning of the Internet.
However, given the lacking rigor in requirements engineering that we found among evaluation studies, we believe that the community would benefit from explicitly adopting the rigorous framework of design science more closely.
}

\subsection{Hypothesis Testing}
Other fields, like the social sciences and safety science, regularly use theories as a guiding concept in their research.
They employ a body of existing theories to formulate hypothesis that they can then test using appropriate research designs.
Of course, there are other ways to create a hypothesis, such as through previous work or through anecdotal evidence.
Only \ExpResearchfocusTestahypothesisMarkedTotal expert-user studies test a hypothesis, of which only one also uses an existing theory or framework.
The remaining ones build hypotheses based on informal observations and related work.
For end-user studies, \UsrResearchfocusTestahypothesisMarkedTotal papers test hypotheses.
In general, work testing hypotheses often overlaps with evaluation and exploratory studies, and only few papers solely focus on testing a hypothesis.

\subsection{Observations and Recommendations}

\noindent\textbf{Key Observations:}
At the moment, research is dominated by exploratory and perspective work, focusing on instances of problems instead of generalizing to a wider societal and organizational setting.
Especially considering our earlier observations on recruitment and a geographic bias in current work, this poses a challenge for our field.
As a field, we have to move beyond purely observing, and conduct work that systematizes, understands, and proposes solutions to the effects we observe.

\noindent\textbf{Key Recommendations:}
To accomplish the further maturation of our field, we suggest that researchers who investigate human factors in computer security adopt the concept of theories (see \Fref{sec:ana-theo}).
Furthermore, we recommend that researchers adopt the formal process of design science~\cite{march1995design,von2004design,hevner2007three}.
While, technically, some work already follows (parts) of this framework, diligently following it can increase the rigor and reproducibility in our work.
This will allow us to build and refine our understanding, and derive and test solutions from this body of understanding in a structured way.

\section{Research Methods}
\label{sec:ana-method}

\begin{figure*}[!t]
    \centering
       \begin{subfigure}[t]{0.5\textwidth}
        \centering
        \includegraphics[width=\columnwidth,trim={0 0cm 0 0cm},clip]{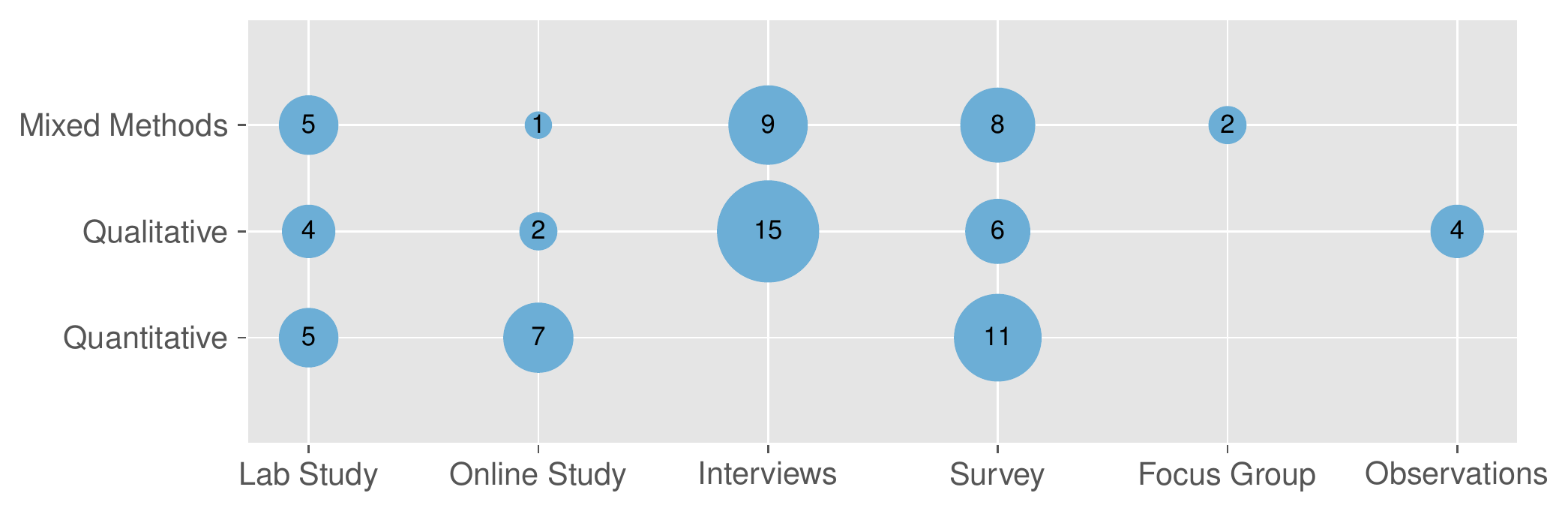}
        \caption{Expert Users}
        \label{fig:rm_exp}
    \end{subfigure}%
     ~
    \begin{subfigure}[t]{0.5\textwidth}
        \centering
        \includegraphics[width=\columnwidth,trim={0 0cm 0 0cm},clip]{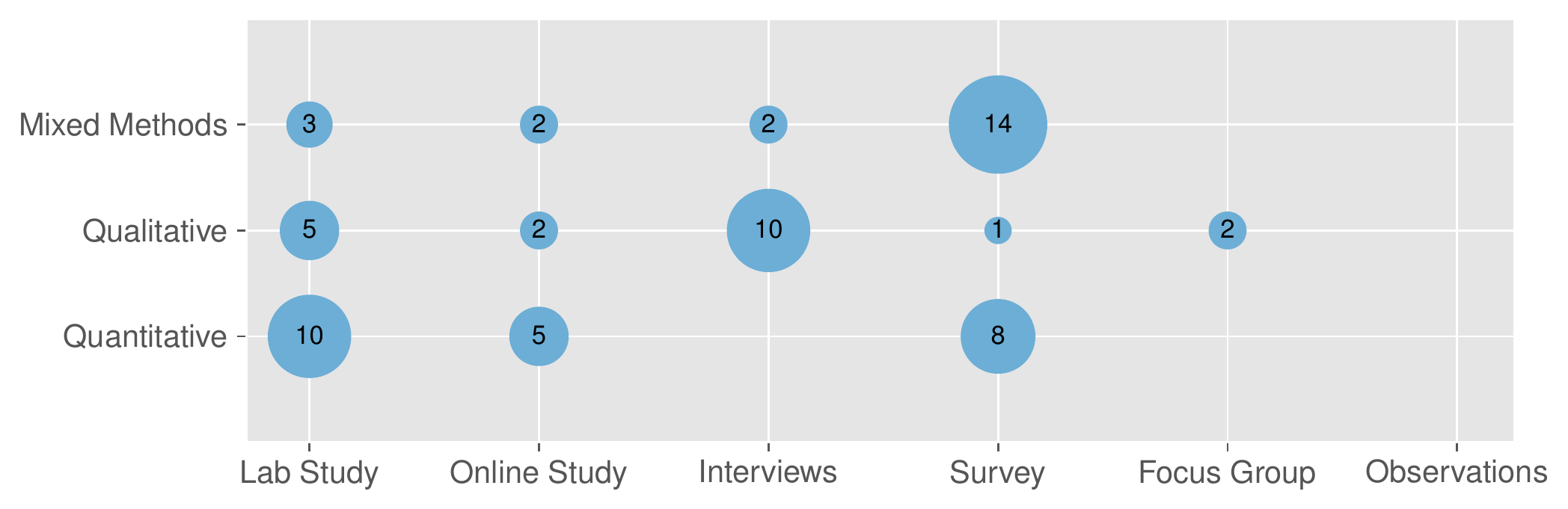}
        \caption{End Users}
        \label{fig:rm_end}
    \end{subfigure}
\caption{Overview of research methods in expert-user (\Fref{fig:rm_exp}) and end-user papers (\Fref{fig:rm_end}).
We find a classical distribution of methods (Surveys more quantitative/mixed methods and Interviews more qualitative/mixed methods).
In expert-user related research we find focus groups as a common instrument to generate the foundation of a questionnaire.
}
    \label{fig:research_methods}
\end{figure*}

In this section, we analyze the research methods that are used to perform user studies, that is, \emph{which research methods} are used to investigate users?
Research methodologies are usually quantitative (statistical evaluation of large datasets), qualitative (extraction of qualitative insights from data not statistically analyzable), or both (mixed methods).
According to Creswell~\cite{creswell2017}, a quantitative approach tests theories by developing hypotheses and collecting data to support or refute the hypotheses.
This is done using an experimental design and instrument-based data collection (like a survey) followed by a statistical analysis.
The qualitative approach, however, seeks to understand the meaning of certain phenomenon from the views of the participants situated in specific contexts.

For mixed methods research, both approaches are combined, either sequentially (elaborate the findings of one method with another method), concurrent (merging data from both to provide a comprehensive analysis), or transformative (an overarching theoretical lens within a design using both data types)~\cite{creswell2017}.

The column labeled ``Research Method'' in \Fref{tab:exp-papers} and \Fref{tab:usr-papers} holds a summary of our findings.
We mark studies using a quantitative approaches (\(\LEFTcircle\)), those following a qualitative approach (\(\RIGHTcircle\)), and those using mixed methods (\(\CIRCLE\)).

In our sample, we find all three research approaches are being used across six common research tools.
However, we notice that quantitative methods are sometimes used for qualitative research and vice-versa, for example, by collecting data for statistic analyses in interviews, or by collecting free-text responses in surveys.
We also find that there is no consistency in explicitly mentioning the methodology used to inform the research design and select an appropriate research tool.

For expert users, interviews and surveys are the most used research method, while focus groups and naturalistic observations are least used.
Intriguingly, especially naturalistic observations do not suffer from a self-reporting bias, as can usually be found in surveys and interviews~\cite{redmiles2017summary,}.
For end users, surveys are the most used method, followed by laboratory studies.
While only two studies conducted focus groups, none of the end-user studies in our sample have employed naturalistic observation as a research method.

In our analysis, we find that the expert-user research has a slightly and not significantly higher number of qualitative research compared to mixed methods research and quantitative research (15 mixed methods, 15 quantitative, 18 qualitative) while the end-user research has a high number of quantitative research (17 mixed methods, 18 quantitative, 13 qualitative).
The research methods used are also dependent on the identified research objectives (see \Fref{sec:ana-obj}).
Research gathering users' perspectives is mostly qualitative or mixed methods research.
Evaluation studies, on the other hand, are mostly quantitative or mixed methods.
Studies that test a hypotheses are almost entirely quantitative, as to be expected.
Finally, exploratory studies are mostly qualitative or mixed methods.

Hence, our results are in line with our earlier observations on research objectives.
With an emphasis on exploratory and perspective gathering research, qualitative methods are common.
Quantitative methods are more prevalent in evaluation studies and hypotheses testing.
Where as understanding user perspective or performing exploratory research requires qualitative methods, as they are applicable when studying novel phenomenon or explaining social factors and dynamics.

\subsection{Observations and Recommendations}

\noindent\textbf{Key Observations:}
At the moment, the choice of research tools is commonly driven by the ultimate goal of a study, instead of being a result of a reflection on these goals.
We also find that naturalistic observations, which, as we mentioned earlier, would be instrumental in understanding secure behavior especially in the day-to-day workings of expert users are not commonly used.

\noindent\textbf{Key Recommendations:}
We suggest that future research considers the trade-off between a study's objective and the available tools more carefully.
Especially for exploratory work, researchers should consider naturalistic observations and technical measurements of behavior~\cite{dekoven2019measuring} more closely, instead of relying on interviews and surveys, which potentially suffer from a self-reporting bias.

\section{Theory}
\label{sec:ana-theo}

The use of theories is a common practice in the social sciences.
According to Van de Ven~\cite{van1989nothing}, theories explain why something is happening by describing and explaining causal relationships.
They help us to see the findings of a particular study as special cases of a more general set of relationships, rather than as isolated pieces of empirical knowledge.
These relationships can then be tested and revised by others.
Gregor~\cite{gregor2006nature} claims that a good theory consists of three elements:
\begin{enumerate*}[label=\emph{(\roman*)}, itemjoin={{, }}, itemjoin*={{, and }}, after={{. }}]
\item \textit{Generalization:} Abstraction and generalization from one situation to another are key aspects of any theory
\item \textit{Causality:} Causality is the relation between cause and effect
\item \textit{Explanation and Prediction:} Explanation is closely linked to human understanding, while predictions allow the theory to be tested and used to guide action
\end{enumerate*}

In summary, theories (should) explain \emph{why} something happens and from this starting point, can be used for prescriptive or design purposes.
Theorizing can bring together different understandings of the problem, thereby ensuring that research contributes to a general class of problems and to a broad variety of organizational and societal settings, instead of a single problem instance.
Especially the last step is instrumental to generalize results and provide a scientific foundation.

\subsection{Theory Use}
\label{ssec:theo-use}
We investigate if and how human factors researchers in computer security have used theories.
In case the authors did not use an established theory, we survey a list of existing theories to identify applicable ones~\cite{TwenteThoeries}, marked with a \(\Circle\) in the tables.
The list of theories was compiled by the Communication Science department at the University of Twente in 2003/2004 for students to better understand theoretical frameworks and aid them in choosing one.

We find 20 papers, seven expert-user papers and thirteen end-user papers that actively use a theory to inform their research, which we mark with \(\CIRCLE\) under the theories section.
A further three papers on expert users and nine on end users mention theories in the context of their findings, which we mark with \(\RIGHTcircle\).

The most commonly used theory is that of mental models, which is being used in six (two expert and four end-user papers) and mentioned in a further three end-user papers.
Mental models are used as a tool to study the ways in which users understand and interact with their environments.
Furthermore, we find a cluster of three papers focusing on activity theory.
Activity Theory is based on the idea that activity is primary~\cite{HashimAT}.
It holds that doing precedes thinking and that goals, images, cognitive models, intentions and abstract notions like ``definition'' emerge out of people doing things.
Apart from these clusters, we find a diverse set of individual theories being used or mentioned in the remaining 26 papers from both samples that use or mention a theory.

We also evaluated the papers to see which theories might have been applicable, based on their research topic.
Mental Models are the most commonly applicable theory, applicable to a further 21 papers, ten for expert users and eleven for end users.
Sensemaking theory~\cite{Weick1995} is promising as well, as it would be applicable to 19 expert-user papers, and two more end-user studies.
The theory of reasoned action~\cite{fishbein1977TRA} holds promise for two expert-user papers and six end-user papers.

Apart from these three theories, the other theories are only applicable to a limited set of papers, as, for example, activity theory is only applicable to the three papers where it is also being used.
There is no one-size-fits-all approach of a set of ``best'' theories to inform human factors in security research.
Instead, we suggest that researchers do not only focus on selecting specific ``heavy hitter'' theories, but instead refer to a more comprehensive list, such as the one by the University of Twente~\cite{TwenteThoeries}, at the beginning of their research projects.

\subsection{Grounded Theory}
\label{ssec:gt}
\begin{figure}[tb!]
    \centering
       \begin{subfigure}[t]{0.20\textwidth}
        \centering
        \includegraphics[width=\columnwidth,trim={0 0cm 0 0cm},clip,angle=0]{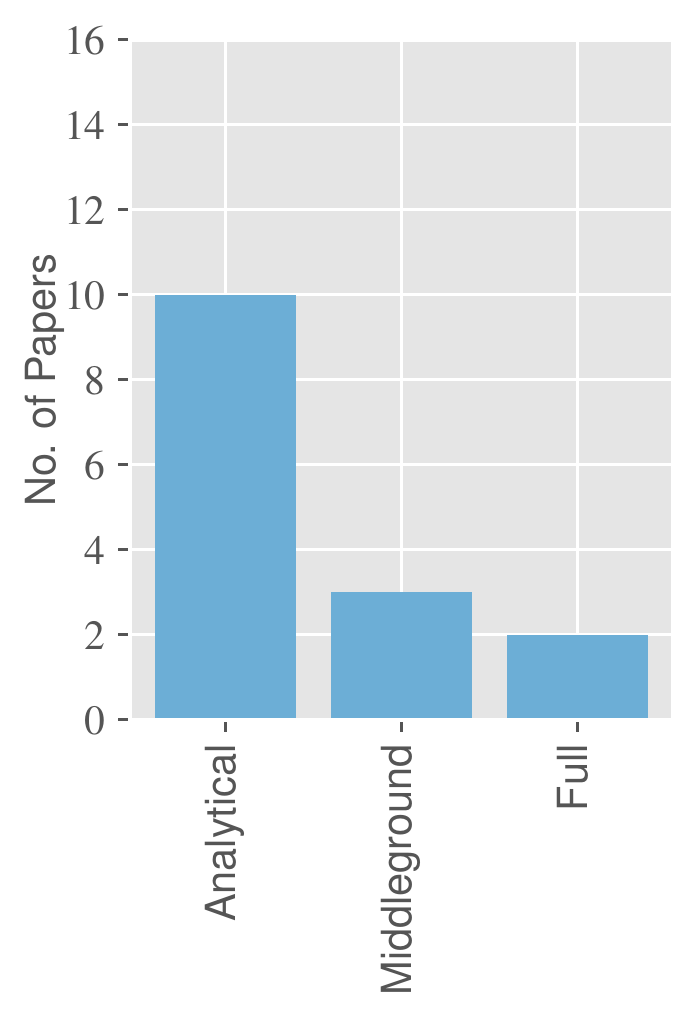}
        \caption{Expert Users}
        \label{fig:gt_exp}
    \end{subfigure}%
     ~
    \begin{subfigure}[t]{0.20\textwidth}
        \centering
        \includegraphics[width=\columnwidth,trim={0 0cm 0 0cm},clip,angle=0]{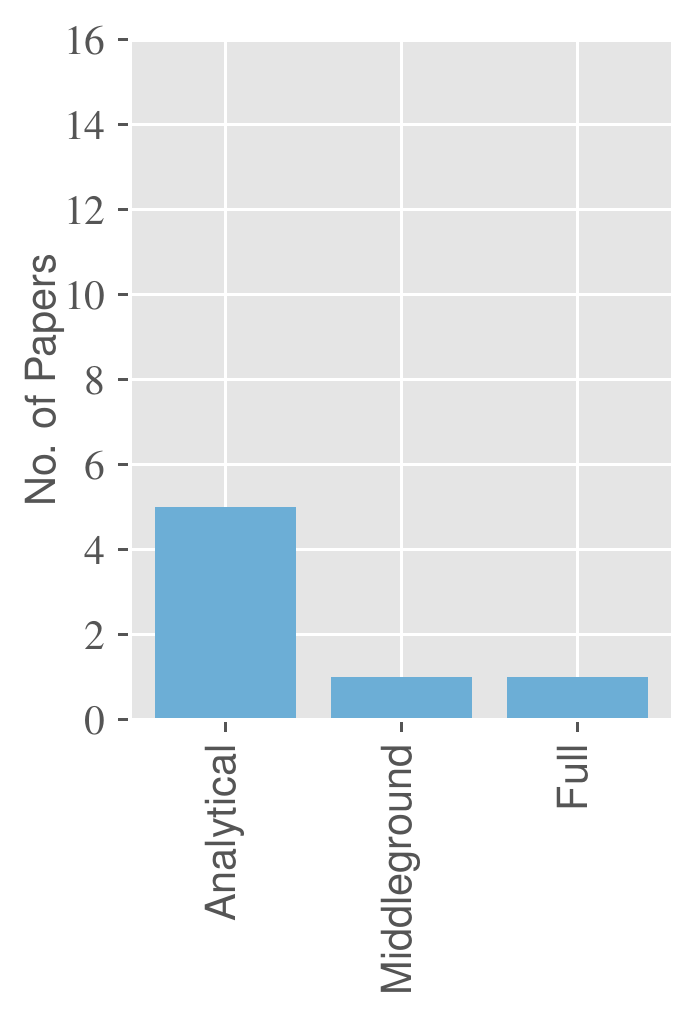}
        \caption{End Users}
        \label{fig:gt_end}
    \end{subfigure}
\caption{Overview of how Grounded Theory is being used in expert-user (\Fref{fig:gt_exp}) and end-user papers (\Fref{fig:gt_end}).
In both samples, the majority of papers claiming to use GT do so analytically, i.e., skip the theory generation step.
Note, that GT is far more prominent in explorative research with expert users, but the distribution between papers fully using GT and those only using it for analytical purposes is comparable.
}
    \label{fig:gt}
\vspace{-1em}
\end{figure}

Grounded Theory (GT), first developed by Corbin and Strauss~\cite{corbin1990}, is a structured method to derive a theory from data, instead of utilizing an existing theory.
It is a common method for exploratory research, especially in new and emerging fields, and when using qualitative data sources.
We surveyed all papers in our sample on their use of GT, independent from their use of other established theories.

We find that more than twice as many (\ExpTheoryandFrameworkGTMarkedTotal compared to \UsrTheoryandFrameworkGTMarkedTotalNumber) papers investigating expert users, rather than end users, leverage grounded theory.
This is in line with our earlier observation that expert studies primarily focus on exploratory research, such as investigating user perspectives on issues or their work environment, or trying to get a first look at a specific issue.
These approaches usually rely on qualitative data and, hence, are amenable to a GT-based methodology.

However, when investigating \emph{how} GT is being used in the literature, we find that the majority of papers do not use GT to develop a new theory (see column ``GT'' under ``Theory/Framework'').
Instead, most studies (\ExpTheoryandFrameworkGTAnalyticalTotalNumber/\ExpTheoryandFrameworkGTMarkedTotal for experts and \UsrTheoryandFrameworkGTAnalyticalTotal/\UsrTheoryandFrameworkGTMarkedTotal for end-user studies) reference GT only for the sake of the coding process, including the calculation of Cohen's kappa for inter-rater reliability, and rules for establishing saturation.
This means that the authors do not follow the full four-step process for GT (open coding, axial coding, selective coding, theory generation) by omitting the last stage.
Instead, these publications provide conclusions around an overview of the discovered codes, often connected to specific quotes from the interviews.
This form of incompletely applying grounded theory as a method to present raw data and enrich it with statistical information to seemingly reach a higher level of validity is also known issue in other fields, for example, management sciences~\cite{suddaby2006editors}.
We mark these \(\Circle\) in the tables.

A further \ExpTheoryandFrameworkGTMiddlegroundTotal papers on expert users, and \UsrTheoryandFrameworkGTMiddlegroundTotal paper for end-users use a middle-ground approach~\cite{sekaran2016research}.
Instead of generating their own theory from the collected data, they utilize an existing theory to explain their findings obtained by the first three steps of GT, or they adapt an existing theory to synthesize their findings.
We mark these \(\RIGHTcircle\) in the tables.
Ultimately, in our sample, only \ExpTheoryandFrameworkGTFullTotal papers on expert users and \UsrTheoryandFrameworkGTFullTotal on end users execute all four steps of GT to contribute to the theory corpus in the field, which we mark \(\CIRCLE\).
In general, these findings align with observations of McDonald et al.\@~\cite{mcdonald2019reliability}, who found uncertainty in the HCI community on when and how to use indicators like inter-rater reliability and a tendency to ``expect'' numeric measures to underline a study's reliability.

\subsection{Observations and Recommendations}

\noindent\textbf{Key Observations:}
At the moment, only a quarter of surveyed human factor papers use theories to guide their research design and result interpretation.
While mental models are a common tool to inform research design, we find no other theory that is consistently \emph{used} across several papers.
Theories that are applicable to a wide range of studies, still go unused (Theory of Reasoned Action, Sensemaking Theory).
This lack of theory is, from a scientific perspective, concerning.
Other authors, for example Muthukrishna and Henrich~\cite{muthukrishna2019problem} see one of the causes for the replication crisis in psychology in an inconsistent and not overarching use of theories in the field.
Grounded Theory, a technique for generating new theories from data is commonly claimed to be used, yet authors do not leverage its potential to generate theories.
Instead, they focus on the analytical aspects of grounded theory to present their data.

\noindent\textbf{Key Recommendations:}
To mature from this state, we encourage the field to adopt the concept of using and improving existing theories, as well as forming new ones.
As already mentioned in \Fref{sec:ana-obj}, theories can help the field to generalize findings in specific situations and use these generalizations to implement and test improvements to the handling of the human factor in IT security.
Given the state of the field, we might indeed be already in a situation similar to the replication crisis of psychology~\cite{muthukrishna2019problem}.
\emph{Grounded Theory}, which can be used for this, is already commonly being used, yet not executed fully.
Hence, we recommend authors adopt the full four-step approach of GT and start to formulate theories.
Given the emerging nature of the field, theories do not yet have to be refined.
Instead, we should start into a process of iteratively testing, validating, and improving findings from earlier work.
We recommend as further research more replication studies, as well as studies replicating findings in diverging socio-economical backgrounds (see \Fref{sec:populationsample}).

\notext{
\section{Existing Theories}
\label{sec:ext-theories}
Next, we describe the three most applicable theories (Sensmaking, Theory of Reasoned Action, Mental Models), as well as a theory in the context of the (missing) intercultural dimension of research (Cultural Dimension Theory.)
We believe that providing these introductions will be beneficial to the community, even though researchers should study existing summaries of theories~\cite{TwenteThoeries} to select the most applicable to their work.

\subsection{Sensemaking Theory}
\label{ssec:st}
Sensemaking Theory is about the process of collectively making sense of a situation.
It refers to the ``ongoing retrospective development of plausible images that rationalize what people are doing''~\cite{WeickSM}.
There are seven aspects to Sensemaking~\cite{Weick1995}:
\begin{enumerate*}[label=\emph{(\roman*)}, itemjoin={{, }}, itemjoin*={{, and }}, after={{. }}]
\item It is grounded in identity construction, that is,  a sensemaker with self-agency
\item Retrospective---reflection on experience and creation of meaning by paying attention to what has already happened
\item Enacting of sensible environments---by taking action, one creates their own environment
\item Social process---involves social function and communication; influenced by the actual, imagined or implied presence of others
\item Ongoing process---processes never start or stop
\item Focused on and by extracted cues---people focus on certain aspects (cues) of an experience and make generalizations for the whole
\item Driven by plausibility rather than accuracy---meanings ascribed to experiences have to seem reasonable and objectivity is not important
\end{enumerate*}

Sensemaking theory can be used to study individuals~\cite{Herrmann2007}, groups~\cite{bird2007sensemaking} and organizations~\cite{maitlis2013sensemaking}.
Other examples of applications include studying responder behaviour in cyber-attacks~\cite{tapanainen2017}, sensemaking questions in crisis response teams~\cite{kalkman2019}, and workplace ``strangeness'' \cite{erbert2016}.

In our evaluation, we found that the sensemaking theory was not used or mentioned in any paper, even though it would have been applicable, especially in expert studies (20 expert and 2 end-user studies).
We find expert-user studies to be over represented here as ~63\% of them focus on gathering the users' perspective, while implicitly considering the expert user as an agent making conscious choices (see \Fref{sec:perspective} regarding the role of the expert).
Hence, as expert users are subjected to abnormal situations more often, the Sensemaking theory can help in understanding how the experts make sense of disruptions, critical events and unexpected situations and subsequently, how this affects their (security-critical) actions.

\subsection{Theory of Reasoned Action}
\label{ssec:tra}
The Theory of Reasoned Action (TRA) is about explaining human behaviour.
TRA suggests that one's behaviour is determined by one's intention (plan of action/expressed willingness) and that intention can be considered the best predictor of human behaviour~\cite{fishbein1977TRA, fishbein1980TRA, TwenteThoeries, LaCaille2013}.
TRA suggests that intention, in turn, is determined by:
\begin{enumerate*}[label=\emph{(\roman*)}, itemjoin={{, }}, itemjoin*={{, and }}, after={{. }}]
\item People's attitudes---beliefs regarding outcomes contribute to attitude formation
\item Subjective norms---the perceived social pressure from important 'others' to perform or not perform the behavior
\end{enumerate*}
The Theory of Planned Behaviour (TPB) extends this concept to include perceived behaviour control as the third determinant of intention~\cite{ajzen1991TPB, LaCaille2013}.
This refers to peoples' own perception of their ability to perform certain behaviours, and of the external factors that may enable or impede them.

This theory is used to explain and predict human behaviours.
Examples of the application of this theory include predicting unethical behaviour~\cite{chang1998tra}, predicting interaction with Facebook page like ads~\cite{kim2015tra}, and a prevention program to reduce cyberbullying~\cite{doane2016tra}.
In our evaluation, TRA seemed to be applicable to a total of 9 papers (2 expert and 7 end-user studies).
Seven of these studies are identified as perspective-gathering studies.
TRA is useful when studying human behaviour from the perspective of intentions and how these intentions are shaped by factors like attitudes, norms and beliefs.

\subsection{Mental Models}
\label{ssec:mm}
Mental models are mental representations that people have to understand a specific phenomenon and predict the consequences of certain actions~\cite{TwenteThoeries,Mathieu2000}. The idea is that the mental model shapes behavior.
The theory offers three main predictions \cite{TwenteThoeries}:
\begin{enumerate*}[label=\emph{(\roman*)}, itemjoin={{, }}, itemjoin*={{, and }}, after={{. }}]
\item Normally, reasoners build models of what is true, not what is false
\item It is easier to reason from one model than from multiple ones
\item Reasoners are led to erroneous conclusions as they tend to focus on one of the many models of multi-model problems
\end{enumerate*}
Mental models are used to study users' understanding of and interaction with their environments.
Examples include understanding users' mental models of the internet~\cite{Kang2015}, of Tor operations~\cite{Gallagher2017} and of privacy through illustration~\cite{Oates2018}.

While the theory is used in only a small number of papers (2 expert and 5 end-user studies), it was found to be applicable in a larger number of studies (11 expert and 12 end-user studies).
This is---again---consistent with the finding that a majority of the papers are investigating the user perspectives and understanding.
Mental models are a useful tool for this because our deductive reasoning depends on our mental models \cite{johnson2001}.

\subsection{Cultural Dimension Theory}
\label{ssec:cdt}

Cultural Dimension Theory (CDT) is a conceptual framework from cross-cultural psychology. It describes the interaction between a society's culture and the behaviour of its members.
Geert Hofstede identified four dimensions his seminal 1984 work ``Culture's Consequences''~\cite{Hofstede1984cc}.
Subsequent research led to the addition of two more dimensions~\cite{Hofstede2011}.
Hofstede's six cultural dimensions are~\cite{Hofstede2011}:
\begin{enumerate*}[label=\emph{(\roman*)}, itemjoin={{, }}, itemjoin*={{, and }}, after={{. }}]
\item Power Distance---the extent to which the less powerful members of the society accept and expect that power is distributed unequally
\item Individualism vs. Collectivism---the degree to which society's members are integrated into groups
\item Masculinity vs. Femininity---the presence and distribution of so-called masculine and feminine roles and values across members in society
\item Uncertainty Avoidance---society's tolerance for ambiguity
\item Long Term vs. Short Term Orientation---some values associated with Long Term Orientation are perseverance, thrift, learning from others and some values associated with Short Term Orientation are traditions are sacrosanct, pride, service to others
\item Indulgence vs Restraint---the degree of freedom in gratification of basic and natural human desires related to enjoying life and having fun as opposed to controlled gratification of needs, regulated by strict social norms
\end{enumerate*}

While it is mostly used in the context if international business or marketing, CDT has found application in other fields.
Some examples include the exploration of attack vector preferences for attackers~\cite{Sample2017}, studying cultural dimensions and cybersecurity development~\cite{Onumo2017} and to examine factors influencing information security policy violations~\cite{al2015}.

In our dataset, we did not come across the CDT being applied or mentioned in any paper.
However, considering that current research appears to be Western-centric in terms of the population sample used and the authors' location, CDT may play an important role in overcoming this limitation.
Studies performed using CDT show how various countries score differently on Hofstede's cultural dimensions index~\cite{hofstede2003cultural}.
If the community decides to move to a more inter-cultural perspective, CDT will be instrumental in making results from international studies comparable by helping researchers place their results within a global cultural context.
In turn, this will help in deriving general theories and thinking about the limits to generalizability.

\noindent\textbf{Key insights:}
We describe four existing theories that are, in addition to grounded theory, feasible starting points for the field to adopt a theory-driven approach.
Especially Cultural Dimension Theory provides necessary tools to overcome the current situation of Western-centric human factors in computer security research.
}

\section{Ethics}
\label{sec:ana-ethics}
In this section we assess the implementation of ethical considerations in research involving human subjects.
Traditionally, this includes whether the study is ethically justifiable, especially in the context of deception studies and whether participants were exposed to unreasonable harm.
However, this point usually also includes whether informed consent was correctly obtained, and the general handling of research data, i.e., whether applicable local privacy laws are followed, and if the authors anonymized the data as soon as feasible during the research project.

Hence, for each paper we identify whether ethical considerations were properly discussed \emph{and} the study has been submitted to an ethics review board\footnote{A common, yet US centric implementation is the well-known Institutional Review Board (IRB)} approval (\(\CIRCLE\)),  whether the authors evaluate the ethics of their research themselves and discuss their review in the paper (\(\RIGHTcircle\)), or whether ethics are not discussed in the publication (\(\Circle\)).

Even in 2018, individual publications still do not involve an ethics committee, but the general trend is towards a thorough consideration of ethical requirements.
Despite this positive trend, it appears that papers investigating expert users initially discussed the ethics of their work less consistently.
A common issue, leading to authors not involving an ethics committee, are cases where the authors' ethics committee is not sufficiently equipped to deal with the specific research plan.
A classical case of this is the 2015 study of Burnett and Feamster~\cite{burnett2015encore}, which measures censorship, but does so raising ethical concerns~\cite{narayanan2015no}.
However, the ethics committee of the researchers' institution signed off on this work, most likely due to the board being unfamiliar with the ethical implications of research at the intersection of human factors and computer science.
Other studies, for example, Dietrich et al.~\cite{Dietrich2018}, did not involve an ethics committee because their host institutions does not have such an entity.

\subsection{Observations and Recommendations}

\noindent\textbf{Key Observations:}
While the field made significant progress in the inclusion of ethical considerations, some institutions still lack the appropriate research infrastructure.
Furthermore, especially for expert-user related work, authors even in 2018, still do not always discuss their work's ethical implications.

\noindent\textbf{Key Recommendations:}
Authors should adopt the habit of evaluating the ethical implications of their work.
In case no ethics board is available, the Menlo report can provide guidance on how to evaluate the ethical implications of one's work~\cite{dittrich2012menlo}.
When considered for publication, authors should be held to these standards, that is, documenting their efforts in handling ethical implications and subjects data rights should be mandatory.
Furthermore, we suggest to address the issue of no capable ethics board being available by introducing a community driven ethics board, capable of reviewing human factors in security studies, for example, by the IEEE and ACM extending their existing bodies.

\begin{figure}[tb!]
    \centering
       \begin{subfigure}[t]{0.20\textwidth}
        \centering
        \includegraphics[width=\columnwidth,trim={0 0cm 0 0cm},clip,angle=0]{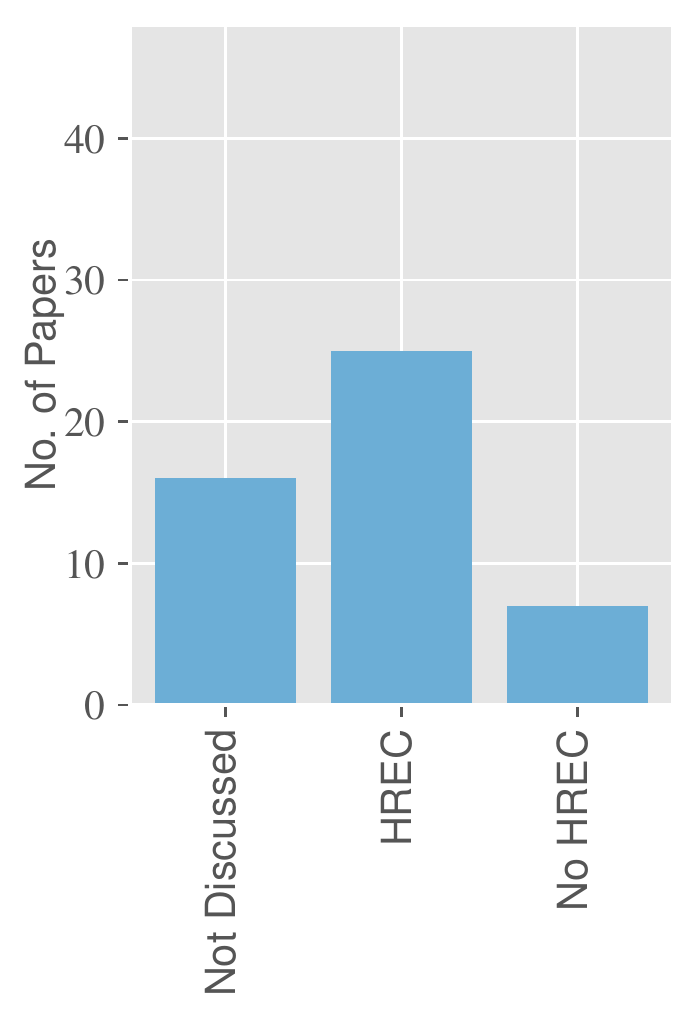}
        \caption{Expert Users}
        \label{fig:ethics_exp}
    \end{subfigure}%
     ~
    \begin{subfigure}[t]{0.20\textwidth}
        \centering
        \includegraphics[width=\columnwidth,trim={0 0cm 0 0cm},clip,angle=0]{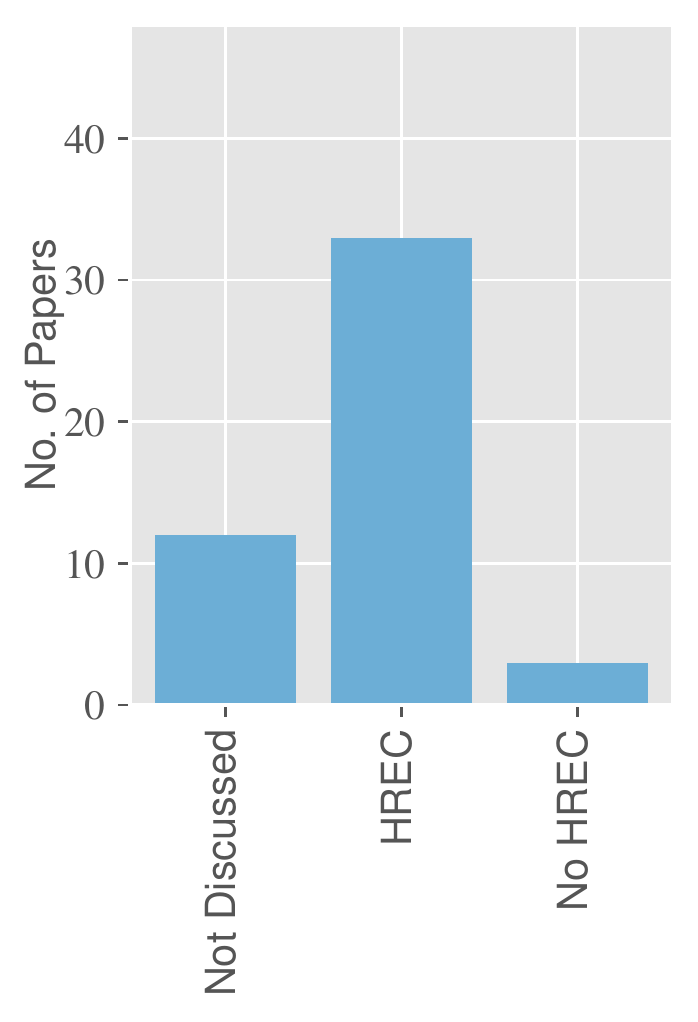}
        \caption{End Users}
        \label{fig:ethics_end}
    \end{subfigure}
\caption{Overview of how ethical considerations are handled in expert-user (\Fref{fig:ethics_exp}) and end-user papers (\Fref{fig:ethics_end}).
We find that the share of papers not discussing ethical implications of their work (informed consent, handling of data etc.) is higher for the expert-user portion of our dataset.
Similarly, the number of papers where no suitable HREC was available is higher in the expert-user sample.
}
\vspace{1em}
    \label{fig:ethics}

\end{figure}

\section{Limitations}
\label{sec:lim}

Our literature survey has several limitations.
Firstly, We do not take into account the research before 2008. 

While a historical perspective going back to the earliest papers nearly 30 years ago might prove useful to understand the origins of the field, a more recent scope is better suited to provide an overview of the state of the art and comprehensive recommendations on how the field can improve further today.

Secondly, instead of searching the standard databases like SCOPUS or Web of Science using particular keywords to find relevant publications, we chose to search all the top-tier computer security venues.
We didn't use any keywords for search but included all the publications from the top-tier venues after 2008.
We made this choice so as to showcase the work of top-tier computer security venues in regards to human factor research.
We understand that this may exclude notable human factors research from outside these top-tier computer venues but we consider those out of scope as we want to learn what the leading security venues are doing.

Thirdly, after an in-depth review of the 48 expert user publications, we were interested in comparing these publications with the end-user publications.
Therefore, in order to have a reasonable number of papers to review, we opted for balancing the two user groups.
For this, we used a random sample of 48 papers from the end-user group.
These papers were chosen so as to match the number of expert user papers per year. 
We understand that this is a random selection but we believe it serves our purpose in answering our research question. 
While we do not review all the end-user papers, we review enough to gather the overall gist of end-user research and to be able to provide an overview of the research in both user groups.

Overall, we systematize a significantly larger body of literature than related surveys, for example, Hamm et al., who cover three conferences over five years~\cite{hamm2019systematic}, or Tahei and Vaniea, who focus only on developers~\cite{tahaei2019survey}.

\section{Conclusion}
\label{sec:conclusion}

In this paper, we present the systematization of how research of the past ten years in the emerging field of human factors in computer security is conducted, with a special focus on expert versus non-expert users.
Although the field is growing, we find that there is an opportunity for the community to adopt methods, rigor, and practices from other fields in which human factors research has matured over the past years.
Most notably, we can learn from safety science in terms of how we treat the human factor, and the social sciences in terms of utilizing theories to streamline our research work, and their experience in the ongoing struggle with WEIRD study populations, which we share.

Moreover, we find that expert users are under-represented in human factors research.
Only around 9\% of all papers have focused on this group, even though their choices and mistakes typically have more impact than those of regular end users.
For the critical population of expert users, our field can benefit from safety science's perspective on human error (see \Fref{sec:perspective}). 
In this field, human error is a ``normal'' probabilistic outcome of a set of organizational and institutional conditions under which users interact with the technology, rather than the failure of an individual. 
Systems have to be build in a way that handles and accounts for the occurrence of these errors.
``Fixing the (expert) user'' is not the path to better security and privacy~\cite{dodier2017paternalistic}.

In terms of methodology, population selection and recruitment we find that currently most work is biased towards samples that are locally accessible to researchers. 
This means that current work is heavily dominated by a U.S. and Europe-centric view (see \Fref{sec:populationsample}).
This current focus of samples may lead to a biased perspective of the work we do, only focusing on the needs, expectations, and behavior of citizens of the global north.
In the pursuit of diversifying the populations that our field studies, for example, utilizing Cultural Dimensions Theory might prove useful.
Similarly, Design Science is a promising framework to formalize the process of designing and evaluating an artifact, that is, starting with requirements gathering from a population, designing it while considering best practices from the literature, and properly evaluating the final artifact.

At the moment human factors research in computer security is still dominated by exploratory and perspective-gathering research (see \Fref{sec:ana-obj}).
Hence, to further advance the field, we suggest to adopt the concept of theorizing from the social sciences and psychology (see \Fref{sec:ana-theo}).
Only a fraction of the published work leverages theories (see \Fref{ssec:theo-use}), even though many of these studies could have benefited from including theories, like Mental Models, Sensemaking Theory or the Theory of Reasoned Action.

Current use of theories is either observational, that is, to improve experimental design in case of Mental Models, or fragmented, not consistently focusing on a specific set of theories.
While several recent publications claim to utilize \emph{Grounded Theory}, we find that work typically does not execute the full process of \emph{Grounded Theory}, which should culminate in true theorizing.
Instead, it is used as an analytical framework to formalize experimental design and the qualitative data analysis process authors conduct (see \Fref{ssec:gt}).

\subsection*{Future research recommendations}
\label{ssec:future}
Considering our research question and sub questions, we can make the following recommendations for future research.

Firstly, in addition to preventing human error, we should also try to understand which behavior leads to secure outcomes, and how we can facilitate that behavior.
For this, we recommend investigating expert users and their interactions with their environment from different qualitative perspectives.
On top of interviews and surveys, we recommend employing different research methods (e.g. naturalistic observations) to study human factors. 
Secondly, we recommend investigating more diverse population samples and also better discussing the external validity and limitations of the findings with regards to the samples studied.
Thirdly, we recommend exploring and using existing theoretical frameworks to inform the research design.
Fourthly, we suggest using and improving upon existing theories as well as forming new ones. This will help in generalizing the results. 
We also recommend more replication studies, specially replicating findings in different socio-economic backgrounds.
Lastly, we suggest that researchers should try to evaluate the ethical considerations of their work in human factors research. 
We also suggest the possibility of creating a community-driven ethics board which can help researchers that do not have an ethics committee available to them. 

Our literature review has has several limitations, as discussed earlier.
We do not claim to have represented the totality of several decades of human factors research, assuming that would even be possible.
We do claim to provide a thorough overview of the research on experts in the past decade and a representative view on work on non-expert user populations for the purpose of making a comparison.
We suggest extending the scope of the review by diving deeper into various user categories to gather specific insights and by investigating other security venues that were excluded in this study.

Over the past decade, human factors research has been increasingly recognized as a key contribution to the field of computer security.
Now, it is time to learn from its own successes and failures as well as observations and experiences from other fields to further mature it during the next decade.

\balance
\printbibliography

@inProceedings{Redmiles2016,
  title    = {I Think They're Trying to Tell Me Something: Advice Sources and Selection for Digital Security},
  author   = {Redmiles, Elissa M. and Malone, Amelia R. and Mazurek, Michelle L.},
  crossref = {IEEESSP2016},
  doi      = {10.1109/SP.2016.24},
  pages    = {272--288},
}

@inProceedings{Egelman2015,
author = {Egelman, Serge and Peer, Eyal},
title = {{Scaling the Security Wall: Developing a Security Behavior Intentions Scale (SeBIS)}},
  crossref = {ACMCHI2015},
pages = {2873--2882},
   doi = {10.1145/2702123.2702249},
}

@InProceedings{Harmonik2018,
author={H{\'a}mornik, Bal{\'a}zs P{\'e}ter and Krasznay, Csaba},
title={A Team-Level Perspective of Human Factors in Cyber Security: Security Operations Centers},
booktitle={Proceedings of the AHFE 2018 International Conference on Human Factors in Cybersecurity},
year={2018},
publisher={Springer},
pages={224--236},
isbn={978-3-319-60585-2}
}

@techreport{Hollnagel2013,
author = {Hollnagel, Erik and Leonhardt, J{\"{o}}rg and Licu, Tony and Shorrock, Steven},
% publisher = {European Organisation for the Safety of Air Navigation (EUROCONTROL)},
title = {{From Safety-I to Safety-II: a white paper}},
year = {2013}
}

@string{journal:JAP = {Journal of Applied Psychology}}

@article{journal:JAP2000-04,
  journal   = journal:JAP,
  date      = {2000-04},
  volume    = {85},
%   issue     = {2},
  publisher = {APA},
 issn = {0021-9010},
}

@article{Mathieu2000,
author = {Mathieu, John E and Goodwin, Gerald F. and Heffner, Tonia S. and Cannon-Bowers, Janis A. and Salas, Eduardo},
title = {{The influence of shared mental models on team process and performance}},
crossref = {journal:JAP2000-04},
pages = {273--283},
doi = {10.1037//0021-9010.85.2.273},
}

@online{Au2019,
author = {{Office of the Australian Information Commissioner}, Australian Government},
booktitle = {OAIC},
title = {{Lessons learned during first 11 months of Notifiable Data Breaches scheme| Office of the Australian Information Commissioner - OAIC}},
url = {https://www.oaic.gov.au/media-and-speeches/news/lessons-learned-during-first-12-months-of-notifiable-data-breaches-scheme},
% urldate = {2019-05-28},
year = {2019}
}

@online{Vizard2019,
author = {Vizard, Michael},
booktitle = {Security Boulevard},
title = {{McAfee Survey Finds IT at Cybersecurity Fault Most}},
url = {https://securityboulevard.com/2019/05/mcafee-survey-finds-it-at-cybersecurity-fault-most/},
% urldate = {2019-05-28},
year = {2019}
}

@online{Rashid2019,
author = {Rashid, Fahmida Y.},
booktitle = {Decipher},
title = {{Digging Deep into the Verizon DBIR}},
url = {https://duo.com/decipher/digging-deep-into-the-verizon-dbir},
% urldate = {2019-05-28},
year = {2019}
}

@online{Hiskey2019,
author = {Hiskey, Michael},
booktitle = {Infosecurity Magazine},
title = {{Before Blaming Hackers, Check Your Configurations}},
url = {https://www.infosecurity-magazine.com/opinions/blaming-hackers-configurations-1-1-1/},
% urldate = {2019-05-28},
year = {2019}
}

@online{Braue2019,
author = {Braue, David},
booktitle = {CSO Online},
title = {{At least 10m records compromised in single Australian data breach despite drop in NDB reports}},
url = {https://www.cso.com.au/article/661702/least-10m-records-compromised-single-australian-data-breach-despite-drop-ndb-reports/},
% urldate = {2019-05-28},
year = {2019}
}

@article{Kitchenham2009,
author = {Kitchenham, Barbara and {Pearl Brereton}, O. and Budgen, David and Turner, Mark and Bailey, John and Linkman, Stephen},
doi = {10.1016/j.infsof.2008.09.009},
issn = {09505849},
journal = {Information and Software Technology},
number = {1},
pages = {7--15},
title = {{Systematic literature reviews in software engineering - A systematic literature review}},
volume = {51},
year = {2009}
}

@book{adams2006layman,
author = {Adams, David},
booktitle = {ATSB Safety Information Paper},
isbn = {9781921092749},
number = {June},
publisher = {ATSB},
title = {{A Layman's Introduction to Human Factors in Aircraft Accident and Incident Investigation}},
% url = {https://www.atsb.gov.au/media/32882/b20060094.pdf},
year = {2006}
}

@inProceedings{Werlinger2008,
author = {Werlinger, Rodrigo and Hawkey, Kirstie and Muldner, Kasia and Jaferian, Pooya and Beznosov, Konstantin},
 title = {The Challenges of Using an Intrusion Detection System: Is It Worth the Effort?},
  crossref = {SOUPS2008},
 pages = {107--118},
 doi = {10.1145/1408664.1408679},
}

@inProceedings{Neto2008,
  author    = {Neto, Afonso Ara{\'{u}}jo and Vieira, Marco},
  title     = {Towards assessing the security of {DBMS} configurations},
  crossref = {DSN2008},
  pages     = {90--95},
  doi       = {10.1109/DSN.2008.4630074},
}

@inProceedings{Epstein2009,
  author    = {Epstein, Jeremy},
 title={A Survey of Vendor Software Assurance Practices},
  crossref = {ACSAC2009},
  pages={528-537},
  doi={10.1109/ACSAC.2009.56},
}

@inProceedings{Bauer2009,
author = {Bauer, Lujo and Cranor, Lorrie Faith and Reeder, Robert W. and Reiter, Michael K. and Vaniea, Kami},
 title={Real life challenges in access-control management},
  crossref = {ACMCHI2009},
   pages = {899--908},
   doi = {10.1145/1518701.1518838},
}

@inProceedings{Fenz2009,
 author = {Fenz, Stefan and Ekelhart, Andreas},
 title = {Formalizing Information Security Knowledge},
  crossref = {ACMASIACCS2009},
    pages = {183--194},
   doi = {10.1145/1533057.1533084},
}

@inProceedings{Johnson2010,
  author = {Johnson, Maritza and Karat, John and Karat, Clare-Marie and Grueneberg, Keith},
 title = {Optimizing a Policy Authoring Framework for Security and Privacy Policies},
  crossref = {SOUPS2010},
 pages = {8:1--8:9},
    doi = {10.1145/1837110.1837121},
}

@inProceedings{Huang2011,
  author = {Huang, Xin and Monrose, Fabian and Reiter, Michael K.},
title={Amplifying limited expert input to sanitize large network traces},
  crossref = {DSN2011},
pages={494-505},
 doi={10.1109/DSN.2011.5958262},
}

@inProceedings{Jaferian2011,
   author = {Jaferian, Pooya and Hawkey, Kirstie and Sotirakopoulos, Andreas and Velez-Rojas, Maria and Beznosov, Konstantin},
 title = {Heuristics for Evaluating IT Security Management Tools},
  crossref = {SOUPS2011},
 pages = {7:1--7:20},
 doi = {10.1145/2078827.2078837},
}

@inProceedings{Xie2012,
   author = {Xie, Jing and Lipford, Heather and Chu, Bei-Tseng},
 title = {Evaluating Interactive Support for Secure Programming},
  crossref = {ACMCHI2012},
pages = {2707--2716},
 doi = {10.1145/2207676.2208665},
}

@inProceedings{Fahl2013,
 author = {Fahl, Sascha and Harbach, Marian and Perl, Henning and Koetter, Markus and Smith, Matthew},
 title = {Rethinking SSL Development in an Appified World},
  crossref = {ACMCCS2013},
pages = {49--60},
 doi = {10.1145/2508859.2516655},
}

@inProceedings{Beckerle2013,
  author = {Beckerle, Matthias and Martucci, Leonardo A.},
 title = {Formal Definitions for Usable Access Control Rule Sets from Goals to Metrics},
  crossref = {SOUPS2013},
 pages = {2:1--2:11},
 doi = {10.1145/2501604.2501606},
}

@inProceedings{Jaferian2014,
   author = {Jaferian, Pooya and Rashtian, Hootan and Beznosov, Konstantin},
 title = {To Authorize or Not Authorize: Helping Users Review Access Policies in Organizations},
  crossref = {SOUPS2014},
 pages = {301--320},
 doi = {},
}

@inProceedings{Oliveira2014,
  author = {Oliveira, Daniela and Rosenthal, Marissa and Morin, Nicole and Yeh, Kuo-Chuan and Cappos, Justin and Zhuang, Yanyan},
 title = {It's the Psychology Stupid: How Heuristics Explain Software Vulnerabilities and How Priming Can Illuminate Developer's Blind Spots},
  crossref = {ACSAC2014},
 pages = {296--305},
  doi = {10.1145/2664243.2664254},
}

@inProceedings{Fahl2014,
 author = {Fahl, Sascha and Acar, Yasemin and Perl, Henning and Smith, Matthew},
 title = {Why Eve and Mallory (Also) Love Webmasters: A Study on the Root Causes of SSL Misconfigurations},
  crossref = {ACMASIACCS2014},
 pages = {507--512},
  doi = {10.1145/2590296.2590341},
}

@inProceedings{Sundaramurthy2015,
 author = {Sundaramurthy, Sathya Chandran and Bardas, Alexandru G. and Case, Jacob and Ou, Xinming and Wesch, Michael and McHugh, John and Rajagopalan, S. Raj},
 title = {A Human Capital Model for Mitigating Security Analyst Burnout},
  crossref = {SOUPS2015},
 pages = {347--359},
  doi = {},
}

@inProceedings{Ion2015,
 author = {Ion, Iulia and Reeder, Rob and Consolvo, Sunny},
 title = {"...No One Can Hack My Mind": Comparing Expert and Non-expert Security Practices},
  crossref = {SOUPS2015},
 pages = {327--346},
  doi = {},
}

@inProceedings{Kang2015,
 author = {Kang, Ruogu and Dabbish, Laura and Fruchter, Nathaniel and Kiesler, Sara},
 title = {"My Data Just Goes Everywhere": User Mental Models of the Internet and Implications for Privacy and Security},
  crossref = {SOUPS2015},
 pages = {39--52},
  doi = {},
}

@inProceedings{Gember-Jacobson2015,
 author = {Gember-Jacobson, Aaron and Wu, Wenfei and Li, Xiujun and Akella, Aditya and Mahajan, Ratul},
 title = {Management Plane Analytics},
  crossref = {IMC2015},
 pages = {395--408},
 doi = {10.1145/2815675.2815684},
}

@inProceedings{Falsina2015,
 author = {Falsina, Luca and Fratantonio, Yanick and Zanero, Stefano and Kruegel, Christopher and Vigna, Giovanni and Maggi, Federico},
 title = {Grab 'N Run: Secure and Practical Dynamic Code Loading for Android Applications},
  crossref = {ACSAC2015},
 pages = {201--210},
 doi = {10.1145/2818000.2818042},
}

@inProceedings{DeLuca2016,
 author = {De Luca, Alexander and Das, Sauvik and Ortlieb, Martin and Ion, Iulia and Laurie, Ben},
 title = {Expert and Non-expert Attitudes Towards (Secure) Instant Messaging},
  crossref = {SOUPS2016},
 pages = {147--157},
 doi = {},
}

@inProceedings{Sundaramurthy2016,
 author = {Sundaramurthy, Sathya Chandran and McHugh, John and Ou, Xinming and Wesch, Michael and Bardas, Alexandru G. and Rajagopalan, S. Raj},
 title = {Turning Contradictions into Innovations or: How We Learned to Stop Whining and Improve Security Operations},
  crossref = {SOUPS2016},
 pages = {237--251},
 doi = {},
}

@inProceedings{Yakdan2016,
author = {Yakdan, Khaled and Dechand, Sergej and Gerhards-Padilla, Elmar and Smith, Matthew},
title = {Helping Johnny to Analyze Malware: A Usability-Optimized Decompiler and Malware Analysis User Study},
  crossref = {IEEESSP2016},
pages = {158-177},
doi = {10.1109/SP.2016.18},
}

@inProceedings{Acar2016,
author={Acar, Yasemin and Backes, Michael and Fahl, Sascha and Kim, Doowon and Mazurek, Michelle L. and Stransky, Christian},
title={You Get Where You're Looking for: The Impact of Information Sources on Code Security},
  crossref = {IEEESSP2016},
pages={289-305},
doi={10.1109/SP.2016.25},
}

@inProceedings{Zanutto2017,
author = {Zanutto, Alberto and Shreeve, Ben and Follis, Karolina and Busby, Jerry and Rashid, Awais},
title = {The Shadow Warriors: In the no man{\textquoteright}s land between industrial control systems and enterprise {IT} systems},
  crossref = {SOUPS2017},
 pages = {},
 doi = {},
}

@inProceedings{Gallagher2017,
author = {Gallagher, Kevin and Patil, Sameer  and Memon, Nasir},
title = {New Me: Understanding Expert and Non-Expert Perceptions and Usage of the Tor Anonymity Network},
  crossref = {SOUPS2017},
pages = {385--398},
 doi = {},
}

@inProceedings{Acar2017,
 author = {Acar, Yasemin and Stransky, Christian and Wermke, Dominik and Mazurek, Michelle L. and Fahl, Sascha},
 title = {Security Developer Studies with Github Users: Exploring a Convenience Sample},
  crossref = {SOUPS2017},
 pages = {81--95},
 doi = {},
}

@inProceedings{Naiakshina2017,
 author = {Naiakshina, Alena and Danilova, Anastasia and Tiefenau, Christian and Herzog, Marco and Dechand, Sergej and Smith, Matthew},
  title = {Why Do Developers Get Password Storage Wrong?: A Qualitative Usability Study},
  crossref = {ACMCCS2017},
 pages = {311--328},
 doi = {10.1145/3133956.3134082},
}

@inProceedings{Nguyen2017,
 author = {Nguyen, Duc Cuong and Wermke, Dominik and Acar, Yasemin and Backes, Michael and Weir, Charles and Fahl, Sascha},
 title = {A Stitch in Time: Supporting Android Developers in WritingSecure Code},
  crossref = {ACMCCS2017},
 pages = {1065--1077},
 doi = {10.1145/3133956.3133977},
}

@inProceedings{Derr2017,
 author = {Derr, Erik and Bugiel, Sven and Fahl, Sascha and Acar, Yasemin and Backes, Michael},
 title = {Keep Me Updated: An Empirical Study of Third-Party Library Updatability on Android},
  crossref = {ACMCCS2017},
 pages = {2187--2200},
 doi = {10.1145/3133956.3134059},
}

@inProceedings{Gamero2017,
 author = {Gamero-Garrido, Alexander and Savage, Stefan and Levchenko, Kirill and Snoeren, Alex C.},
 title = {Quantifying the Pressure of Legal Risks on Third-party Vulnerability Research},
  crossref = {ACMCCS2017},
 pages = {1501--1513},
  doi = {10.1145/3133956.3134047},
}

@inProceedings{Acar2017_2,
author = {Acar, Yasemin and Backes, Michael and Fahl, Sascha and Garfinkel, Simson and Kim, Doowon and Mazurek, Michelle L. and Stransky, Christian},
title = {Comparing the Usability of Cryptographic APIs},
  crossref = {IEEESSP2017},
pages = {154-171},
doi = {10.1109/SP.2017.52},
}

@inProceedings{Haney2018,
 author = {Haney, Julie M. and Lutters, Wayne G.},
 title = {"It's Scary...It's Confusing...It's Dull": How Cybersecurity Advocates Overcome Negative Perceptions of Security},
  crossref = {SOUPS2018},
 pages = {411--425},
doi = {},
}

@inProceedings{Mu2018,
 author = {Mu, Dongliang and Cuevas, Alejandro and Yang, Limin and Hu, Hang and Xing, Xinyu and Mao, Bing and Wang, Gang},
 title = {Understanding the Reproducibility of Crowd-reported Security Vulnerabilities},
  crossref = {USENIXSEC2018},
pages = {919--936},
doi = {},
}

@inProceedings{Stock2018,
 author = {Stock, Ben and Pellegrino, Giancarlo and Li, Frank and Backes, Michael and Rossow, Christian},
 title = {Didn?t You Hear Me? {--} Towards More Successful Web Vulnerability Notifications},
 crossref = {NDSS2018},
 doi = {10.14722/ndss.2018.23171},
}

@inProceedings{Dietrich2018,
  author = {Dietrich, Constanze and Krombholz, Katharina and Borgolte, Kevin and Fiebig, Tobias},
 title = {Investigating System Operators' Perspective on Security Misconfigurations},
  crossref = {ACMCCS2018},
 pages = {1272--1289},
 doi = {10.1145/3243734.3243794},
}

@inProceedings{Skirpan2018,
 author = {Skirpan, Michael Warren and Yeh, Tom and Fiesler, Casey},
 title = {What's at Stake: Characterizing Risk Perceptions of Emerging Technologies},
  crossref = {ACMCHI2018},
  pages = {70:1--70:12},
  doi = {10.1145/3173574.3173644},
}

@inProceedings{Votipka2018,
author = {Votipka, Daniel and Stevens, Rock and Redmiles, Elissa M. and Hu, Jeremy and Mazurek, Michelle L.},
title = {Hackers vs. Testers: A Comparison of Software Vulnerability Discovery Processes},
  crossref = {IEEESSP2018},
  pages = {374-391},
 doi={10.1109/SP.2018.00003},
}

@inProceedings{Gorski2018,
 author = {Gorski, Peter Leo and Iacono, Luigi Lo and Wermke, Dominik and Stransky, Christian and Moeller, Sebastian and Acar, Yasemin and Fahl, Sascha},
 title = {Developers Deserve Security Warnings, Too: On the Effect of Integrated Security Advice on Cryptographic API Misuse},
  crossref = {SOUPS2018},
  pages = {265--280},
 doi={},
}

@inProceedings{Merrill2018,
author = {Merrill, Nick and Chuang, John},
 title = {From Scanning Brains to Reading Minds: Talking to Engineers About Brain-Computer Interface},
  crossref = {ACMCHI2018},
  pages = {323:1--323:11},
 doi = {10.1145/3173574.3173897},
}

@inProceedings{Wermke2018,
 author = {Wermke, Dominik and Huaman, Nicolas and Acar, Yasemin and Reaves, Bradley and Traynor, Patrick and Fahl, Sascha},
 title = {A Large Scale Investigation of Obfuscation Use in Google Play},
  crossref = {ACSAC2018},
 pages = {222--235},
 doi = {10.1145/3274694.3274726},
}

@inProceedings{Adams2018,
 author = {Adams, Devon and Bah, Alseny and Barwulor, Catherine and Musaby, Nureli and Pitkin, Kadeem and Redmiles, Elissa M.},
title = {Ethics Emerging: the Story of Privacy and Security Perceptions in Virtual Reality},
  crossref = {SOUPS2018},
pages = {427--442},
 doi = {},
}

@inProceedings{Naiakshina2018,
 author = {Naiakshina, Alena and Danilova, Anastasia and Tiefenau, Christian and Smith, Matthew},
 title = {Deception Task Design in Developer Password Studies: Exploring a Student Sample},
  crossref = {SOUPS2018},
 pages = {297--313},
 doi = {},
}

@inProceedings{Assal2018,
 author = {Assal, Hala and Chiasson, Sonia},
title = {Security in the Software Development Lifecycle},
  crossref = {SOUPS2018},
pages = {281--296},
 doi = {},
}

@inProceedings{Oliveira2018,
author = {Oliveira, Daniela Seabra and Lin, Tian and Rahman, Muhammad Sajidur and Akefirad, Rad and Ellis Donovan and Perez, Eliany and Bobhate, Rahul and DeLong, Lois A. and Cappos, Justin and Brun, Yuriy},
title = {{API} Blindspots: Why Experienced Developers Write Vulnerable Code},
  crossref = {SOUPS2018},
pages = {315--328},
 doi = {},
}

@inProceedings{Thomas2018,
 author = {Thomas, Tyler W. and Tabassum, Madiha and Chu, Bill and Lipford, Heather},
 title = {Security During Application Development: An Application Security Expert Perspective},
  crossref = {ACMCHI2018},
 pages = {262:1--262:12},
doi = {10.1145/3173574.3173836},
}

@inProceedings{Simko2018,
  author = {Simko, Lucy and Zettlemoyer, Luke and Kohno, Tadayoshi},
 title = {Recognizing and Imitating Programmer Style: Adversaries in Program Authorship Attribution},
  crossref = {PETS2018},
 pages  =  {127--144},
doi = {},
}

@inProceedings{Hansch2018,
   author = {H\"{a}nsch, Norman and Schankin, Andrea and Protsenko, Mykolai and Freiling, Felix and Benenson, Zinaida},
  title = {Programming Experience Might Not Help in Comprehending Obfuscated Source Code Efficiently},
  crossref = {SOUPS2018},
 pages = {341--356},
doi = {},
}

@inProceedings{Haney2018_2,
author = {Haney, Julie M. and Theofanos, Mary and Acar, Yasemin and Prettyman, Sandra Spickard},
title = {``We make it a big deal in the company": Security Mindsets in Organizations that Develop Cryptographic Products},
  crossref = {SOUPS2018},
pages = {357--373},
doi = {},
}

@inProceedings{Forget2008,
 author = {Forget, Alain and Chiasson, Sonia and van Oorschot, P. C. and Biddle, Robert},
title = {``We make it a big deal in the company": Security Mindsets in Organizations that Develop Cryptographic Products},
  crossref = {SOUPS2008},
pages = {1--12},
 doi = {10.1145/1408664.1408666},
}

@inProceedings{Egelman2008,
 author = {Egelman, Serge and Cranor, Lorrie Faith and Hong, Jason},
 title = {You'Ve Been Warned: An Empirical Study of the Effectiveness of Web Browser Phishing Warnings},
  crossref = {ACMCHI2008},
 pages = {1065--1074},
 doi = {10.1145/1357054.1357219},
}

@inProceedings{Mcdonald2009,
 author = {Mcdonald, Aleecia M. and Reeder, Robert W. and Kelley, Patrick Gage and Cranor, Lorrie Faith},
 title = {A Comparative Study of Online Privacy Policies and Formats},
  crossref = {PETS2009},
 pages = {37--55},
 doi = {10.1007/978-3-642-03168-7_3},
}

@inProceedings{Klasnja2009,
  author = {Klasnja, Predrag and Consolvo, Sunny and Jung, Jaeyeon and Greenstein, Benjamin M. and LeGrand, Louis and Powledge, Pauline and Wetherall, David},
 title = {"When I Am on Wi-Fi, I Am Fearless": Privacy Concerns \& Practices in Eeryday Wi-Fi Use},
  crossref = {ACMCHI2009},
 pages = {1993--2002},
  doi = {10.1145/1518701.1519004},
}

@inProceedings{Karlof2009,
 author = {Karlof, Chris and Tygar, J. D. and Wagner, David},
 title = {Conditioned-safe Ceremonies and a User Study of an Application to Web Authentication},
  crossref = {SOUPS2009},
 pages = {38:1--38:1},
 doi = {10.1145/1572532.1572578},
}

@inProceedings{Ion2010,
 author = {Ion, Iulia and Langheinrich, Marc and Kumaraguru, Ponnurangam and \v{C}apkun, Srdjan},
 title = {Influence of User Perception, Security Needs, and Social Factors on Device Pairing Method Choices},
  crossref = {SOUPS2010},
 pages = {6:1--6:13},
doi = {10.1145/1837110.1837118},
}

@inProceedings{Zhu2011,
 author = {Zhu, Feng and Carpenter, Sandra and Kulkarni, Ajinkya and Kolimi, Swapna},
 title = {Reciprocity Attacks},
  crossref = {SOUPS2011},
 pages = {9:1--9:14},
 doi = {10.1145/2078827.2078839},
}

@inProceedings{Shin2011,
 author = {Shin, Dongwan and Lopes, Rodrigo},
 title = {An Empirical Study of Visual Security Cues to Prevent the SSLstripping Attack},
  crossref = {ACSAC2011},
 pages = {287--296},
 doi = {10.1145/2076732.2076773},
}

@inProceedings{Sae_Bae2012,
 author = {Sae-Bae, Napa and Ahmed, Kowsar and Isbister, Katherine and Memon, Nasir},
 title = {Biometric-rich Gestures: A Novel Approach to Authentication on Multi-touch Devices},
  crossref = {ACMCHI2012},
 pages = {977--986},
doi = {10.1145/2207676.2208543},
}

@inProceedings{Ruoti2013,
  author = {Ruoti, Scott and Kim, Nathan and Burgon, Ben and van der Horst, Timothy and Seamons, Kent},
 title = {Confused Johnny: When Automatic Encryption Leads to Confusion and Mistakes},
  crossref = {SOUPS2013},
 pages = {5:1--5:12},
 doi = {10.1145/2501604.2501609},
}

@inProceedings{Denning2013,
 author = {Denning, Tamara and Lerner, Adam and Shostack, Adam and Kohno, Tadayoshi},
 title = {Control-Alt-Hack: the design and evaluation of a card game for computer security awareness and education},
  crossref = {ACMCCS2013},
 pages = {915--928},
 doi = {10.1145/2508859.2516753},
}

@inProceedings{Wash2014,
 author = {Wash, Rick and Rader, Emilee and Vaniea, Kami and Rizor, Michelle},
title = {Out of the Loop: How Automated Software Updates Cause Unintended Security Consequences},
  crossref = {SOUPS2014},
pages = {89--104},
 doi = {},
}

@inProceedings{Aviv2014,
 author = {Aviv, Adam J. and Fichter, Dane},
 title = {Understanding Visual Perceptions of Usability and Security of Android's Graphical Password Pattern},
  crossref = {ACSAC2014},
pages = {286--295},
 doi = {10.1145/2664243.2664253},
}

@inProceedings{Mohamed2014,
author = {Mohamed, Manar and Sachdeva, Niharika and Georgescu, Michael and Gao, Song and Saxena, Nitesh and Zhang, Chengcui and Kumaraguru, Ponnurangam and van Oorschot, Paul C. and Chen, Wei-Bang},
  title = {A Three-way Investigation of a game-CAPTCHA: Automated Attacks, Relay Attacks and Usability},
  crossref = {ACMASIACCS2014},
 pages = {195--206},
 doi = {10.1145/2590296.2590298},
}

@inProceedings{Ur2015,
 author = {Ur, Blase and Noma, Fumiko and Bees, Jonathan and Segreti, Sean M. and Shay, Richard and Bauer, Lujo and Christin, Nicolas and Cranor, Lorrie Faith},
 title = {``I Added '!' at the End to Make It Secure'': Observing Password Creation in the Lab},
  crossref = {SOUPS2015},
 pages = {123--140},
 doi = {},
}

@inProceedings{Angulo2015,
 author = {Angulo, Julio and Ortlieb, Martin},
title = {"WTH..!?!" Experiences, Reactions, and Expectations Related to Online Privacy Panic Situations},
  crossref = {SOUPS2015},
 pages = {19--38},
 doi = {},
}

@inProceedings{Chanchary2015,
 author = {Chanchary, Farah and Chiasson, Sonia},
 title = {User Perceptions of Sharing, Advertising, and Tracking},
  crossref = {SOUPS2015},
  pages = {53--67},
 doi = {},
}

@inProceedings{Bianchi2015,
 author = {Bianchi, Antonio and Corbetta, Jacopo and Invernizzi, Luca and Fratantonio, Yanick and Kruegel, Christopher and Vigna, Giovanni},
 title = {What the App is That? Deception and Countermeasures in the Android User Interface},
 crossref = {IEEESSP2015},
 pages = {931--948},
 doi = {10.1109/SP.2015.62},
}

@inProceedings{Hupperich2015,
 author = {Hupperich, Thomas and Maiorca, Davide and K\"{u}hrer, Marc and Holz, Thorsten and Giacinto, Giorgio},
 title = {On the Robustness of Mobile Device Fingerprinting: Can Mobile Users Escape Modern Web-Tracking Mechanisms?},
  crossref = {ACSAC2015},
 pages = {191--200},
   doi = {10.1145/2818000.2818032},
}

@inProceedings{Fagan2016,
author = {Fagan, Michael and Khan, Mohammad Maifi Hasan},
title = {Why Do They Do What They Do?: A Study of What Motivates Users to (Not) Follow Computer Security Advice},
  crossref = {SOUPS2016},
pages = {59--75},
   doi = {},
}

@inProceedings{Mathur2016,
 author = {Mathur, Arunesh and Engel, Josefine and Sobti, Sonam and Chang, Victoria and Chetty, Marshini},
 title = {``They Keep Coming Back Like Zombies": Improving Software Updating Interfaces},
  crossref = {SOUPS2016},
 pages = {43--58},
   doi = {},
}

@inProceedings{Tischer2016,
author={Tischer, Matthew and Durumeric, Zakir and Foster Sam and Duan, Sunny and Mori, Alec and Bursztein, Elie and Bailey, Michael},
title={Users Really Do Plug in USB Drives They Find},
  crossref = {IEEESSP2016},
pages={306-319},
   doi={10.1109/SP.2016.26},
}

@inProceedings{Ruoti2017,
 author = {Ruoti, Scott and Monson, Tyler and Wu, Justin and Zappala, Daniel and Seamons, Kent},
 title = {Weighing Context and Trade-offs: How Suburban Adults Selected Their Online Security Posture},
  crossref = {SOUPS2017},
   pages = {211--228},
   doi={},
}

@inProceedings{Lastdrager2017,
 author = {Lastdrager, Elmer and Gallardo, In{\'e}s Carvajal and Hartel, Pieter and Junger, Marianne},
 title = {How Effective is Anti-phishing Training for Children?},
  crossref = {SOUPS2017},
  pages = {229--239},
   doi={},
}

@inProceedings{Tian2017,
 author = {Tian, Yuan and Zhang, Nan and Lin, Yueh-Hsun and Wang, XiaoFeng and Ur, Blase and Guo, XianZheng and Tague, Patrick},
 title = {Smartauth: User-centered Authorization for the Internet of Things},
  crossref = {USENIXSEC2017},
 pages = {361--378},
   doi={},
}

@inProceedings{Liu2017,
 author = {Liu, Jian and Wang, Chen and Chen, Yingying and Saxena, Nitesh},
 title = {VibWrite: Towards Finger-input Authentication on Ubiquitous Surfaces via Physical Vibration},
  crossref = {ACMCCS2017},
  pages = {73--87},
   doi = {10.1145/3133956.3133964},
}

@inProceedings{Shirvanian2017,
 author = {Shirvanian, Maliheh and Saxena, Nitesh},
 title = {CCCP: Closed Caption Crypto Phones to Resist MITM Attacks, Human Errors and Click-Through},
  crossref = {ACMCCS2017},
 pages = {1329--1342},
   doi = {10.1145/3133956.3134013},
}

@inProceedings{Zhang2017,
 author = {Zhang, Linghan and Tan, Sheng and Yang, Jie},
 title = {Hearing Your Voice is Not Enough: An Articulatory Gesture Based Liveness Detection for Voice Authentication},
  crossref = {ACMCCS2017},
 pages = {57--71},
   doi = {10.1145/3133956.3133962},
}

@inProceedings{Chatterjee2017,
 author = {Chatterjee, Rahul and Woodage, Joanne and Pnueli, Yuval and Chowdhury, Anusha and Ristenpart, Thomas},
 title = {The TypTop System: Personalized Typo-Tolerant Password Checking},
  crossref = {ACMCCS2017},
 pages = {329--346},
 doi = {10.1145/3133956.3134000},
}

@inProceedings{Abu_salma2017,
author={Abu-Salma, Ruba and Sasse, M. Angela and Bonneau, Joseph and Danilova, Anastasia and Naiakshina, Alena and Smith, Matthew},
title={Obstacles to the Adoption of Secure Communication Tools},
  crossref = {IEEESSP2017},
pages={137-153},
 doi={10.1109/SP.2017.65},
}

@inProceedings{Murillo2018,
 author = {Murillo, Ambar and Kramm, Andreas and Schnorf, Sebastian and De Luca, Alexander},
 title = {"If I Press Delete, It's Gone": User Understanding of Online Data Deletion and Expiration},
  crossref = {SOUPS2018},
 pages = {329--339},
 doi={},
}

@inProceedings{Rashidi2018,
 author = {Rashidi, Yasmeen and Ahmed, Tousif and Patel, Felicia and Fath, Emily and Kapadia, Apu and Nippert-Eng, Christena and Su, Norman Makoto},
 title = {"You Don'T Want to Be the Next Meme": College Students' Workarounds to Manage Privacy in the Era of Pervasive Photography},
  crossref = {SOUPS2018},
 pages = {143--157},
 doi={},
}

@inProceedings{Sambasivan2018,
 author = {Sambasivan, Nithya and Checkley, Garen and Batool, Amna and Ahmed, Nova and Nemer, David and Gayt\'{a}n-Lugo, Laura Sanely and Matthews, Tara and Consolvo, Sunny and Churchil, Elizabeth},
 title = {``Privacy is Not for Me, It's for Those Rich Women": Performative Privacy Practices on Mobile Phones by Women in South Asia},
  crossref = {SOUPS2018},
 pages = {127--142},
 doi={},
}

@inProceedings{Habib2018,
  author = {Habib, Hana and Colnago, Jessica and Gopalakrishnan, Vidya and Pearman, Sarah and Thomas, Jeremy and Acquisti, Alessandro and Christin, Nicolas and Cranor, Lorrie Faith},
 title = {Away from Prying Eyes: Analyzing Usage and Understanding of Private Browsing},
  crossref = {SOUPS2018},
 pages = {159--175},
 doi={},
}

@inProceedings{Habib2018_2,
 author = {Habib, Hana and Emami-Naeini, Pardis and Devlin, Summer and Oates, Maggie and Swoopes, Chelse and Bauer, Lujo and Christin, Nicolas and Cranor, Lorrie Faith},
 title = {User Behaviors and Attitudes Under Password Expiration Policies},
  crossref = {SOUPS2018},
 pages = {13--30},
 doi={},
}

@inProceedings{Zou2018,
 author = {Zou, Yixin and Mhaidli, Abraham H. and McCall, Austin and Schaub, Florian},
 title = { ``I've Got Nothing to Lose": Consumers' Risk Perceptions and Protective Actions After the Equifax Data Breach},
  crossref = {SOUPS2018},
 pages = {197--216},
 doi={},
}

@inProceedings{Karunakaran2018,
author = {Karunakaran, Sowmya and Thomas, Kurt and Bursztein, Elie and Comanescu, Oxana},
title = {Data Breaches: User Comprehension, Expectations, and Concerns with Handling Exposed Data},
  crossref = {SOUPS2018},
pages = {217--234},
 doi={},
}

@inProceedings{Park2018,
author = {Park, Cheul Young and Faklaris, Cori and Zhao, Siyan and Sciuto, Alex and Dabbish, Laura and Hong, Jason},
title = {Share and Share Alike? An Exploration of Secure Behaviors in Romantic Relationships},
  crossref = {SOUPS2018},
pages = {83--102},
 doi={},
}

@inProceedings{Neupane2018,
 author = {Neupane, Ajaya and Satvat, Kiavash and Saxena, Nitesh and Stavrinos, Despina and Bishop, Haley Johnson},
 title = {Do Social Disorders Facilitate Social Engineering?: A Case Study of Autism and Phishing Attacks},
  crossref = {ACSAC2018},
 pages = {467--477},
 doi = {10.1145/3274694.3274730},
}

@inProceedings{Gao2018,
  author = {Gao, Xianyi and Yang, Yulong and Liu, Can and Mitropoulos, Christos and Lindqvist, Janne and Oulasvirta, Antti},
 title = {Forgetting of Passwords: Ecological Theory and Data},
  crossref = {USENIXSEC2018},
 pages = {221--238},
 doi = {-},
}

@inProceedings{Schwarz2018,
   author    = {Schwarz, Michael and Lipp, Moritz and Gruss, Daniel},
  title     = {JavaScript Zero: Real JavaScript and Zero Side-Channel Attacks},
  crossref = {NDSS2018},
 pages = {},
 doi = {},
}

@inProceedings{Colnago2018,
   author = {Colnago, Jessica and Devlin, Summer and Oates, Maggie and Swoopes, Chelse and Bauer, Lujo and Cranor, Lorrie and Christin, Nicolas},
 title = {``It's Not Actually That Horrible'': Exploring Adoption of Two-Factor Authentication at a University},
  crossref = {ACMCHI2018},
 pages = {456:1--456:11},
doi = {10.1145/3173574.3174030},
}

@inProceedings{Das2018,
 author = {Das, Sauvik and Lo, Joanne and Dabbish, Laura and Hong, Jason I.},
 title = {Breaking! A Typology of Security and Privacy News and How It's Shared},
  crossref = {ACMCHI2018},
 pages = {1:1--1:12},
 doi = {10.1145/3173574.3173575},
}

@inProceedings{Reeder2018,
 author = {Reeder, Robert W. and Felt, Adrienne Porter and Consolvo, Sunny and Malkin, Nathan and Thompson, Christopher and Egelman, Serge},
 title = {An Experience Sampling Study of User Reactions to Browser Warnings in the Field},
  crossref = {ACMCHI2018},
 pages = {512:1--512:13},
 doi = {10.1145/3173574.3174086},
}

@inProceedings{Oates2018,
 author = {Oates, Maggie and Ahmadullah, Yama and Marsh, Abigail and Swoopes, Chelse and Zhang, Shikun and Balebako, Rebecca and Cranor, Lorrie Faith},
  title = "Turtles, Locks, and Bathrooms: Understanding Mental Models of Privacy Through Illustration",
  crossref = {PETS2018},
 pages = {5--32},
 doi = {},
}

@inProceedings{Redmiles2018,
  author = {Redmiles, Elissa M. and Zhu, Ziyun and Kross, Sean and Kuchhal, Dhruv and Dumitras, Tudor and Mazurek, Michelle L.},
 title = {Asking for a Friend: Evaluating Response Biases in Security User Studies},
  crossref = {ACMCCS2018},
 pages = {1238--1255},
 doi = {10.1145/3243734.3243740},
}

@inProceedings{Lebeck2018,
  author    = {Lebeck, Kiron and Ruth, Kimberly and Kohno, Tadayoshi and Roesner, Franziska},
  title     = {Towards Security and Privacy for Multi-user Augmented Reality: Foundations with End Users},
  crossref = {IEEESSP2018},
 pages     = {392--408},
 doi       = {10.1109/SP.2018.00051},
}

@misc{TwenteThoeries,
  title = {Communication Theories},
  howpublished = {\url{https://www.utwente.nl/en/bms/communication-theories/}},
  year = {2003}
}

@inproceedings{HashimAT,
 author = {Hashim, Nor H. and Jones, M.},
 title = {Activity Theory: A framework for qualitative analysis},
 booktitle = {Proceedings of the 4th International Qualitative Research Convention (QRC)},
 year = {2007},
 location = {Malaysia},
 pages = {10:1--10:14},
 doi = {10.1145/1837110.1837124},
 publisher = {ACM},
}

@article{WeickSM,
author = {Weick, Karl E. and Sutcliffe, Kathleen M. and Obstfeld, David},
title = {Organizing and the Process of Sensemaking},
journal = {Organization Science},
volume = {16},
number = {4},
pages = {409-421},
year = {2005},
doi = {10.1287/orsc.1050.0133},
}

@book{Weick1995,
  title={Sensemaking in organizations},
  author={Weick, Karl E.},
  volume={3},
  year={1995},
  publisher={SAGE}
}

@article{maitlis2013sensemaking,
  title={Sensemaking and emotion in organizations},
  author={Maitlis, Sally and Vogus, Timothy J. and Lawrence, Thomas B},
  journal={Organizational Psychology Review},
  volume={3},
  number={3},
  pages={222--247},
  year={2013},
  publisher={SAGE}
}

@article{bird2007sensemaking,
  title={Sensemaking and identity: The interconnection of storytelling and networking in a women's group of a large corporation},
  author={Bird, Shelley},
  journal={The Journal of Business Communication},
  volume={44},
  number={4},
  pages={311--339},
  year={2007},
  publisher={SAGE}
}

@article{Herrmann2007,
  title={Stockholders in cyberspace: Weick’s sensemaking online},
  author={Herrmann, Andrew F.},
  journal={The Journal of Business Communication},
  volume={44},
  number={1},
  pages={13--35},
  year={2007},
  publisher={SAGE}
}

@article{fishbein1977TRA,
  title={Belief, attitude, intention, and behavior: An introduction to theory and research},
  author={Fishbein, Martin and Ajzen, Icek},
  journal={Journal of Business Venturing},
  year={1977}
}

@book{fishbein1980TRA,
  title={Understanding attitudes and predicting social behavior},
  author={Fishbein, Martin and Ajzen, Icek},
  year={1980},
  publisher={Englewood Cliffs, NJ: Prentice-Hall}
}

@Inbook{LaCaille2013,
author={LaCaille, Lara},
editor={Gellman, Marc D. and Turner, J. Rick},
title={Theory of Reasoned Action},
bookTitle={Encyclopedia of Behavioral Medicine},
year={2013},
publisher={Springer},
pages={1964--1967},
isbn={978-1-4419-1005-9},
doi={10.1007/978-1-4419-1005-9_1619},
}

@article{ajzen1991TPB,
  title={The Theory of Planned Behavior},
  author={Ajzen, Icek},
  journal={Organizational behavior and human decision processes},
  volume={50},
  number={2},
  pages={179--211},
  year={1991},
  publisher={Elsevier}
}

@article{chang1998tra,
  title={Predicting unethical behavior: a comparison of the theory of reasoned action and the theory of planned behavior},
  author={Chang, Man Kit},
  journal={Journal of business ethics},
  volume={17},
  number={16},
  pages={1825--1834},
  year={1998},
  publisher={Springer}
}

@article{kim2015tra,
  title={Norms in social media: The application of theory of reasoned action and personal norms in predicting interactions with Facebook page like ads},
  author={Kim, Soojung and Lee, Joonghwa and Yoon, Doyle},
  journal={Communication Research Reports},
  volume={32},
  number={4},
  pages={322--331},
  year={2015},
  publisher={TANDF}
}

@article{doane2016tra,
  title={Reducing cyberbullying: A theory of reasoned action-based video prevention program for college students},
  author={Doane, Ashley N. and Kelley, Michelle L and Pearson, Matthew R},
  journal={Aggressive Behavior},
  volume={42},
  number={2},
  pages={136--146},
  year={2016},
  publisher={Wiley Online Library}
}

@book{Hofstede1984cc,
  title={Culture's consequences: International differences in work-related values},
  author={Hofstede, Geert},
  volume={5},
  year={1984},
  publisher={SAGE}
}

@article{Hofstede2011,
  title={Dimensionalizing cultures: The Hofstede model in context},
  author={Hofstede, Geert},
  journal={Online readings in psychology and culture},
  volume={2},
  number={1},
  pages={8},
  year={2011},
  publisher={International Association for Cross-Cultural Psychology}
}

@inproceedings{Sample2017,
  title={Cultural exploration of attack vector preferences for self-identified attackers},
  author={Sample, Char and Cowley, Jennifer and Hutchinson, Steve},
  booktitle={Proceedings of the 11th International Conference on Research Challenges in Information Science (RCIS)},
  pages={305--314},
  year={2017},
  publisher={IEEE}
}

@inproceedings{Onumo2017,
  title={An Empirical Study of Cultural Dimensions and Cybersecurity Development},
  author={Onumo, Aristotle and Cullen, Andrea and Ullah-Awan, Irfan},
  booktitle={Proceedings of the 5th International Conference on Future Internet of Things and Cloud (FiCloud)},
  pages={70--76},
  year={2017},
  publisher={IEEE}
}

@article{al2015,
  title={An examination of factors that influence the number of information security policy violations in Qatari organizations},
  author={Al-Mukahal, Hasan M. and Alshare, Khaled},
  journal={Information \& Computer Security},
  volume={23},
  number={1},
  pages={102--118},
  year={2015},
  publisher={Emerald Group Publishing Limited}
}

@inproceedings{tapanainen2017,
  title={Mindfulness in Cyber Security-Examining Responder Behaviors in Cyber-Attacks},
  author={Tapanainen, Tommi J.},
  booktitle={Proceedings of the 23rd Americas Conference on Information Systems (AMCIS)},
  year={2017}
}

@article{kalkman2019,
  title={Sensemaking questions in crisis response teams},
  author={Kalkman, Jori Pascal},
  journal={Disaster Prevention and Management: An International Journal},
  year={2019},
  publisher={Emerald Publishing Limited}
}

@article{erbert2016,
  title={Organizational Sensemaking: Interpretations of Workplace “Strangeness”},
  author={Erbert, Larry A.},
  journal={International Journal of Business Communication},
  volume={53},
  number={3},
  pages={286--305},
  year={2016},
  publisher={SAGE}
}

@article{corbin1990,
  title={Grounded theory research: Procedures, canons, and evaluative criteria},
  author={Corbin, Juliet M. and Strauss, Anselm},
  journal={Qualitative sociology},
  volume={13},
  number={1},
  pages={3--21},
  year={1990},
  publisher={Springer}
}

@book{creswell2017,
  title={Research design: Qualitative, quantitative, and mixed methods approaches},
  author={Creswell, John W. and Creswell, J. David},
  year={2017},
  publisher={SAGE}
}

@article{johnson2001,
  title={Mental models and deduction},
  author={Johnson-Laird, Philip N.},
  journal={Trends in cognitive sciences},
  volume={5},
  number={10},
  pages={434--442},
  year={2001},
  publisher={Elsevier}
}

@inProceedings{czyz2016,
  author={Czyz, Jakub and Luckie, Matthew J. and Allman, Mark and Bailey, Michael},
 title={Don't Forget to Lock the Back Door! A Characterization of IPv6 Network Security Policy},
  crossref = {NDSS2016},
  pages = {},
 doi       = {10.14722/ndss.2016.23047},
}

@article{hofstede2003cultural,
  title={Cultural dimensions},
  author={Hofstede, Geert},
 journal={\url{www.geert-hofstede.com}},
  year={2003}
}

@article{silva2019,
  title={Algorithms, platforms, and ethnic bias},
  author={Silva, Selena and Kenney, Martin},
  journal={Communications of the ACM},
  volume={62},
  number={11},
  pages={37--39},
  year={2019},
  publisher={ACM}
}

@article{taylor1911,
  title={The principles of scientific management},
  author={Taylor, Frederick W.},
  journal={New York},
  volume={202},
  year={1911}
}

@article{dekker2016,
  title={Examining the asymptote in safety progress: a literature review},
  author={Dekker, Sidney and Pitzer, Corrie},
  journal={International Journal of Occupational Safety and Ergonomics},
  volume={22},
  number={1},
  pages={57--65},
  year={2016},
  publisher={TANDF}
}

@online{Addingtoolbox,
 author={Pupulidy, Ivan},
  title = {Understanding and Adding to the Investigation Toolbox},
  url = {https://www.safetydifferently.com/understanding-and-adding-to-the-investigation-toolbox/},
  year = {2017}
}

@book{dekker2019foundations,
  title={Foundations of Safety Science: A Century of Understanding Accidents and Disasters},
  author={Dekker, Sidney},
  year={2019},
  publisher={Routledge}
}

@article{swuste2010safety,
  title={Safety metaphors and theories, a review of the occupational safety literature of the US, UK and The Netherlands, till the first part of the 20th century},
  author={Swuste, Paul and van Gulijk, Coen and Zwaard, Walter},
  journal={Safety science},
  volume={48},
  number={8},
  pages={1000--1018},
  year={2010},
  publisher={Elsevier}
}

@book{burnham2010accident,
  title={Accident prone: a history of technology, psychology, and misfits of the machine age},
  author={Burnham, John},
  year={2010},
  publisher={University of Chicago Press}
}

@inproceedings{dekker2010system,
  title={In the system view of human factors, who is accountable for failure and success?},
  author={Dekker, Sidney},
  booktitle={Proceedings of the Human Factors and Ergonomics Society Europe Chapter Annual Meeting},
  year={2010}
}

@book{van1999risks,
  title={{Risks, Disasters and Management: A Comparative Study of Three Passenger Transport Systems}},
  author={van Poortvliet, Abraham},
  year={1999},
  publisher={Eburon}
}

@article{frederick2000blame,
  title={Blame the worker},
  author={Frederick, James and Lessin, Nancy},
  journal={Multinational Monitor},
  volume={21},
  number={11},
  pages={10},
  year={2000}
}

@article{guarnieri1992landmarks,
  title={Landmarks in the history of safety},
  author={Guarnieri, Michael},
  journal={Journal of Safety Research},
  volume={23},
  number={3},
  pages={151--158},
  year={1992},
  publisher={Elsevier}
}

@article{reason1998achieving,
  title={Achieving a safe culture: theory and practice},
  author={Reason, James},
  journal={Work \& Stress},
  volume={12},
  number={3},
  pages={293--306},
  year={1998},
  publisher={Taylor \& Francis}
}

@book{perrow1984normal,
  title={Normal accidents: living with high-risk technologies},
  publisher = {New York, NY, Basic Books},
  author={Perrow, Charles},
  year={1984}
}

@article{weick1987organizational,
  title={Organizational culture as a source of high reliability},
  author={Weick, Karl E.},
  journal={California management review},
  volume={29},
  number={2},
  pages={112--127},
  year={1987},
  publisher={SAGE}
}

@article{dekker2008resilience,
  title={Resilience Engineering: New directions for measuring and maintaining safety in complex systems},
  author={Dekker, Sidney and Hollnagel, Erik and Woods, David and Cook, Richard},
  journal={Lund University School of Aviation Technical Report},
  year={2008}
}

@book{dekker2018just,
  title={Just culture: restoring trust and accountability in your organization},
  author={Dekker, Sidney},
  year={2018},
  publisher={CRC Press}
}

@inproceedings{whitten1999johnny,
  title={Why Johnny Can't Encrypt: A Usability Evaluation of PGP 5.0.},
  author={Whitten, Alma and Tygar, J. Doug},
  booktitle={Proceedings of the 8th USENIX Security Symposium (USENIX Security)},
  volume={348},
  pages={169--184},
  year={1999},
  publisher = {USENIX},
}

@inProceedings{Krombholz2017,
  title    = {``I Have No Idea What I'm Doing''-On the Usability of Deploying HTTPS},
  author   = {Krombholz, Katharina and Mayer, Wilfried and Schmiedecker, Martin and Weippl, Edgar},
  crossref = {USENIXSEC2017},
  pages    = {1339--1356},
}

@inproceedings{vredenburg2002survey,
  title={A survey of user-centered design practice},
  author={Vredenburg, Karel and Mao, Ji-Ye and Smith, Paul W and Carey, Tom},
  crossref={ACMCHI2002},
  year={2002},
}

@article{von2004design,
  title={Design science in information systems research},
  author={Von Alan, R. Hevner and March, Salvatore T. and Park, Jinsoo and Ram, Sudha},
  journal={MIS quarterly},
  volume={28},
  number={1},
  pages={75--105},
  year={2004},
  publisher={Springer}
}

@article{hevner2007three,
  title={A three cycle view of design science research},
  author={Hevner, Alan R.},
  journal={Scandinavian journal of information systems},
  volume={19},
  number={2},
  pages={4},
  year={2007}
}

@article{march1995design,
  title={Design and natural science research on information technology},
  author={March, Salvatore T. and Smith, Gerald F.},
  journal={Decision support systems},
  volume={15},
  number={4},
  pages={251--266},
  year={1995},
  publisher={ELSEVIER}
}

@inproceedings{burnett2015encore,
  title={Encore: Lightweight measurement of web censorship with cross-origin requests},
  author={Burnett, Sam and Feamster, Nick},
  pages={653--667},
  crossref={ACMSIGCOMM2015},
}

@article{narayanan2015no,
  title={No encore for encore? ethical questions for web-based censorship measurement},
  author={Narayanan, Arvind and Zevenbergen, Bendert},
  journal={Ethical Questions for Web-Based Censorship Measurement},
  year={2015}
}

@article{van1989nothing,
  title={Nothing is quite so practical as a good theory},
  author={Van de Ven, Andrew H.},
  journal={Academy of management Review},
  volume={14},
  number={4},
  pages={486--489},
  year={1989},
  publisher={Academy of Management Briarcliff Manor, NY}
}

@article{gregor2006nature,
  title={The nature of theory in information systems},
  author={Gregor, Shirley},
  journal={Management Information Systems quarterly},
  pages={611--642},
  year={2006},
  publisher={JSTOR}
}

@inProceedings{mcdonald2019reliability,
  title={Reliability and Inter-rater Reliability in Qualitative Research: Norms and Guidelines for CSCW and HCI Practice},
  author={McDonald, Nora and Schoenebeck, Sarita and Forte, Andrea},
  booktitle={Proceedings of the 2019 ACM CHI Conference on Human Factors in Computing Systems (CHI)},
  volume={3},
  year={2019},
  publisher={ACM}
}

@inproceedings{golla2018site,
  title={What was that site doing with my Facebook password?: Designing Password-Reuse Notifications},
  author={Golla, Maximilian and Wei, Miranda and Hainline, Juliette and Filipe, Lydia and D{\"u}rmuth, Markus and Redmiles, Elissa M. and Ur, Blase},
  crossref = {ACMCCS2018},
  pages={1549--1566},
  doi = {10.1145/3243734.3243767}
}

@inproceedings{herley2017sok,
  title={Sok: Science, security and the elusive goal of security as a scientific pursuit},
  author={Herley, Cormac and Van Oorschot, Paul C},
  crossref={IEEESSP2017},
  pages={99--120},
}

@inproceedings{redmiles2019well,
  title={How well do my results generalize? comparing security and privacy survey results from mturk, web, and telephone samples},
  author={Redmiles, Elissa M. and Kross, Sean and Mazurek, Michelle L.},
  crossref={IEEESSP2019},
  pages={227--244},
}

@techreport{redmiles2017summary,
  title={A summary of survey methodology best practices for security and privacy researchers},
  author={Redmiles, Elissa M. and Acar, Yasemin and Fahl, Sascha and Mazurek, Michelle L.},
  year={2017}
}

@inproceedings{dekoven2019measuring,
  title={Measuring Security Practices and How They Impact Security},
  author={DeKoven, Louis F and Randall, Audrey and Mirian, Ariana and Akiwate, Gautam and Blume, Ansel and Saul, Lawrence K and Schulman, Aaron and Voelker, Geoffrey M and Savage, Stefan},
  booktitle={Proceedings of the 2019 Internet Measurement Conference (IMC)},
  pages={36--49},
  year={2019},
  organization={ACM}
}

@book{sekaran2016research,
  title={Research methods for business: A skill building approach},
  author={Sekaran, Uma and Bougie, Roger},
  year={2016},
  publisher={John Wiley \& Sons}
}

@misc{suddaby2006editors,
  title={From the editors: What grounded theory is not},
  author={Suddaby, Roy},
  year={2006},
  publisher={Academy of Management Briarcliff Manor, NY}
}

@incollection{fiebig2018learning,
  title={Learning from the Past: Designing Secure Network Protocols},
  author={Fiebig, Tobias and Lichtblau, Franziska and Streibelt, Florian and Kr{\"u}ger, Thorben and Lexis, Pieter and Bush, Randy and Feldmann, Anja},
  booktitle={Cybersecurity Best Practices},
  pages={585--613},
  year={2018},
  publisher={Springer}
}

@inproceedings{hamm2019systematic,
 author = {Hamm, Peter and Harborth, David and Pape, Sebastian},
 title = {A Systematic Analysis of User Evaluations in Security Research},
 booktitle = {Proceedings of the 14th International Conference on Availability, Reliability and Security (ARES)},
%  series = {ARES '19},
 year = {2019},
 isbn = {978-1-4503-7164-3},
 pages = {91:1--91:7},
 articleno = {91},
 numpages = {7},
 doi = {10.1145/3339252.3340339},
 publisher = {ACM},
}

@inproceedings{dodier2017paternalistic,
  title={From paternalistic to user-centred security: Putting users first with value-sensitive design},
  author={Dodier-Lazaro, Steve and Abu-Salma, R and Becker, I and Sasse, MA},
  booktitle={Proceedings of the ACM CHI Workshop on Values in Computing},
  year={2017},
}

@inproceedings{tahaei2019survey,
  title={A Survey on Developer-Centred Security},
  author={Tahaei, Mohammad and Vaniea, Kami},
  booktitle={Proceedings of the IEEE European Symposium on Security and Privacy Workshops (EuroS\&PW)},
  pages={129--138},
  year={2019},
}

@inproceedings{frik2019privacy,
  title={Privacy and security threat models and mitigation strategies of older adults},
  author={Frik, Alisa and Nurgalieva, Leysan and Bernd, Julia and Lee, Joyce and Schaub, Florian and Egelman, Serge},
  booktitle={Fifteenth Symposium on Usable Privacy and Security (SOUPS)},
  year={2019}
}

@book{liabook,
  title={The Practice of System and Network Administration: Volume 1: DevOps and other Best Practices for Enterprise IT (3rd Edition)},
  author={Thomas A. Limoncelli, Christina J. Hogan, Strata R. Chalup},
  year={2017},
  publisher={Addison-Wesley}
}

@article{henrich2010most,
  title={Most people are not WEIRD},
  author={Henrich, Joseph and Heine, Steven J and Norenzayan, Ara},
  journal={Nature},
  volume={466},
  number={7302},
  pages={29--29},
  year={2010},
  publisher={Nature Publishing Group}
}

@article{muthukrishna2019problem,
  title={A problem in theory},
  author={Muthukrishna, Michael and Henrich, Joseph},
  journal={Nature Human Behaviour},
  volume={3},
  number={3},
  pages={221--229},
  year={2019},
  publisher={Nature Publishing Group}
}

@techreport{dittrich2012menlo,
  title={The Menlo Report: Ethical principles guiding information and communication technology research},
  author={Dittrich, David and Kenneally, Erin and others},
  year={2012},
  institution={US Department of Homeland Security}
}

@misc{guillory2020,
author ={Devin Guillory},
title = {Combating Anti-Blackness in the AI Community},
url = {https://arxiv.org/abs/2006.16879},
% urldate = {2019-05-28},
    eprint={2006.16879},
    archivePrefix={arXiv},
    primaryClass={cs.CY},
year = {2020}
}

@article{zimmermann2019,
  title={Moving from a ‘human-as-problem” to a ‘human-as-solution” cybersecurity mindset},
  author={Zimmermann, Verena and Renaud, Karen},
  journal={International Journal of Human-Computer Studies},
  volume={131},
  pages={169--187},
  year={2019},
  publisher={Elsevier}
}
\balance

\end{document}